\documentclass[11pt,a4paper]{article}
\usepackage{amsmath,amssymb,bm,bbm}
\usepackage[utf8]{inputenc}
\usepackage{graphicx}
\usepackage{color}
\usepackage[dvipsnames]{xcolor}
\numberwithin{equation}{section}
\allowdisplaybreaks[3]
\usepackage{braket}
\usepackage{subcaption}
\usepackage[linktocpage=true]{hyperref}
\hypersetup{
    colorlinks=true,
    linkcolor=blue,
    citecolor=blue,
    pdfpagemode=FullScreen,
    }

\edef\restoreparindent{\parindent=\the\parindent\relax}
\usepackage{parskip}
\restoreparindent

\usepackage{tikz}
\usetikzlibrary{arrows,arrows.meta,intersections, calc,positioning,decorations.pathreplacing,decorations.pathmorphing,shapes}
\usetikzlibrary{patterns}
\usetikzlibrary{decorations.markings}
\usetikzlibrary{knots}

\def\d{{\rm d}}

\def\CD{{\cal D}}

\def\CF{{\cal F}}

\def\CI{{\cal I}}

\def\CN{{\cal N}}
\def\CO{{\cal O}}

\def\d{\mathrm{d}}

\def\bz{\bar{z}}

\def\Del{\Delta}
\def\bh{\bar{h}}
\def\hlambda{\hat{\lambda}}

\begin{document}
\begin{titlepage}

\renewcommand{\thefootnote}{\fnsymbol{footnote}}
\begin{flushright}
\begin{tabular}{l}
YITP-22-99\\
\end{tabular}
\end{flushright}

\vfill
\begin{center}

\noindent{\Large \textbf{ Late-time correlation functions in dS$_3$/CFT$_2$ correspondence}}

\vspace{1.5cm}

\noindent{Heng-Yu Chen,$^{a}$ Shi Chen$^{a}$ and Yasuaki Hikida$^b$}

\bigskip

\vskip .6 truecm

\centerline{\it $^a$Department of Physics, National Taiwan University, Taipei 10617, Taiwan }



\medskip

\centerline{\it $^b$Center for Gravitational Physics and Quantum Information, Yukawa Institute for Theoretical Physics, }
\centerline{\it Kyoto University, Kitashirakawa Oiwakecho, Sakyo-ku, Kyoto 606-8502, Japan}

\end{center}

\vfill
\vskip 0.5 truecm
\begin{abstract}

We compute the  late-time correlation functions on three-dimensional de Sitter spacetime for a higher-spin gravity theory. For this, we elaborate on the formulation to obtain the wave functional of universe from a dual conformal field theory, which is used to compute the late-time correlation functions. We argue that the relation to direct bulk Feynman diagram computations in the in-in formulation. We furthermore provide a precise prescription to construct a higher-spin dS$_3$ holography as an analytic continuation of Gaberdiel-Gopakumar duality for AdS$_3$. Part of results here were already reported in an earlier letter.
We explain the details of their derivations and extend the analysis to more generic cases in this paper. Previously, we have examined two- and three-point functions and a simple four-point correlator at the leading order in Newton constant. Here we also evaluate more complicated four-point correlators. Finally, we study late-time correlators in an alternative limit of dS$_3$/CFT$_2$ with critical level coset, such as, two-point correlator on conical defect geometry.
We also examine one-loop corrections to two-point correlator on dS$_3$.

\end{abstract}
\vfill
\vskip 0.5 truecm

\setcounter{footnote}{0}
\renewcommand{\thefootnote}{\arabic{footnote}}
\end{titlepage}

\newpage

\hrule
\tableofcontents

\bigskip
\hrule
\bigskip

\section{Introduction and summary}

Our universe experienced an inflationary era at the early time, during which its geometry can be approximated by de Sitter (dS) spacetime. In order to understand the beginning of our universe, it is important to formulate quantum gravity on dS spacetime, which has been understood very little so far. In the case of anti-de Sitter (AdS) spacetime, the situation is quite different mainly due to AdS/CFT correspondence proposed by Maldacena \cite{Maldacena:1997re,Gubser:1998bc,Witten:1998qj}. AdS/CFT correspondence is supposed to formulate quantum gravity on AdS spacetime by its boundary conformal field theory (CFT), and there have been a lot of developments since its discovery. It is thus natural to expect that an explicit formulation of dS/CFT correspondence  \cite{Strominger:2001pn,Witten:2001kn,Maldacena:2002vr} would be helpful for our purpose. 
Recently, the late-time correlation functions on three-dimensional dS (dS$_3$) have been analyzed by making use of the holography in \cite{Chen:2022ozy}. The late-time correlators on dS are expected to be useful to understand what happened during the inflation era  \cite{Maldacena:2002vr,Maldacena:2011nz,Arkani-Hamed:2015bza,Arkani-Hamed:2018kmz}. 
In \cite{Chen:2022ozy}, only partial results were reported and derivations were explained quite briefly. In this paper, we present our full results on late-time correlation functions on dS spacetime and explain our methods and calculations in more details.

Currently, dS/CFT correspondence is not as well understood compared with its AdS/CFT counterpart. One of the main reasons is that only very few concrete calculable examples have been constructed so far. In \cite{Anninos:2011ui} it was proposed that a higher-spin gravity on dS$_4$ is dual to a 3d Sp$(N)$ vector model. The proposal may be regarded as an ``analytic continuation'' of Klebanov-Polyakov duality \cite{Klebanov:2002ja} between higher-spin gravity on AdS$_4$ and 3d O$(N)$ vector model. For previous works, see, e.g., \cite{Sundborg:2000wp,Mikhailov:2002bp,Witten,Sezgin:2002rt}.
Recently, an explicit class of dS$_3$/CFT$_2$ correspondence was proposed in \cite{Hikida:2021ese,Hikida:2022ltr,Chen:2022ozy} between a (higher-spin) gravity on dS$_3$ and a certain 2d CFT with an imaginary central charge. In these papers, the relation to Gaberdiel-Gopakumar duality \cite{Gaberdiel:2010pz} involving a higher-spin gravity on AdS$_3$ was also clarified. The higher-spin theory is given explicitly by 3d Prokushkin-Vasiliev theory \cite{Prokushkin:1998bq}, which includes higher-spin gauge fields with spin $s=2,3,\ldots$ and two complex scalars with mass:
\begin{align}
\ell^2_\text{AdS} m^2 = -1 + \hat \lambda^2 \, . \label{PVmass}
\end{align}
Here $\ell_\text{AdS}$ denotes the AdS radius and a dimensionless parameter $\hat{\lambda}$ is introduced. The dual 2d CFT is supposed to be a coset
\begin{align}
    \frac{\text{SU}(N)_k \times \text{SU}(N)_1}{\text{SU}(N)_{k+1}} \,  \label{cosetCFT}
\end{align}
and the classical limit of gravity theory corresponds to taking the 't Hooft limit of the coset:
\begin{align}
 N, k \to \infty, \quad   \lambda = \frac{N}{N+k} \to \text{fixed} \, . \label{tHooft}
\end{align}
The 't Hooft parameter $\lambda$ is identified with  $\hat \lambda$ appeared in the dual gravity theory. In the bulk, we map from AdS$_3$ to dS$_3$ mainly by replacing $\ell_\text{AdS}$ by $- i \ell$ with $\ell$ being the dS radius. It was shown that the asymptotic symmetry near the future infinity is given by a W-algebra with an imaginary central charge \cite{Ouyang:2011fs} (see also \cite{Strominger:2001pn,Maldacena:2002vr}). 
We thus identify the dual CFT with fixed $\lambda$ but taking an analytic continuation such that the central charge becomes purely imaginary as $- i c^{(g)}$ with real $c^{(g)}$. Making use of the simplicity and explicitness of the lower-dimensional theories, we could investigate various properties of dS/CFT correspondence.

In AdS/CFT correspondence, the scattering amplitudes or more strictly transition amplitudes of gravity theory between the boundaries of AdS$_{d+1}$ can be obtained from the correlation functions of its dual CFT \cite{Gubser:1998bc,Witten:1998qj}. It was similarly proposed that the correlation functions of fields in a gravity theory located at the points on its late time boundary of dS$_{d+1}$ can be computed from the correlation functions of the dual CFT \cite{Maldacena:2002vr}.
For this, we prepare the wave functional of the universe by integrating over bulk fields $\psi_j$ (including a metric $g$) subject to boundary conditions $\psi_j = \psi_j^0$ (in particular, $g =h$) at late time $t = t_\infty$ as 
\begin{align}
\Psi [\psi_j^0 ] = \int \mathcal{D} \psi_j \exp \left(i S[\psi_j] \right)
\end{align}
where  $S[\psi_j]$ is the action of the gravity theory.
The proposal by \cite{Maldacena:2002vr} is that the wave functional can be computed from certain dual CFT as
\begin{align}
\Psi [\psi_j^0 ] = \left \langle \exp \left( \ell^d \int d^d \vec x \psi_j^0 \mathcal{O}^j \right) \right\rangle \, . \label{wfmap}
\end{align}
Here $\mathcal{O}^j$ are CFT operators dual to $\psi_j$. 
The correlation functions of $\psi_j^0$ are then computed as expectation values as
\begin{align}
\langle\, \psi_1^0 (\vec x_1)\cdots \psi_{m}^0 (\vec x_m)\, \rangle = \int [\CD \psi_j^0]\, \left|\Psi \left[\psi_l^0\right]\right|^2\,  \psi_1^0 (\vec x_1) \cdots \psi_m^0 (\vec x_m) \, . \label{expectationvalue}
\end{align}
In our previous work \cite{Chen:2022ozy}, we briefly explained how to compute late-time correlation functions of bulk theory from an analytic continuation of the coset CFT \eqref{cosetCFT}. We also provided final results on two- and three-point correlators. Furthermore, we examined a simple four-point correlator. In this paper, we explain in more details our methods and how to derive these results. We consider more complicated four-point correlators as well. 

Some earlier works on the bulk late-time correlators have been made in the in-in formalism and relation to bulk correlators on AdS backgrounds have been revealed  \cite{Sleight:2020obc,Sleight:2021plv} (see also \cite{DiPietro:2021sjt}). 
In particular, they found certain phase factors should be associated when corresponding Feynman diagrams for an identical theory on AdS and dS backgrounds are compared. We provide arguments on the relation to our results. For instance, we obtain the phase factors associated with the correlators, which reproduce the same phase factors up to a subtlety associate with massless limit of higher-spin fields. We see that the two formulations are consistent with each other. As an advantage, our formalism would allow us to compute non-trivial correlators, which are difficult to compute directly from the bulk theory due to, e.g., the lack of knowledge of quantum aspects on the (higher-spin) gravity theory.

\subsection{Future directions}

Our formalism is generic enough to apply to other cases. In order to illustrate this, we compute some interesting examples, which could be useful to reveal important properties of dS/CFT correspondence.
There are several open questions we would like to address in near future. In this paper, we only explain how to apply our method to compute bulk correlators related to these questions, and we postpone detailed analysis based on the correlators to separate publications.

It is important to consider bulk late-time correlations on more generic asymptotic dS$_3$ geometry. In \cite{Hikida:2021ese,Hikida:2022ltr}, dS$_3$ conical geometry was constructed based on the analysis for AdS$_3$ case in \cite{Castro:2011iw}.
In this paper, we compute bulk late-time correlations of scalar fields on dS$_3$ conical geometry, though its bulk interpretation will be examined elsewhere.
Generically, it is important problem to read off  the properties of the asymptotic dS$_3$ geometry from late-time two-point functions of bulk scalar field.
In particular, we are interested in the properties of higher-spin black hole with asymptotic dS$_3$ condition, see, e.g., \cite{Krishnan:2013zya,Emparan:2022ijy}.
For asymptotic AdS$_3$ case, refer, e.g., \cite{Gutperle:2011kf,Ammon:2012wc}.

Throughout this paper, we mainly consider the leading order effects in the Newton constant $G_N$. It is important to systematically analyze higher order gravitational effects, and it is exciting if we can deal with even finite $G_N$ effects, see, e.g., \cite{Goodhew:2020hob,DiPietro:2021sjt,Hogervorst:2021uvp}.%
\footnote{In particular, it is known that bulk unitarity strongly constrain bulk correlators at late time or wave functional of universe. It is an important problem to improve our analysis by applying these results.}
In order to show that our method can be applied to higher orders in $G_N$, we examine a type of one-loop corrections to two-point correlator on dS$_3$ at late time, though the full detailed analysis would be presented elsewhere. We analyze the expansions of four-point correlators up to the next non-trivial order as well.

There are other examples, which will not be addressed within this paper.
For instance, the original Gaberdiel-Gopakumar duality has been extended to include supersymmetry or matrix-valued fields in order to see the relation to superstring theory. Concretely, $\mathcal{N}=3$ supersymmetry with matrix-valued extension was considered in \cite{Creutzig:2011fe,Creutzig:2013tja,Creutzig:2014ula}. The $\mathcal{N}=4$ supersymmetry and its relation to strings on $\text{AdS}_3 \times \text{S}^3 \times \text{T}^4$ was analyzed in \cite{Gaberdiel:2013vva,Gaberdiel:2014cha}. It is interesting to consider an analytic continuation of these extended cases.
It is also worthwhile extending the current analysis to the higher dimensional example of \cite{Anninos:2011ui}.

\subsection{Organization}

This paper is organized as follows.
In the next section, we explain the generic method to compute late-time correlators on dS$_{d+1}$ by using the wave functional of universe as in \eqref{expectationvalue}. In particular, we examine what kind of phase factors arise from the map from AdS$_{d+1}$ to dS$_{d+1}$ for operators dual to bulk fields. We also compare our analysis to generic arguments with bulk Feynman diagrams and the in-in formulation \cite{Sleight:2020obc,Sleight:2021plv}. 
In section \ref{sec:hsdscft}, we introduce our version of dS$_3$/CFT$_2$ correspondence by providing a prescription to perform an analytic continuation of Gaberdiel-Gopakumar duality in \cite{Gaberdiel:2010pz}. We also compute two- and three-point correlators on dS$_3$ at late time.
In section \ref{sec:4pt}, which contains some of the main results in this paper, we present the computations of four-point correlators on dS$_3$ at late time.
In a previous paper \cite{Chen:2022ozy}, a simple four-point correlator has been analyzed. Here we explain the computation in more details and generalize the analysis to more complicated four-point correlators.
In section \ref{sec:kto-2}, we consider alternative limit of dS$_3$/CFT$_2$ correspondence, which was introduced in \cite{Hikida:2021ese,Hikida:2022ltr}. With the setup, we deal with two-point correlators on dS$_3$ conical defect.
In section \ref{Sec:dS loop}, we examine the higher order corrections to bulk correlators in $G_N$, such as, one-loop corrections on two-point function on dS$_3$. We also analyze the structure of $1/c^2$-expansions of four-point functions of scalar operators in the dual CFT.
In appendix \ref{app:cb}, we collect detailed analysis on conformal block and partial wave expansions.
In appendix \ref{sec:bulk}, we review the embedding formalism for AdS$_{d+1}$ and summarize some bulk analyses specific to three dimensions, including holographic dual configurations for three-point functions and conformal blocks involving holomorphic conserved tensors.

\section{General strategy}

The main purpose of this work is to compute correlation functions of higher-spin gravity on dS$_3$ at late time from dual CFT$_2$. For this, we utilize the wave functional prescription in \cite{Maldacena:2002vr}, which will be explained to some details in this section.
Before focusing on an explicit setting, we consider a bulk theory defined on $(d+1)$-dimensional dS spacetime and including massive scalar fields and (massive) symmetric tensor fields in order to make argument as generic as possible.
We assume that dual CFT exists as well.

\subsection{Analytic continuation from \texorpdfstring{AdS$_{d+1}$}{AdSd+1} to \texorpdfstring{dS$_{d+1}$}{dSd+1}}
\label{sec:ST}

We start from the physical quantities on Euclidean AdS space and then discuss the map to those on Lorentzian dS spacetime.
In the bulk theory on Euclidean AdS$_{d+1}$, we include at least one scalar field $\phi^\text{AdS}$ and symmetric tensor fields $\sigma_{i_1 \cdots i_s}^\text{AdS}$.
We denote the dual scalar operator by $\mathcal{O}_\text{AdS}$ and the dual symmetric traceless tensor by $J^{i_1 \cdots  i_s}_\text{AdS}$. We denote the mass of a bulk field as $m$, then the conformal dimension of dual operator is $\Delta = \Delta_\pm$ depending on the boundary condition assigned to the given bulk field.
Here $\Delta_\pm$ satisfy
\begin{align}
\ell_\text{AdS}^2\, m^2 = - (\Delta_+ \Delta_- + s) \, , \qquad \Delta_- = d -  \Delta_+ 
\end{align}
with the AdS radius $\ell_\text{AdS}$.
We represent the spin of the field by $s$ and set $s=0$ for the scalar field. 
We will later consider the case where the dual symmetric tensor currents are conserved, and in that case $\Delta_+ = s + d -2$. For the metric of Euclidean AdS$_{d+1}$, we adopt the Poincar\'e coordinates as
\begin{align}
d s^2  = \ell_\text{AdS}^2\, \frac{d y^2 + d \vec{x} ^2}{y^2} \label{AdSmetric}
\end{align}
with $y \geq 0$. The boundary of AdS space is located at $y=0$.
The bulk fields on AdS$_{d+1}$ behave near the boundary $y \to +  0$  as
\begin{align}
    \begin{aligned}
        \phi ^\text{AdS}(y,\vec x) &\sim \phi _+^\text{AdS} (\vec x)\, y ^{ \Delta_+ } + \phi _-^\text{AdS} (\vec x)\, y ^{ \Delta_- }   \, , \\
        \sigma_{i_1 \cdots i_s} ^\text{AdS}(y, \vec x) &\sim \sigma_{i_1 \cdots  i_s}^{+,\text{AdS}} (\vec x)\, y ^{\Delta_+ - s} + \sigma_{i_1 \cdots  i_s}^{-,\text{AdS}} (\vec x)\, y ^{ \Delta_- - s}\ .
    \end{aligned}
\end{align}
It is convenient to define the $d$-dimensional boundary metric as
\begin{align}
    d s^2 = d \vec x^2 \, , \label{bmetric}
\end{align}
such as to be independent of the AdS radius $\ell_\text{AdS}$.
With the boundary metric above, the boundary couplings between the bulk fields and the CFT operators are given by
\begin{align}
\ell^d_\text{AdS} \int d^d \vec x\, \phi^\text{AdS}_\pm\,\mathcal{O}^\pm_\text{AdS} \, , \quad \ell^d_\text{AdS} \int d^d \vec x\, \sigma^{\pm,\text{AdS}}_{i_1 \cdots i_s}\, J^{i_1 \cdots i_s}_{\pm,\text{AdS}} \, . \label{coupling}
\end{align}

We then consider the bulk theory on Lorentzian dS spacetime including at least a scalar field $\phi$ and symmetric tensor fields $\sigma_{i_1 \cdots i_s}$ as in the case of Euclidean AdS space. We also denote the dual scalar operator by $\mathcal{O}$ and the dual symmetric tensor operators by $J^{i_1 \cdots i_s}$.
The conformal dimension of dual operator is given by:
\begin{align}
\ell^2\, m^2 = \Delta_+ \Delta_- + s  \, , \qquad \Delta_- = d -  \Delta_+ \, ,
\end{align}
where $m$ corresponds to the mass of bulk field.
We express the Poincar\'e patch of Lorentzian dS$_{d+1}$ with the metric
\begin{align}
 d s^2  = \ell ^2\,\frac{- d \eta^2 + d \vec{x} ^2}{\eta^2} \, , \label{dSmetric}
\end{align}
where we consider the region with $- \infty < \eta  \leq 0$.
The future infinity is located at $\eta \to -0$.
We may relate the dS metric to the AdS one given in \eqref{AdSmetric} by:
\begin{align}
y =  i \eta \ , \qquad \ell_\text{AdS} = - i \ell \, . \label{map}
\end{align}
In the following, we will see how such an analytic continuation acts on bulk fields and CFT operators.

We begin with the simpler case with scalar field and then move to more complicated case with symmetric tensor fields.
Near the future infinity $\eta \to - 0$, the bulk scalar field behaves as
\begin{align}
&\phi (\eta,\vec x) \sim \phi_+ (\vec x)\, ( - \eta )^{ \Delta_+ } + \phi_- (\vec x)\, ( - \eta) ^{ \Delta_- }   \, . \label{sasym}
\end{align}
As mentioned in \cite{Maldacena:2002vr}, $\phi^\text{AdS}$ may be identified with $\phi$ under the map \eqref{map}, which leads to
\begin{align}
\phi_\pm = e^{-i\frac{\pi}{2}\Delta_\pm}\,  \phi_\pm^\text{AdS} \, . \label{sAdStodS}
\end{align}
In AdS/CFT correspondence, the boundary couplings between bulk fields and CFT operators are given by \eqref{coupling}. On the other hand, in dS/CFT correspondence, we have the boundary coupling as in \eqref{wfmap} instead. The two types of boundary couplings can be identified if we relate the corresponding dual CFT operators as:
\begin{align}
 \mathcal{O}_\pm = e^{i \frac{\pi}{2}(\Delta_\pm - d)}\, \mathcal{O}_\pm^\text{AdS} \, . \label{sphase}
\end{align}
In the case of spin-$s$ tensor fields, their behaviors near the future infinity are given by
\begin{align}
    &\sigma_{i_1 \cdots i_s} (\eta, \vec x) \sim \sigma^+_{i_1 \cdots  i_s}  (\vec x)\, ( -\eta ) ^{\Delta_+ - s} + \sigma^-_{i_1 \cdots  i_s}  (\vec x)\, ( -\eta) ^{ \Delta_- - s}  \, .
\end{align}
Here notice that the indices of bulk fields are  lowered and raised by the bulk metric $g_{\mu \nu} \sim \ell^{2}$ and $g^{\mu \nu} \sim \ell^{-2}$. On the other hand, the boundary metric is defined as in \eqref{bmetric}, which is independent of $\ell$. 
It is thus convenient to define boundary operators, whose indices are raised and lowered by the boundary metric independent of $\ell$, see, e.g., \cite{Anninos:2011ui} for a similar argument.
Therefore, we assign
\begin{align}
    \ell^{s}\, \sigma^\pm_{i_1 \cdots  i_s}  (\vec x) = ( - i)^{ \Delta_\pm - s }\, \ell_\text{AdS}^{ s}\, \sigma^{\pm,\text{AdS}}_{i_1 \cdots  i_s}  (\vec x)  \, ,
 \end{align}
which leads to
\begin{align}
    \sigma^{\pm}_{i_1 \cdots  i_s}  (\vec x) = e^{-i\frac{\pi}{2}\Delta_\pm}\, \sigma^{\pm , \text{AdS}}_{i_1 \cdots  i_s}  (\vec x)  \, , \qquad
    J^\pm_{i_1 \cdots  i_s}  (\vec x) = e^{i \frac{\pi}{2}(\Delta_\pm - d)}\, J^{\pm , \text{AdS}}_{i_1 \cdots  i_s}  (\vec x) \, .
\end{align}
For instance, a spin-$s$ conserved current have conformal dimension $\Delta_+ = s + d -2$, and hence symmetric tensor currents are related as
\begin{align}
    J^+_{i_1 \cdots  i_s}  (\vec x) = e^{i\frac{\pi}{2} (s-2)}\, J^{+ , \text{AdS}}_{i_1 \cdots  i_s}  (\vec x) \, . \label{jreld}
\end{align}
In particular, the energy momentum tensor with $s=2$ does not receive any phase factor and the standard convention of energy momentum tensor is preserved by this prescription.

\subsection{Late-time correlators from the wave functional of universe}
\label{sec:maldacena}

In the previous subsection, we obtained the phase factors for operators dual to bulk fields by making use of the map from AdS to dS as in \eqref{map}. In this subsection, we explain how to compute correlation functions of the bulk fields at late time from these dual operators, see, e.g., appendix A of \cite{Arkani-Hamed:2018kmz} as well. As explained in the introduction, we utilize the wave functional of universe to compute the correlation functions on dS at late time. The proposal of \cite{Maldacena:2002vr} is that the wave functional can be evaluated from a certain dual CFT as in \eqref{wfmap}.
In AdS/CFT correspondence, the boundary values of bulk fields are fixed, and they can be regarded as the sources in dual CFT \cite{Gubser:1998bc,Witten:1998qj}.
However, in dS/CFT correspondence, boundary values $\psi_j^0$ at late time $t = t_\infty$ are not constants but path-integral variables. Therefore, it is natural to expand the wave functional by the boundary fields $\psi^0_j$ as:
\begin{align}
\begin{aligned}
\Psi [\psi_l^0] &= \exp \left[  \sum_{m  \geq 2} \int d^d \vec x_1 \cdots d^d \vec x_{m}   C_{m} (\{ \vec x_l\}) \prod_{j=1}^m \psi_j^0 (\vec x_j) \right]\, . \label{exp}
\end{aligned}
\end{align}
Thus, it is better to state that the coefficient functions $C_{m} (\{ \vec x_l\})$ could be evaluated by the dual CFT living on the late boundary of dS. 
{ We will specify the precise relation between $C_{m} (\{ \vec x_l\})$ and the $m$-point CFT correlation function in the concrete example considered later.}
The correlation functions of $\psi_j^0$ are computed as expectation values as in \eqref{expectationvalue}.

There are a lot of computations of late-time correlation functions of bulk theory such as in \cite{Maldacena:2002vr,Maldacena:2011nz,Arkani-Hamed:2015bza,Arkani-Hamed:2018kmz}, see also \cite{Shukla:2016bnu}. They are however analyzed usually in momentum basis. This is because their prime interest is to deduce the information after the inflation ends, where scale invariance is broken and only translation invariance remains, {thus the momentum representation is useful.} Generically, CFT correlation functions in momentum basis are much more difficult to treat than those in coordinate basis. In order to avoid unnecessary complication, we mainly work in coordinate basis in this paper. If one wishes, the expressions obtained here could be mapped to those in momentum basis by performing Fourier transformations in principle.

In the following sections, we explicitly compute bulk correlation functions on dS$_3$ at late time by applying the method explained above. We will systematically consider the order-by-order expansion of these correlation functions in the Newton constant $G_N$ (see \eqref{cs-charge} below for the relation with other parameters).
In this subsection, we concentrate on the leading order contributions.  We will discuss the higher order effects in subsection \ref{sec:higherorder}.
At the leading non-trivial order in $G_N$, bulk correlators can be expressed by CFT correlators as follows.
The two-point function of scalar field is obtained through functional inverse as:%
\footnote{The overall minus sign in front of the two point function comes from the convergence of momentum Gaussian path integral, while the sign of the ${\rm Re}\langle\dots\rangle$ come from the branch choice of the analytic continuation in \eqref{sasym}. \label{ft:sign}}
\begin{align}
        &\langle\,  \phi  (\vec x_1)\, \phi  (\vec x_2)\, \rangle 
            =
            -\frac{1}{2\, \text{Re}\,  \langle\, \mathcal{O}_i (\vec x_1)\, \mathcal{O}_i  (\vec x_2) \, \rangle} \, . \label{ms2pt}
\end{align}
For symmetric tensor fields, they are written as 
\begin{align}
       & \langle\,  \sigma_{i_1  \cdots i_s} (\vec x_1)\, \sigma^{j_1 \cdots j_s} (\vec x_2)\, \rangle
            =
            - \frac{\Pi^{j_1 \cdots j_s}_{i_1 \cdots i_s} (\hat x_{12})}{2\, \text{Re} \, \langle\, J_s (\vec x_1)\, J_s (\vec x_2)\,  \rangle} \, .  \label{mt2pt}
\end{align}
Here we have used the expression
\begin{align}
    \langle\,  J_{i_1  \cdots i_s} (\vec x_1)\, J^{j_1 \cdots j_s} (\vec x_2)\, \rangle =  \Pi^{j_1 \cdots j_s}_{i_1 \cdots i_s} (\hat x_{12})\,  \langle\, J_s (\vec x_1)\, J_s (\vec x_2)\,  \rangle
\end{align}
with the unit vector $\hat x = \vec x /|\vec x|$ and the bi-tensorial projector into symmetric traceless tensor $\Pi^{j_1 \cdots j_s}_{i_1 \cdots i_s} (\hat x) $, which captures the index structure. Here we have also set $\vec x_{12} = \vec x_1 - \vec x_2$.
We should take care of defining the inverse of two-point functions in coordinate basis, which will be explicitly given later.
Similarly, the three-point function we will use is
\begin{align}
\begin{aligned}
   & \langle\, \phi_1 (\vec x_1)\, \phi_2  (\vec x_2)\, \sigma_{i_1 \cdots i_s} ( \vec x_3)\, \rangle \\
    & \quad
        = \int\prod_{l=1}^3 \d^d \vec x_l '
        \frac{2\, \text{Re}\,\langle\,  \mathcal{O}_1 (\vec x_1 ')\, \mathcal{O}_2 (\vec x_2 ')\, J_{j_1 \cdots j_s} (\vec x_3 ')\,  \rangle \, \Pi^{j_1 \cdots j_s}_{i_1 \cdots i_s} (\vec x_{33'}) } { \left[ \prod_{i=1}^2(-2\,   \text{Re}\,  \langle\, \mathcal{O}_i (\vec x_i)\, \mathcal{O}_i (\vec x_i ')\,  \rangle ) \right](-2\, \text{Re}\, \langle\, J_s (\vec x_3)\, J_s (\vec x_3 ')\,  \rangle )} \ . \label{m3pt}
\end{aligned}
\end{align}
We also consider scalar four-point function, which can be decomposed into:
\begin{align} \label{4pt}
    \langle\, \phi_1 (\vec x_1)\, \phi_2  (\vec x_2) \, \phi_3 (\vec x_3)\, \phi_4 (\vec x_4)\, \rangle 
        =\int\prod_{l=1}^4 d^d \vec x_l '
        \frac{ \langle \mathcal{O}^4 \rangle_c + \langle \mathcal{O} ^4\rangle _d } {  \prod_{i=1}^4 (-2\,   \text{Re} \, \langle\, \mathcal{O}_i (\vec x_i)\, \mathcal{O}_i (\vec x_i ') \, \rangle ) } \ .
\end{align}
Here we have split the contribution in the following:
\begin{align}
   \langle\, \mathcal{O}^4 \,\rangle_c 
        &=
        2\, \text{Re}\, \langle\,  \mathcal{O}_1 (\vec x_1 ')\, \mathcal{O}_2 (\vec x_2 ') \,  \mathcal{O}_3 (\vec x_3 ')\, \mathcal{O}_4 (\vec x_4 ')\, \rangle_{c} \, , \label{O4c}\\
    \langle\, \mathcal{O}^4 \,\rangle_d \label{O4d} 
        &=\sum \int d^d \vec y_1 d^d  \vec y_2
        \\ &  \times
        \frac{2\text{Re}\,\langle \, \mathcal{O}_1 (\vec x_1 ') \, \mathcal{O}_2 (\vec x_2 ') \, J_{j_1 \cdots j_s} (\vec y_1) \,  \rangle\,  \Pi^{j_1 \cdots j_s}_{i_1 \cdots i_s} (\vec y_{12}) \, 2\text{Re}\,\langle \, J^{i_1 \cdots i_s} (\vec y_2 ) \, \mathcal{O}_3 (\vec x_3 ') \, \mathcal{O}_4 (\vec x_4 ') \, \rangle  } {  2\text{Re} \, \langle\, J_s (\vec y_1) \, J_s ( \vec y_2) \, \rangle }  + \text{perm}\, . \nonumber
\end{align}
The term $\langle \, \mathcal{O} ^4 \,\rangle  _c$, which is manifestly crossing invariant without summing up different channels, corresponds to the contributions exclusively from the four-point bulk contact diagrams. 
On the other hand for $\langle \, \mathcal{O} ^4 \,\rangle  _d$, we have to sum up all intermediate exchange bulk fields for all channels.
It should be noted that only appropriate real combinations appear here as we take the product of the wave functional $\Psi[\psi_l^0]$ and its complex conjugate $\bar{\Psi}[\psi_l^0]$ when extracting bulk correlation functions.

Let us remark here on a subtlety to define the coefficient functions $C_m (\{\vec{x_l}\})$ by an analytic continuation from correlation functions of CFT$_d$ dual to gravity theory on AdS$_{d+1}$. Here we assume that the CFT has a large $N$ structure.
We consider a four-point function of scalar operators 
\begin{align}
 \mathcal{A}_4 \equiv  \left \langle  \mathcal{O}_1^\text{AdS} (\vec x_1 ) \mathcal{O}_2^\text{AdS} (\vec x_2 )\mathcal{O}_3^\text{AdS} (\vec x_3 )\mathcal{O}_4^\text{AdS} (\vec x_4 ) \right \rangle \, .
\end{align}
Its generalization to other correlation functions is straightforward.
We expand the four-point function by $1/N$ as
\begin{align}
    \mathcal{A}_4 = \mathcal{A}^{(0)}_4 + \mathcal{A}^{(1)}_4 + \cdots \, .
\end{align}
The leading order contribution comes from the products of two-point functions
\begin{align}
       \mathcal{A}^{(0)}_4  & = \left \langle  \mathcal{O}_1^\text{AdS} (\vec x_1 )  \mathcal{O}_2^\text{AdS} (\vec x_2 )  \right \rangle  \left \langle  \mathcal{O}_3^\text{AdS} (\vec x_3 )  \mathcal{O}_4^\text{AdS} (\vec x_4 )  \right \rangle   + \text{perm} \, .
\end{align}
This is a kind of definition of large $N$ structure. We would like to read off coefficient functions $C_4(\{\vec x_l\})$ in \eqref{exp} from the computation of four-point function $\mathcal{A}_4$. However, the bulk correlators are computed via \eqref{expectationvalue}, and the products of two-point functions can already be obtained from two-point coefficients $C_2(\{\vec x_l\})$ in \eqref{exp}.
Therefore, in order to obtain $C_4(\{\vec x_l\})$ from an analytic continuation of $\mathcal{A}_4$, the leading order contribution $\mathcal{A}_N^{(0)}$ should be extracted beforehand. The next leading order can decomposed as
\begin{align}
    \mathcal{A}^{(1)}_4  =  \mathcal{A}^{(1)}_{4,c}   +  \mathcal{A}^{(1)}_{4,d} \, .
    \label{decA1}
\end{align}
Here we define $\mathcal{A}^{(1)}_{4,d}$ by
\begin{align} \label{A1d}
  \mathcal{A}^{(1)}_{4,d}
        & =\sum \int d^d \vec y_1 d^d  \vec y_2 
        \\ &   \times
        \frac{  \langle \, \mathcal{O}_1 ^\text{AdS}(\vec x_1) \, \mathcal{O}_2 ^\text{AdS}(\vec x_2) \, J^\text{AdS}_{j_1 \cdots j_s} (\vec y_1) \,  \rangle\, \Pi^{j_1 \cdots j_s}_{i_1 \cdots i_s} (\vec y_{12})\langle \, J^{i_1 \cdots i_s}_\text{AdS} (\vec y_2 ) \, \mathcal{O}_3 ^\text{AdS}(\vec x_3) \, \mathcal{O}_4^\text{AdS} (\vec x_4) \, \rangle  } {   \langle\, J_s^\text{AdS} (\vec y_1) \, J_s^\text{AdS} ( \vec y_2) \, \rangle }  + \text{perm.}\, ,
\nonumber
\end{align}
where the sum is over all intermediate operators. As expressed in \eqref{O4d}, $\langle\, \mathcal{O}^4 \,\rangle_d$ can be computed only with two- and three-point coefficient functions. Therefore, the four-point coefficient functions $C_4(\{\vec x_l\})$ should be obtained by an analytic continuation of the rest $\mathcal{A}_{4,c}^{(1)}$.

\begin{figure}
  \centering
  \includegraphics[width=13cm]{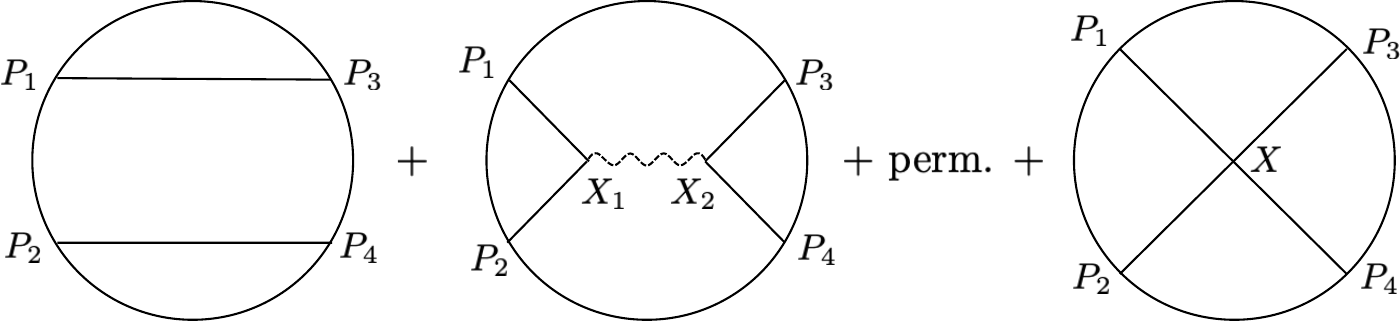}
 \caption{Bulk interpretation for the decompositions of four-point functions.}
  \label{fig:Witten4pts}
\end{figure}
From the bulk viewpoints, the above procedure may be explained as follows. As seen in fig.~\ref{fig:Witten4pts}, the four-point function can be computed by the sum of Witten diagrams. A type of diagrams consists of the product of two bulk propagators, which correspond to the product of two-point functions. Another type consists of the exchange diagrams of spin-$s$ fields, which roughly correspond to the conformal partial waves of spin-$s$ current exchange. The rest is the four-point contact diagram, which basically corresponds to the term we just extracted. The precise relation can be read off from the decomposition of conformal partial waves for bulk exchange diagrams as explained in appendix \ref{app:embedding}.

\subsection{Relation to the in-in formulation}
\label{sec:formalisms}

In the previous subsection, we explain our method to utilize the wave functional of universe for evaluating correlation functions on bulk dS at late time from dual CFT. 
Recently, there are developments on a different approach to obtain correlation functions by making use of bulk Feynman diagram computations \cite{Sleight:2020obc,Sleight:2021plv}, see also \cite{DiPietro:2021sjt}. In their analysis, they compute bulk correlation functions in the in-in formulation, whose good review may be found in \cite{Weinberg:2005vy}.
They put an identical theory on AdS and dS and compare the contributions coming from corresponding Feynman diagrams. In particular, they concluded that only a phase factor should be associated with each contribution along with the analytic continuation.

In this subsection, we compare the two formulations in order to clarify their similarity and difference.
In the in-in formulation, we compute correlation function by
\begin{align}
\langle \Omega (t_0) |\, \bar T \left( e^{i \int_{t_0}^{t_\infty} \d t' H (t')} \right) \psi_1^0 (x_1 ) \cdots \psi_m^0 (x_n)\,  T \left( e^{- i \int_{t_0}^{t_\infty} \d t' H(t')} \right) \,| \Omega (t_0) \rangle \, .
\end{align}
Here $T$ and $\bar T$ denote time and anti-time orderings. Moreover, correlators are among bulk fields at the late time $t = t_\infty$ and the Bunch-Davies boundary condition is assigned at the early time $t=t_0$. The vacuum corresponding to the boundary condition is denoted as $\Omega (t_0)$. 
Note that there are two types of interactions corresponding to time and anti-time orderings.
On the other hand, the formalism adopted in this paper utilize the wave functional of universe as explained in subsection \ref{sec:maldacena}.
The point here is that we have to sum over the both contributions from the coefficient functions of $\Psi[\psi_l^0]$ and $\bar \Psi[\psi_l^0]$.

\begin{figure}
  \centering
  \includegraphics[width=4cm]{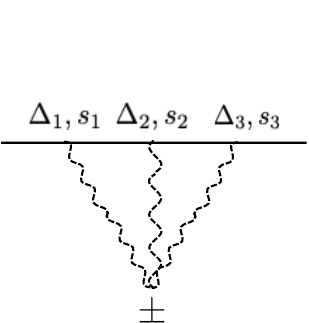}
 \caption{Feynman diagrams for three-point correlators on dS at late time. The $\pm$ indicates the (anti-)time ordering of inserted interaction terms in the Hamiltonian.}
  \label{fig:3pt}
\end{figure}

We begin with three-point function at the tree level.
In this case, the correspondence between the two formalisms is rather clear. In the in-in formulation, there are only two types of contact interactions, where one comes in the anti-time ordered insertions and the other comes in the time ordered insertions, see fig.~\ref{fig:3pt}. 
As in \cite{Sleight:2021plv}, we put $+$ or $-$ in order to distinguish them. In the formulation with wave functional, there are two types of contributions from $\Psi$ and $\bar \Psi$ as mentioned above. In other words, a diagram in fig.~\ref{fig:3pt} corresponds to $C_3(\{\vec x\})$, and the other corresponds to the complex conjugate of $C_3(\{\vec x\})$. Precisely speaking, in order to match with the scalar correlators on dS at late time, we have to multiply the inverse of two-point functions, see subsection \ref{sec:maldacena}.

\begin{figure}
  \centering
  \includegraphics[width=9cm]{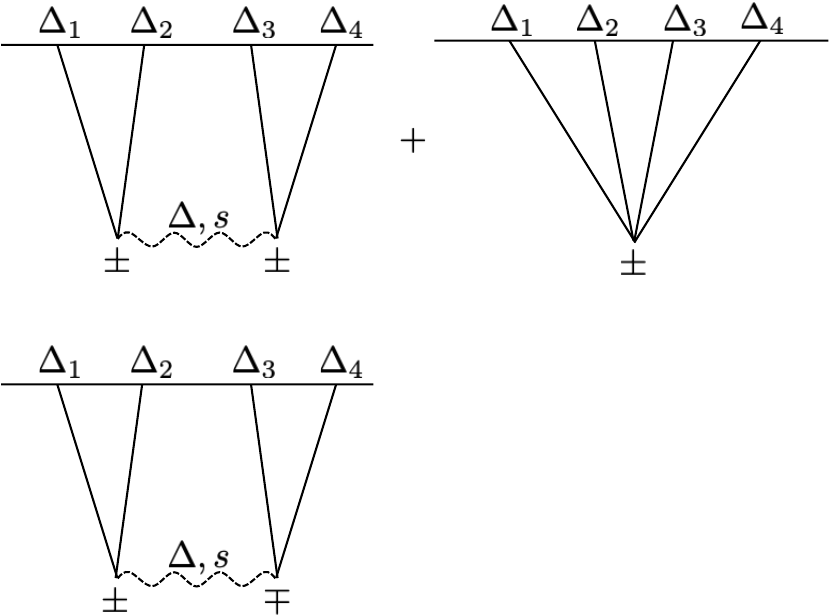}
 \caption{Feynman diagrams for four-point correlators on dS at late time. }
  \label{fig:4pt}
\end{figure}

Interesting phenomena arise from four-point functions; here we only consider four-point functions of external scalar operators and ignore the contributions coming from the products of two-point functions.  
The sums of Feynman diagrams in the first line of fig.~\ref{fig:4pt} 
come from different analytic continuations of bulk AdS computations, which correspond to $\langle \mathcal{O} \mathcal{O} \mathcal{O} \mathcal{O} \rangle$ and its complex conjugate, respectively. As discussed in subsection \ref{sec:maldacena}, the four-point function of scalar operators can be obtained from the sum of contact and exchange diagrams. The contact diagram with $\pm$ corresponds to $C_4(\{\vec x\})$ or its complex conjugate. The remaining contributions coming from the $\pm$ and $\pm$ exchange diagrams can be constructed from the lower-point coefficient functions.
The Feynman diagrams in the second line of fig.~\ref{fig:4pt}
are made of bulk two-point functions between $\pm$ and $\mp$ interactions. 
We can interpret the various contributions when expanding \eqref{O4d} as different combinations of analytic continuations from AdS space.
The bulk exchange diagrams correspond to conformal partial waves with exchange of boundary spin-$s$ current.
They correspond to the products of $\langle \mathcal{O} \mathcal{O} J\rangle$
and its complex conjugate $\overline{\langle \mathcal{O} \mathcal{O} J\rangle}$ divided properly by two-point coefficient functions.
We have thus listed the all Feynman diagrams corresponding to the contributions after expanding the terms in \eqref{O4c} and \eqref{O4d}.

\subsection{Higher order contributions in  \texorpdfstring{$G_N$}{GN}}
\label{sec:higherorder}

So far, we have restricted ourselves to the leading order contributions in the Newton constant $G_N$ to bulk correlation functions. However, the holographic formulation with the wave functional of universe can be applied to higher order contributions as well. Note that, as in the case of four-point functions, careful treatments have to be performed, which is explained in this section with the example of one-loop corrections to two-point bulk correlator. In subsection \ref{sec:oneloop0}, we will evaluate it for an explicit example.

Let us compute two-point correlator of scalar field $\phi$ on dS$_{d+1}$ at late time by applying \eqref{expectationvalue} with the wave functional of universe \eqref{exp}. 
At the leading order in $G_N$, the two-point function is given by \eqref{ms2pt}. There could be several types of one-loop contributions to two-point correlator, but here we only focus on one particular type.
We define the following integral as
\begin{align}
\begin{aligned}
 &  \mathcal{F}_2 (x_1 ' , x_2 ')
        =\sum_{s=2}^\infty \int d^d \vec y_1 d^d  \vec y_2 d^d \vec y_3 d^d  \vec y_4  \\
& \quad \times 
        \frac{      2\, \text{Re}\,\langle \, \mathcal{O} (\vec x_1 ') \, \mathcal{O} (\vec y_1) \, J_{j_1 \cdots j_s} (\vec y_2) \, \rangle \,  \Pi^{j_1 \cdots j_s}_{i_1 \cdots i_s} (\vec y_{23}) \,  2\, \text{Re}\,\langle \, J^{i_1 \cdots i_s} (\vec y_3 ) \, \mathcal{O}_3 (\vec y_4) \, \mathcal{O}_4 (\vec x_2 ') \, \rangle  } {  ( -2\, \text{Re} \, \langle\, J_s (\vec y_2) \, J_s ( \vec y_3) \, \rangle  )(- 2\, \text{Re} \, \langle\, \mathcal{O} (\vec y_1) \, \mathcal{O} ( \vec y_4) \, \rangle  )}  \, . \label{C211}
\end{aligned}
\end{align}
This integral naturally arises when considering so-called split representation of one-loop correction to the bulk to bulk scalar propagator.
The particular type of contribution to the two-point function considered here can be evaluated from the inversion formula:
\begin{align}
        \int d ^d \vec x_1 ' d^d \vec x_2 '  \frac{ \mathcal{F}_2 (x_1 ' , x_2 ')}{
            (- 2\, \text{Re}\,  \langle\, \mathcal{O} (\vec x_1)\, \mathcal{O}  (\vec x_1 ') \, \rangle )   (- 2\, \text{Re}\,  \langle\, \mathcal{O} (\vec x_2)\, \mathcal{O}  (\vec x_2 ' ) \, \rangle ) } \, . \label{C212}
\end{align}
From \eqref{expectationvalue} with \eqref{exp}, we can then read off other types of contributions.

\begin{figure}
  \centering
  \includegraphics[width=10cm]{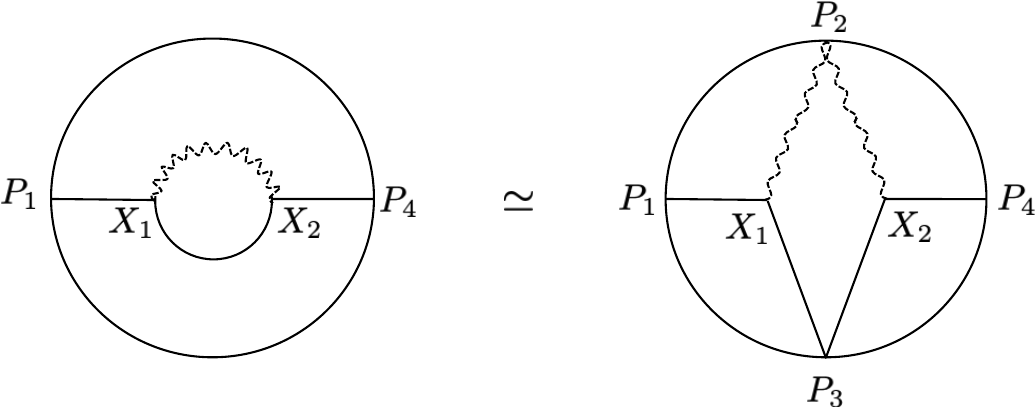}
 \caption{Split representation of a contribution to one-loop corrections in two-point function. }
  \label{fig:2pt}
\end{figure}

The above expression implies another subtlety when $C_2 (\{ \vec x_l \})$ is obtained from an analytic continuation of AdS case.
For CFT$_d$ dual to AdS$_{d+1}$, we may be able to obtain an exact expression of two-point function 
\begin{align}
\langle \mathcal{O}^\text{AdS} (\vec x_1)  \mathcal{O}^\text{AdS} (\vec x_2)  \rangle \label{AdS2pt}
\end{align}
in terms of $N$, which would be related to $1/N^2 \propto G_N$ (or $1/N \propto G_N$ in a higher-spin holography as our case, see \eqref{cs-charge}). However, a part of contributions at the next leading order can be computed by
\begin{align}
\begin{aligned}
 &   \sum_{s=2}^\infty \int d^d \vec y_1 d^d  \vec y_2 d^d \vec y_3 d^d  \vec y_4  \\
& \quad \times 
        \frac{  \langle \, \mathcal{O}^\text{AdS} (\vec x_1 ') \, \mathcal{O} ^\text{AdS}(\vec y_1) \,J_{j_1 \cdots j_s}^\text{AdS} (\vec y_2) \,  \rangle \,  \Pi^{j_1 \cdots j_s}_{i_1 \cdots i_s} (\vec y_{23})  \langle \, J^{i_1 \cdots i_s} _\text{AdS}(\vec y_3 ) \, \mathcal{O}_3 (\vec y_4) \, \mathcal{O}_4 (\vec x_2 ') \, \rangle  } {   \langle\, J_s^\text{AdS} (\vec y_2) \, J_s ^\text{AdS}( \vec y_3) \, \rangle    \langle\, \mathcal{O}^\text{AdS} (\vec y_1) \, \mathcal{O} ^\text{AdS}( \vec y_4) \, \rangle  }  \, , \label{AdScorrection}
\end{aligned}
\end{align}
which corresponds to evaluating the Witten diagram in fig.~\ref{fig:2pt}.
If we define $C_2 (\{ \vec x_l \})$  from  \eqref{AdS2pt} with higher order corrections in $1/N$, then the contribution in  \eqref{AdScorrection} would be double-counted in \eqref{C212} with \eqref{C211}. Thus we have to remove the contribution in \eqref{AdScorrection} from  \eqref{AdS2pt}  to define $C_2 (\{ \vec x_l \})$ via an analytic continuation. We can see that the situation is quite analogous to the case of four-point correlator discussed above. In the same way, we can see that all higher order contributions in $1/N$ to the two-point function \eqref{AdS2pt} have to be removed in order to avoid double-counting.

\section{Higher-spin \texorpdfstring{$\text{dS}_3/\text{CFT}_2$}{dS3/CFT2} correspondence}
\label{sec:hsdscft}

In the previous section, we explained the method to compute bulk correlation functions at late-time from dual CFT using the formalism with the wave functional of universe. In this section, we introduce a prescription to construct a concrete example of dS$_3$/CFT$_2$ correspondence as an analytic continuation of Gaberdiel-Gopakumar duality \cite{Gaberdiel:2010pz}. In the next subsection, we review the known results on the higher-spin AdS$_3$ holography. In subsection \ref{sec:dSCFT}, we explain our prescription to perform the analytic continuation from the AdS$_3$ holography to dS$_3$ holography.
We also examine two- and three-point correlators as simple examples.
In the succeeding sections, we examine more interesting examples including four-point correlators.

\subsection{Review of Gaberdiel-Gopakumar duality}
\label{sec:GG}

It was proposed in \cite{Gaberdiel:2010pz} that classical higher-spin gravity on AdS$_3$ is dual to the coset model \eqref{cosetCFT} at the 't Hooft limit \eqref{tHooft}. The higher-spin theory is the one constructed by Prokushkin and Vasiliev in \cite{Prokushkin:1998bq}, which includes higher-spin gauge fields with spin $s \, (= 2,3,\ldots , \infty)$ and two massive complex scalar fields. The gauge sector can be described by a Chern-Simons theory with the action
\begin{align}
 S = S_\text{CS} [A] - S_\text{CS} [\bar A] \, , \quad
 S_\text{CS} = \frac{k_\text{CS}}{4 \pi} \int \text{tr} \left( A \wedge d A + \frac{2}{3} A \wedge A \wedge A \right) \, . \label{CSaction}
\end{align}
Here the 
gauge fields $A$ and $\bar A$ are holomorphic and anti-holomorphic one-forms and they take values in higher-spin gauge algebra $\mathfrak{hs}[\hlambda]$. The gauge algebra is characterized such as to be truncated to $\mathfrak{sl}(N')$ if we set $\hlambda = \pm N'$.
We denote the generators by $V^s_m$. The index $s$ is related to the spin of higher-spin gauge field as $s =2,3,\ldots$. The other index $m$ runs over $m = - s + 1 , - s+2 , \ldots , s-1$. In particular, they satisfy
\begin{align}
    [V^2_m , V^s_n] = (m (s-1) - n) V^s_{m+n} \, ,
\end{align}
see, e.g., \cite{Gaberdiel:2011wb} for other commutation relations.
The coupling is given by the Chern-Simons level $k_\text{CS}$, and it is related to the gravity parameters as  
\begin{align}\label{cs-charge}
k_\text{CS} =  \frac{\ell_\text{AdS}}{4G_N} \, .
\end{align}
The asymptotic symmetry near the AdS boundary was examined in \cite{Gaberdiel:2011wb,Henneaux:2010xg,Campoleoni:2010zq,Campoleoni:2011hg}.
The symmetry algebra is identified with an infinite dimensional higher-spin algebra denoted by $W_\infty[\hat \lambda]$ with the central charge
\begin{align}\label{AdS-c}
    c = 6 k_\text{CS} = \frac{3 \ell_\text{AdS}}{2 G_N} \, , 
\end{align}
which is the same as the famous Brown-Henneaux value \cite{Brown:1986nw}. 
The parameter $\hat \lambda$ is the same as the one in \eqref{PVmass}, and the algebra can be truncated to {\bf $W_{N'}$}-algebra with currents of spin $s \, (=2,3,\ldots, N')$ if we set {\bf $\hat \lambda = \pm N'$}.
The massive sector of the Prokushkin-Vasiliev theory consists of two scalar fields $\phi_\pm$ with the same mass given in \eqref{PVmass} as
$\ell ^2_\text{AdS}\, m^2  = -1 + { \hat{\lambda}} ^2 $.
Here we assign $0 < \hat{\lambda} < 1$.
The coupling to the gauge fields is described by the equations of motion \cite{Prokushkin:1998bq}, see \cite{Ammon:2011ua} for a good review.

The CFT dual to the higher-spin theory on AdS$_3$ was proposed in \cite{Gaberdiel:2010pz} to be the coset CFT \eqref{cosetCFT}, which describes a unitary minimal model with respect to $W_N$-algebra. The central charge of the model is given by
\begin{align}
c = (N-1) \left(1 - \frac{N(N+1)}{(N+k) (N+k+1)} \right) \, . \label{central}
\end{align}
The classical limit of higher-spin gravity corresponds to the 't Hooft limit of the CFT given by \eqref{tHooft}.
The coset CFT \eqref{cosetCFT} at the 't Hooft limit is known to possess the symmetry of $W_\infty[\lambda]$-algebra with the central charge \eqref{central} \cite{Gaberdiel:2011wb,Gaberdiel:2012ku}. 
In such a limit the central charge becomes:
\begin{equation}
\lim_{\rm 't Hooft} c \to N(1-\lambda^2) + \CO(N^0) \, , \quad \lambda =\frac{N}{N+k} \to \text{fixed} \, .     \label{thooftf} 
\end{equation}
The 't Hooft parameter $\lambda$ defined above is identified with $\hat \lambda$ appearing in the higher-spin theory, and in the following we only use $\lambda$ instead of $\hat \lambda$.
We will thus take \eqref{thooftf} to be the defining relation between the central charge $c$ and $N$ in the 't Hooft limit \eqref{tHooft} and we will mostly be working in this limit throughout unless otherwise stated. Furthermore using \eqref{AdS-c}, we can directly relate the large $c$/large $N$ expansion with the perturbative expansion in $G_N$ in AdS$_3$.

The states of the coset CFT \eqref{cosetCFT} are labeled by three Young diagrams $(\Lambda_+ , \omega ; \Lambda_-)$. There are selection rules and field identifications, and not all labels are allowed, see \cite{DiFrancesco:1997nk} for more details. In this case, $\omega$ is uniquely fixed by the other two labels $(\Lambda_+;\Lambda_-)$. The state labeled by $(\Lambda_+;\Lambda_-)$ has conformal weight
\begin{align}
    h_{(\Lambda_+ ; \Lambda_-)} = n + h^{(k)}_{\Lambda_+} + h^{(1)}_{\omega} - h^{(k+1)}_{\Lambda_-} \, , \quad h_\Lambda^{(k)} = \frac{C_2 (\Lambda)}{k+N} \, , \quad C_2 (\Lambda) = \frac12 (\Lambda , \Lambda + 2 \rho) \, . \label{conformalweight}
\end{align}
Here the positive integer $n$ represents how the denominator representation is embedded into the numerator one. Moreover, $C_2(\Lambda)$ is the eigenvalue of the quadratic Casimir operator and $\rho$ is the Weyl vector.
The fundamental states are given by those labeled by $(\square;0)$ and $(0;\square)$ (and their conjugate representations) and other states may be generated by fusing these fundamental states. We denote the corresponding complex scalar operators by 
$\mathcal{O}_\pm(z , \bar z )$, whose conformal dimensions are 
\begin{align}\label{Delta-0}
\Delta_\pm^{(0)} = 2 h_\pm = 1 \pm \lambda
\end{align}
at the 't Hooft limit \eqref{tHooft}.
They are dual to bulk scalar fields with standard $(+)$ and alternative $(-)$ boundary conditions (see, e.g., \cite{Ammon:2011ua}). 
The states labeled by generic Young diagrams $(\Lambda_+;\Lambda_-)$ are interpreted as bound states of these scalar fields.
More precisely, to compare CFT results with gravity computations later, let us introduce the expansion of the scaling dimensions in terms of the inverse central charge $1/c$:
\begin{equation}\label{Delta-ex}
\Delta_{\pm} = \sum_{r= 0}^{\infty} \Delta_{\pm}^{(r)} c^{-r} \, , \quad c \to \infty \, .  
\end{equation}
As explained earlier, this is equivalent to an expansion in $G_N$ hence \eqref{Delta-0} should strictly be taken as the leading classical value. We will apply such an expansion when expanding the higher-point correlation functions in the next section.

The coset CFT \eqref{cosetCFT} has the symmetry of $W_N$-algebra as proven in \cite{Arakawa:2018iyk} and in the 't Hooft limit, the algebra is generated by conserved currents $J_{(s)}^\text{AdS}(z)$ (and anti-holomorphic currents $\bar J_{(s)}^\text{AdS}(\bar z)$) with $s = 2,3,\ldots , \infty$. These conserved currents are dual to higher-spin gauge fields $A$ (and $\bar A$) in the bulk theory.
The gauge configuration corresponding to the AdS background is given by
\begin{align}
A= e^\rho\, V_1^2\, d z + V^2_0\, d \rho \, , \qquad \bar A = e^\rho\, V_{-1}^2\, d z - V^2_0\, d \rho \,. \label{AAdS}
\end{align}
Here $\rho$ relates to the radial coordinate $y$ in the Poincare metric \eqref{AdSmetric} via $y=e^{-\rho}$ and $(z, \bar z)$ are the complex coordinates for the plane parallel to the boundary.
We consider a small fluctuation from the holomorphic AdS background gauge field $A$, while keeping anti-holomorphic $\bar{A}$ fixed as (see (3.20) of \cite{Ammon:2011ua})
\begin{align}
\begin{aligned}
&A_z =   e^\rho\, V_1^2 + \frac{1}{c B^{(s)}}\, J_{(s)}^\text{AdS} (z)\, e^{- (s-1) \rho}\, V^s_{ -(s-1)} \, , \\
&A_{\bar z } = - \mu_{(s)} ^\text{AdS} (z)\, e^{(s-1) \rho}\, V^s_{s-1} + \cdots \, . 
\end{aligned} \label{AAdSfluc}
\end{align}
Here we use the convention with
\begin{align}
    \qquad B^{(s)} 
        =  \frac{1}{2 ^{2s} \pi^\frac{5}{2}}\, \frac{\sin (\pi \lambda)}{ \lambda (1 - \lambda^2)} \,
    \frac{\Gamma (s) \, \Gamma (s - \lambda)\, \Gamma (s + \lambda)}{ \Gamma \left(s - \frac12\right)} \, , 
\end{align}
{where $\mu_{(s)}^{\rm AdS}(z)$ acts as the source term for $J_{(s)}^{\rm AdS}(z)$.}
The generators of $W_\infty [\lambda]$ near the AdS boundary are given by $J_{(s)}^\text{AdS}$, which can be identified with the generators of symmetry algebra in the dual CFT.

Solving the bulk equations of motion, the authors in  \cite{Ammon:2011ua}
obtained three-point functions of scalar-scalar-higher spin current  as
\begin{align}
    \langle\, \mathcal{O}_\pm ^\text{AdS}(z_1)\, \bar{ \mathcal{O}}_\pm ^\text{AdS}(z_2)\, J_{(s)}^\text{AdS} (z_3)\, \rangle  = C_\pm^{(s)} 
    \left ( \frac{z_{12}}{z_{13} z_{23}} \right) ^s\,  \langle \, \mathcal{O}_\pm ^\text{AdS}(z_1) \, \bar{\mathcal{O}}_\pm ^\text{AdS}(z_2) \, \rangle \label{3pt}
\end{align}
with
\begin{align}
C_\pm^{(s)} = \frac{\eta^s_\pm}{2 \pi} \, \frac{\Gamma (s)^2}{\Gamma (2 s-1)} \frac{\Gamma (s \pm \lambda) }{ \Gamma (1 \pm \lambda)} \, . \label{3pteta}
\end{align}
The phase factors can be chosen arbitrary but here we set $\eta^s_+ = 1$ and $\eta^s_- = (-1)^s$ as in \cite{Hikida:2017byl}.
The three-point function with $\bar J_{(s)}^\text{AdS}$ has an extra factor $(-1)^s$ along with $z$ replaced by $\bar z$. The higher-spin currents are normalized as
\begin{align}
    \langle\, J_{(s)}^\text{AdS} (z)\, J_{(s)}^\text{AdS} (0)\, \rangle 
        =
        \frac{c B^{(s)}}{z^{2s}} \, , \quad 
    \langle\, \bar J_{(s)}^\text{AdS} (\bar z)\, \bar J_{(s)}^\text{AdS} (0)\, \rangle 
        =
        \frac{c B^{(s)}}{\bar z^{2s}} \, .     
    \label{2pt}
\end{align}
They are evaluated by tree-level computations in the bulk theory, thus they are valid at the leading order in $1/c$. It is also possible to confirm them from the dual CFT viewpoints \cite{Ammon:2011ua,Hikida:2017byl}, see also appendix \ref{sec:sym}.

We should also mention about the two-point functions of scalar operators. The coordinate dependence is fixed by conformal symmetry and the overall normalization is just a convention. However, as in the case of two-point functions of higher-spin currents \eqref{2pt}, the canonical convention in the bulk theory leads to 
\begin{align}
 \langle\, \mathcal{O}_\pm ^\text{AdS}(z_1)\, \bar{\mathcal{O}}_\pm ^\text{AdS}(z_2) \, \rangle \sim \frac{ c  }{|z_{12}|^{4 h_\pm}} \ ,
 \label{2ptnorm0}
\end{align}
up to an multiplication of real number. We will use a different normalization as defined below, but the original factor is quite important to perform an analytic continuation from AdS$_3$ to dS$_3$, see \cite{Anninos:2011ui} for a similar argument for higher-spin dS$_4$ holography.

\subsection{Analytic continuation from \texorpdfstring{AdS$_3$}{AdS3} to \texorpdfstring{dS$_3$}{dS3}}
\label{sec:dSCFT}

We would like to introduce a dS$_3$/CFT$_2$ correspondence by taking a proper analytic continuation of Gaberdiel-Gopakumar duality reviewed in the previous subsection. For the bulk theory, we consider the Prokushkin-Vasiliev theory on dS$_3$, which is obtained by the map $\ell_\text{AdS} = - i  \ell $ as in \eqref{map}. Because of this, we have to change the level of Chern-Simons theory as \cite{Witten:1988hc} 
\begin{align}
k_\text{CS} =  \frac{\ell_\text{AdS}}{4G_N} =  - i  \frac{\ell}{4G_N} = - i \kappa \label{cg}
\end{align}
with $\kappa \in \mathbb{R}$.
The symmetry algebra at the future infinity of dS$_3$ is  $W_\infty [ \lambda]$ with pure imaginary central charge 
as \cite{Strominger:2001pn,Ouyang:2011fs}
\begin{align}
c = - i \frac{3 \kappa}{2 G_N}  \equiv  - i c^{(g)} \, . \label{cg2}
\end{align}
The bulk theory also includes two massive complex scalars with the mass
\begin{align}
    \ell^2 m^2 = 1 - { \lambda}^2 \, ,
\end{align}
which may be obtained from \eqref{PVmass} by the map \eqref{map}.
We use $0 < \lambda < 1$ as before, then we have positive $\ell^2 m^2$.
We assign the standard and alternative boundary conditions near the future infinity, thus the conformal dimensions of dual scalar operators are $\Delta^{(0)}_\pm = 1 \pm  \lambda$ as before.

We would like to claim that the dual CFT$_{2}$ for dS$_{3}$ is given by an analytic continuation of the coset \eqref{cosetCFT} as in \cite{Chen:2022ozy}.
The analytic continuation is different from the one taken in \cite{Hikida:2021ese,Hikida:2022ltr} and {their differences} will be explained in section \ref{sec:kto-2}. 
As mentioned above in the 't Hooft limit, the symmetry organizing the CFT reduces to $W_\infty[\lambda]$ and has two parameters $\lambda$ and $c$.
The central charge is now taken to be purely imaginary value $c = - i c^{(g)}$ with $c^{(g)} \in \mathbb{R}$. 
The original coset is labeled by two parameters $N,k$, and these parameters can be written in terms of $\lambda,c$ instead via \eqref{tHooft} and \eqref{central}. 
In practice, we compute physical quantities by the coset CFT (or its dual description of Toda field theory as derived in \cite{Creutzig:2021ykz}), which may be expressed even with finite $N,k$. We then rewrite them in terms of $\lambda,c$ via \eqref{tHooft} and \eqref{central} and expand them in terms of $1/c$.

\subsubsection{Two-point correlators}

We would like to begin by computing the two-point bulk correlators on dS$_3$ at late time. We will first consider the scalar fields and then move to the symmetric tensor fields. For this, we use the formulas given in \eqref{ms2pt} and \eqref{mt2pt}. In the coordinate basis, taking the inverse of two-point function is actually quite non-trivial.
We first define the inverse function $K_{h , \bar h} (x,y)$ which satisfies
\begin{align}
\int d ^2 y\, K_{h , \bar h}(x,y)\, \frac{1}{(y -z)^{2h}(\bar y - \bar z)^{2\bar h}} = \delta^{(2)} (x - z) \, .
\end{align}
Explicitly we have, 
\begin{align}
K_{h , \bar h} (x, y) \equiv \frac{1}{\pi}k_{h , \bar h} \frac{1}{(x-y)^{2 - 2h} (\bar x - \bar y)^{2 -2 h}} \, , \quad 
k_{h , \bar h} =  \frac{\Gamma (2 -2 \bar h)}{\Gamma (2 h -1)} = (-1)^{2 (h - \bar h)} \frac{\Gamma (2 -2h)}{\Gamma (2 \bar h -1)} \, , \label{K}
\end{align}
see, e.g., \cite{Osborn:2012vt} for more extensive definitions. We have followed the notations there.
Notice that we can equivalently view $K_{h,\bh}(x,y)$ as the two-point correlation function of the so-called shadow operator, see \eqref{defshadow} below.

Let us consider the two-point correlation functions of scalar fields at late time in dS$_3$.
As discussed around \eqref{sphase}, the scalar operators dual to the bulk dS$_{3}$ fields are related to those dual to the bulk AdS$_{3}$ fields as
\begin{align}
\mathcal{O}_\pm = e^{i \frac{\pi}{2}(2 h_\pm - 2)}  \mathcal{O}_\pm^\text{AdS} \, , \qquad
\tilde{\mathcal{O}}_\pm = e^{i \frac{\pi}{2}(2 h_\pm - 2)}  \bar{\mathcal{O}}_\pm^\text{AdS} \, , \label{omap}
\end{align}
where we have set $\Delta_\pm^{(0)} = 2 h_\pm = 1 \pm \lambda$ and $d=2$.
Note that $ \bar{\mathcal{O}}_\pm^\text{AdS} $ is complex conjugate of $ \mathcal{O}_\pm^\text{AdS}$, while $\tilde{\mathcal{O}}_\pm $ is not complex conjugate of $\mathcal{O}_\pm$.
The relations \eqref{omap} and the canonical normalization \eqref{2ptnorm0} imply that the two-point coefficient functions are given by
\begin{align}
    \begin{aligned}
        \langle\, \mathcal{O}_\pm (z_1)\, \tilde{\mathcal{O}}_\pm (z_2) \, \rangle 
            &= 
            \left. e^{2 i \pi h_\pm  }\, \langle\, \mathcal{O}_\pm ^\text{AdS}(z_1) \, \bar{\mathcal{O}}_\pm ^\text{AdS}(z_2) \,  \rangle \right|_{c \to - i c^{(g)}} \\
            &= - \frac{i c^{(g)}}{c} \,e^{2 i \pi h_\pm  }\, \langle\, \mathcal{O}_\pm ^\text{AdS}(z_1)\, \bar{\mathcal{O}}_\pm ^\text{AdS}(z_2)\,  \rangle \, .
    \end{aligned}
\end{align}
In the following, we will set the overall normalization factors of the AdS/dS two-point correlation functions to absorb $c/c^{(g)}$ respectively, such that we are keeping only the overall phase after the analytic continuations \eqref{omap}:
\begin{align}
    \langle\, \mathcal{O}_\pm ^\text{AdS}(z_1)\, \bar{\mathcal{O}}_\pm ^\text{AdS}(z_2)\,  \rangle = \frac{1}{|z_{12}|^{4 h_\pm}} \, , \quad
    \langle\, \mathcal{O}_\pm (z_1)\, \tilde{\mathcal{O}}_\pm (z_2)\,  \rangle = - i e^{2 i \pi h_\pm  }\,\frac{1}{|z_{12}|^{4 h_\pm}} \, . \label{2ptnorm}
\end{align}
Now we can obtain the bulk two-point correlators at late time from the wave functional formalism explicitly by applying the inversion formula \eqref{ms2pt}. Here the second expression in \eqref{2ptnorm} plays the role of $C_2(\{\vec{x}_l\})$, and we obtain
\begin{align}
\langle\,  \phi_\pm (z_1)\, \bar \phi _\pm (z_2)\, \rangle 
            =  \frac{1}{2 \sin (2 \pi h_\pm)} K_{h_\pm , h_\pm} (z_1,z_2) 
            = a_{h_\pm} \frac{1}{\pi}\frac{\Gamma(2- 2 h_\pm)}{\Gamma(2 h_\pm -1)} \frac{1}{|z_{12}|^{4-4 h_\pm}\,} \, .
\end{align}
Here we have used $K_{h_\pm , h_\pm} (z_1,z_2)$ defined in \eqref{K}.
Compared with pure AdS$_{3}$ case, we have extra phase factor
\begin{align}
            a_{h_\pm} = \frac{1}{2 \sin (2 \pi h_\pm)} \, . \label{aDelta}
\end{align}
The factor of $2\sin(2\pi h_{\pm})$ in the denominator of \eqref{aDelta} arises as we need to include the contributions from $\Psi[\psi_l^0]$ and its complex conjugate when extracting the bulk two-point correlators after the analytic continuation.
Note that the same factor $a_{h_\pm} = c_{2 h_\pm}^\text{dS-AdS}$ was inserted in order to relate bulk propagators on AdS$_{d+1}$ and dS$_{d+1}$ as in (2.15) of \cite{Sleight:2021plv}. In \cite{Sleight:2021plv}, the authors included the factor such that the both two-point propagators have the correct normalization at small distance limit. We also remark that the identical factor  $c_{\Delta}^\text{dS-AdS}$ can be used also for symmetric tensor fields with $\Delta$ and spin $s$.

We next move to the case of higher-spin gauge fields.
The gauge configuration corresponds to dS$_{3}$ background is given by 
\begin{align}
A = i\eta^{-1} (   V_1^2 dz +i  V_0^2d\eta) \,  , \qquad \bar A = i \eta^{-1}  (V_{-1}^2 d\bz -i  V_0^2  d\eta) \,  , \label{AdS}
\end{align}
where we have set $y = e^{-\rho}$ and change as $y = i \eta$ in \eqref{AAdS}.
A small fluctuation {corresponding to the inclusion of holomorphic higher spin fields} can be expressed as
\begin{align}
\begin{aligned}
&A_z =  i \eta^{-1} V_1^2 + \left. \frac{i}{c B^{(s)}}\,  J_{(s)}(z) \eta^{s-1} V^s_{ -(s-1)} \right|_{c \to - i c^{(g)}} = i \eta^{-1} V_1^2 - \frac{1}{c^{(g)} B^{(s)}}\,  J_{(s)}(z)  \eta^{s-1} V^s_{ -(s-1)}  \, , \\
&A_{\bar z }= - i \mu_{(s)}  (z)\,   \eta^{- (s-1)}\, V^s_{s-1} + \cdots \, .
\end{aligned}
\end{align}
Assuming the identification of $A_z$ under the map from AdS$_3$ to dS$_3$, we have the relation
\begin{align}
J_{(s)}(z)  = e^{ i\frac{\pi}{2}(s -2 )} J_{(s)}^\text{AdS} (z) \, . \label{jmap}
\end{align}
The result is consistent with the one in \eqref{jreld} based on generic argument. In a similar manner, we can obtain
\begin{align}
\tilde J_{(s)}(\bar z)  = e^{ i\frac{\pi}{2}(s -2 )} \bar J_{(s)}^\text{AdS} (\bar z) \, . \label{jmapbar}
\end{align}
We again remark that while $\bar J_{(s)}^\text{AdS} (\bar z) $ is complex conjugate of $J^\text{AdS}_{(s)}(z) $ but  $\tilde J_{(s)} (\bar z) $ is not complex conjugate of $J_{(s)}(z) $ due to the same phase assignments for the both.

Using the relation \eqref{jmap}, the two-point coefficient functions are written as
\begin{align}
\begin{aligned}
\langle\, J_{(s)} (z_1)\, J_{(s)} (z_2)\, \rangle  &=  \left.  (-1)^s \,
\langle\,  J_{(s)}^\text{AdS} (z_1)\, J_{(s)}^\text{AdS} (z_2) \, \rangle \right|_{c \to - i c^{(g)}} \\
&= (-1)^{s+1}\frac{i c^{(g)}}{c}\,  \, 
\langle \, J_{(s)}^\text{AdS} (z_1)\, J_{(s)}^\text{AdS} (z_2)\, \rangle \, .
\end{aligned}
\end{align}
We should add also the complex conjugate of the two-point function of $\tilde{J}_{s}(\bar z)$:
\begin{align}
&\langle\, \tilde J_{(s)} (\bar z_1) \, \tilde J_{(s)} (\bar z_2) \, \rangle  = (-1)^{s} \frac{i c^{(g)}}{c}\, 
\langle \, \bar J_{(s)}^\text{AdS} (\bar z_1) \, \bar J_{(s)}^\text{AdS} (\bar z_2) \, \rangle \, ,
\end{align}
where we have used \eqref{jmapbar}.
In contrast with the scalar two-point functions, we do not change the normalization factors here to absorb $c/c^{(g)}$.
Applying the formula \eqref{mt2pt} and \eqref{K}, the bulk correlators of the higher-spin fields at late time are obtained as
\begin{align}
\begin{aligned}
    \langle \mu_{(s)} (z_1) \mu_{(s)} (z_2) \rangle &= (-1)^{s+1} \frac{ i }{2} \frac{1}{c^{(g)} B^{(s)}} K_{s,0} (z_1 ,z_2) \\ &=  a_{(s)} \frac{1}{c^{(g)} B^{(s)} \pi \Gamma(2 s -1) }   \frac{1}{(z_{12})^{2-2s} (\bar z_{12})^{2}}\, .
   \end{aligned}
\end{align}
The factor
\begin{align}
a_{(s)} = (-1)^{s+1}\frac{i}{2} \label{as}
\end{align}
arises due to our prescription of analytic continuation as in the case of scalar field. It does not match with the generic expression \eqref{aDelta} since the factor diverges at the massless limit with $\Delta = s + d-2 = s \in \mathbb{N}$ for $d=2$. This is however not a contradiction, since the massless limit is quite delicate. For instance, let us consider a spin-one massive field on Lorentzian dS$_{d+1}$ spacetime. Its bulk two-point function may be found in (3.18)-(3.20) of \cite{Allen:1985wd}, but it diverges at the massless limit with $m \to 0$. In order to obtain correct two-point function of massless spin-one gauge field, we may have to work with the action of massless gauge field from the beginning as in section 4 of \cite{Allen:1985wd}.
It is natural to expect that the factor \eqref{as} can be obtained by working with massless higher-spin gauge fields from the beginning.
It is an important future problem to reproduce the factor \eqref{as} from the method of \cite{Sleight:2020obc,Sleight:2021plv} for the case as well.

\subsubsection{Three-point correlators}

As in the AdS$_3$ case, here 
we consider the basic three-point correlators in dS$_3$ at the late time consisting of two scalar fields and one higher-spin gauge field.
Using the maps \eqref{omap} and \eqref{jmap}, the three-point coefficient functions are related to \eqref{3pt} as:
\begin{align}
\begin{aligned}
\langle \, \mathcal{O}_\pm (z_1) \, \tilde{ \mathcal{O}}_\pm (z_2) \, J_{(s)}(z_3)\, \rangle 
&= - \left. e^{ i \frac{\pi}{2}(4 h_\pm  + s)}
\langle\, \mathcal{O}_\pm ^\text{AdS}(z_1)\, \bar{ \mathcal{O}}_\pm ^\text{AdS}(z_2)\, J_{(s)}^\text{AdS} (z_3)\, \rangle \right|_{c \to - i c^{(g)}} \\
&= i  \, e^{ i \frac{\pi}{2}(4 h_\pm  + s)} C_\pm^{(s)} \left( \frac{z_{12}}{z_{13}z_{23}}\right)^s \frac{1}{|z_{12}|^{4 h_\pm}} \, .
\end{aligned} \label{ooj}
\end{align}
In the same way, we find 
\begin{align}
\begin{aligned}
\langle \, \tilde{\mathcal{O}}_\pm (z_1) \, \mathcal{O}_\pm (z_2) \, \tilde J_{(s)}(\bar z_3)\, \rangle  =    i \,  e^{ i \frac{\pi}{2}(4 h_\pm  + s)} C_\pm^{(s)} \left( \frac{\bar z_{12}}{\bar z_{13} \bar z_{23}}\right)^s \frac{1}{|z_{12}|^{4 h_\pm}}
\end{aligned}\label{oojbar}
\end{align}
from the maps \eqref{omap} and \eqref{jmapbar}. 
The sum of \eqref{ooj} and the complex conjugate of \eqref{oojbar} yields
\begin{align}\label{Total 3pt}
&  2 \sin \left( (4 h_\pm +s ) \frac{\pi}{2} \right) C_\pm^{(s)} \left( \frac{ z_{12}}{ z_{13} z_{23}}\right)^s \frac{1}{|z_{12}|^{4 h_\pm}} \, .
\end{align}
Notice that the complex conjugation of \eqref{oojbar} only depend on the holomorphic coordinates $z_{ij}$ up to the scalar two-point function just like \eqref{ooj}.

In order to obtain the corresponding dS bulk correlators from the coefficient function $C_3(\{ \vec{x}_l\})$,  which is proportional to \eqref{ooj}, we will use the expression \eqref{m3pt}.
For this, it is convenient to first consider the inversion formula \cite{Osborn:2012vt}
\begin{align}
\begin{aligned}
\frac{k_{h , \bar h}}{\pi} \int d^2 y \frac{1}{(z -y)^{2 - 2 h} (\bar z - \bar y)^{2 - 2 \bar h} } \mathcal{F}^{h , \bar h}_{12} (y , \bar y)  
= \frac{\Gamma(1 - h - h_{12}) \Gamma(1 - \bar h + \bar h_{12})}{\Gamma (\bar h + \bar h_{12}) \Gamma (h - h_{12})}\mathcal{F}^{1-h , 1- \bar h}_{12} (y , \bar y) \, ,
\end{aligned}
\end{align}
with
\begin{align}
\mathcal{F}^{h , \bar h}_{12} (z , \bar z)
= \frac{1}{z_{12}^{h_1 + h_2 - h} (z_1 - z)^{h+h_{12}} (z_2 - z)^{h - h_{12}}} \frac{1}{\bar z_{12}^{\bar h_1 + \bar h_2 - \bar h} (\bar z_1 - \bar z)^{\bar h+\bar h_{12}} (\bar z_2 - \bar z)^{\bar h - \bar h_{12}}} \, . \label{calFhh}
\end{align}
Here we set $h_{12} = h_1 - h_2$ and so on.
Substituting kinematic dependence in \eqref{Total 3pt}, we can now perform its inversion in turns. First we apply the two-point inversion formula:
\begin{align}
\begin{aligned}
&  \int d^2 z_1 ' d^2 z_2 ' K_{h_+ ,h_+} (z_1 , z_1 ') K_{h_+ ,h_+} (z_2 , z_2 ') \left( \frac{ z_{1'2'}}{ z_{1'3'} z_{2'3'}}\right)^s \frac{1}{|z_{1'2'}|^{4 h_\pm}} \\
  & \qquad\qquad  =  (-1)^{s+1} \frac{\Gamma (\mp \lambda) \Gamma (1 \mp \lambda)}{\Gamma (1 -s \pm \lambda) \Gamma(s \pm \lambda )}  \left( \frac{ z_{12}}{ z_{13'} z_{23'}}\right)^s \frac{1}{|z_{12}|^{4 h_\mp}} \, ,
  \end{aligned} \label{inverse1}
 \end{align}
and next we evaluate
\begin{align}
\begin{aligned}
\int d^2 z_3 ' K_{s ,0} (z_3 , z_3 ') \left( \frac{ z_{12}}{ z_{13'} z_{23'}}\right)^s \frac{1}{|z_{12}|^{4 h_\mp}} 
= \frac{(-1)^{s+1}}{\Gamma(s)^2} \left( \frac{ z_{12}}{ z_{13} z_{23}}\right)^{1-s} \left( \frac{ \bar z_{12}}{ \bar z_{13} \bar z_{23}}\right) \frac{1}{|z_{12}|^{4 h_\mp}} \, . 
\end{aligned} \label{inverse2}
 \end{align}
Restoring the pre-factors in \eqref{Total 3pt}, as well as the normalization factors in two-point functions \eqref{aDelta} and \eqref{as}, we finally obtain 
\begin{align}
\begin{aligned}
&\langle\,  \phi_\pm (z_1)\, \bar \phi _\pm (z_2)\, \mu_{(s)} (z_3) \rangle \\
&\quad= \lambda_{h_\pm,h_\pm,s} \frac{\Gamma (\mp \lambda) \Gamma (1 \mp \lambda)}{\Gamma(s)^2 \Gamma (1 -s \pm \lambda) \Gamma(s \pm \lambda )}  \frac{C_\pm^{(s)}}{c^{(g)} B^{(s)}} \left( \frac{ z_{12}}{ z_{13} z_{23}}\right)^{1-s} \left( \frac{\bar z_{12}}{ \bar z_{13} \bar z_{23}}\right) \frac{1}{|z_{12}|^{4 h_\mp}} \label{3pt-inv}
\end{aligned}
\end{align}
with
\begin{align} \label{phase3pt}
   \lambda_{h_\pm,h_\pm,s}  = 
   2  a_{h_\pm}^2 a_{(s)}  \sin \left( (4 h_\pm +  s ) \frac{\pi}{2} \right)  \, .
\end{align}
Almost the same factor appears in (3.24) of \cite{Sleight:2021plv} up to the subtlety associated to the massless limit of higher-spin fields mentioned above, our result here thus serves as a confirmation in an explicit model.

\section{Four-point correlators dual to \texorpdfstring{AdS$_3$}{AdS3} and \texorpdfstring{dS$_3$}{dS3}}
\label{sec:4pt}

We next turn to computations of the bulk four-point correlation functions of the following types:
\begin{align}
\langle\, \phi_- (\infty)\,  \phi_+ (1)\, \bar \phi_+ (z)\, \bar \phi_- (0 )\,\rangle \, , \quad
\langle\, \phi_\pm (\infty) \, \bar \phi_\pm (1) \, \phi_\pm (z) \, \bar \phi_\pm(0 )\, \rangle 
\label{4ptdS}
\end{align}
on dS$_3$ at the late time.
More precisely, we first consider the corresponding four-point correlation functions computed exactly in the 2d coset CFTs dual to higher-spin AdS$_{3}$ gravity, and we then obtain their leading non-trivial $1/c$-order contribution. In 2d CFTs, we normally consider the expansions of correlation functions in terms of Virasoro conformal blocks, and in the limit $c \to \infty$ while keeping fixed all the external scaling dimensions (sometimes called as ``LLLL'' limit), it is well-known that these can be replaced by global conformal blocks, see \cite{Perlmutter:2015iya, Bombini:2018jrg} for recent progress. As mentioned earlier, this also corresponds to the expansion in $G_N$ for the bulk correlation functions, such that at the leading order we only need to include tree-level interactions. We then perform the suitable analytic continuations to obtain the corresponding results in dS$_{3}$.

\subsection{The case dual to \texorpdfstring{AdS$_3$}{AdS3}}\label{Sec:4pt-AdS3}

\subsubsection{Four-point function \texorpdfstring{$G_{-+}(z)$}{Gmp}}

Let us start with the first correlator in \eqref{4ptdS} which is of simpler structure.
This case was already analyzed in \cite{Chen:2022ozy}, but here we explain the analysis in more details.
We utilized the four-point function of operators dual to  the bulk scalars $\phi_+^\text{AdS}$ and $\phi_-^\text{AdS}$ on AdS$_3$. They are evaluated explicitly in \cite{Papadodimas:2011pf} as
\begin{align}
\begin{aligned}
 G_{-+}(z) &= \langle\, \mathcal{O}^\text{AdS}_- (\infty) \, \mathcal{O}^\text{AdS}_+ (1)\, \bar{\mathcal{O}}^\text{AdS}_+ (z)\, \bar{\mathcal{O}}^\text{AdS}_- (0 )\, \rangle  \\
&=  \frac{|z|^{\frac{2}{N} }}{|1-z|^{2\Delta_+}} \left| 1 + \frac{1 -z}{Nz}\right|^2 \, ,
\end{aligned} \label{G-+}
\end{align}
and the expression is exact for all finite $k,N$. Here we also list the exact expressions for $\Delta_\pm$:
\begin{align}
   \Delta_+ = \frac{(N-1) (2 N +1 + k)}{N (N+k)} \, , \quad
   \Delta_- = \frac{(N-1) k}{N (N+k+1)} \, . \label{Deltapm}
\end{align}
Let us now consider the systematic $1/c$-expansion of $G_{-+}(z)$.
As explained above, the symmetry organizing the CFT is $W_\infty[\lambda]$, which has two parameters $\lambda$ and $c$. 
As mentioned above, the coset \eqref{cosetCFT} has two parameters $N,k$, 
and we rewrite physical quantities instead in terms  $\lambda, c$.
Explicitly we have
\begin{align}
    N = \frac{c}{1 - \lambda^2} + \frac{1 + \lambda + \lambda^2}{1 + \lambda} + \mathcal{O} (c^{-1}) \, .
\end{align}
Therefore, taking $N$ large but keeping $\lambda = N/(N+k)$ finite and fixed in the 't Hooft limit, the $1/N$-expansion coincides with $1/c$-expansion at the leading order, but there will be corrections at higher order expansions as we will illustrate. The conformal dimensions $\Delta_\pm$ can be also expanded in the form of \eqref{Delta-ex} as: 
\begin{align} \label{Deltaexp}
    \Delta_+ &= 1 + \lambda + \frac{\lambda ^2 - 1}{c} + \mathcal{O} (c^{-2}) = \Delta_+^{(0)}+\frac{\Delta^{(1)}_+}{c} + \mathcal{O} (c^{-2}) \, , \\
    \Delta_- &= 1 - \lambda - \frac{(\lambda^2 -1)^2}{c} + \mathcal{O} (c^{-2}) =\Delta_-^{(0)}+\frac{\Delta^{(1)}_-}{c}  + \mathcal{O} (c^{-2}) \, .\label{Deltaexp2}
\end{align}
The leading $1/c$-expansion of $G_{-+}(z)$ \eqref{G-+} was obtained in \cite{Hikida:2017byl} (see also appendix \ref{sec:sym}).
In terms of global conformal block expansion, we have
\begin{align}
\begin{aligned}
G_{- +} (z) &= \frac{1}{|1 - z|^{2\Delta_+^{(0)}+2 \Delta_+^{(1)} / c }} \\
& + \frac{1}{c}\frac{1}{|1 - z|^{2 \Delta_+^{(0)} }}\left[ \sum_{s=2}^\infty (-1)^s\, \frac{C^{(s)}_- C^{(s)}_+}{B^{(s)} }\, (1-z)^s\, {}_2 F_1 (s ,s ; 2 s ; 1-z) + \text{c.c.} \right] + \mathcal{O} (c^{-2}) \, . 
\end{aligned}
\label{G-+exp}
\end{align}
Note that 
the hypergeometric function ${}_2 F_1 (s ,s ; 2 s , 1-z)$ (and its complex conjugate) corresponds to the global conformal block with intermediate operator $J_{(s)}$ (and $\bar J_{(s)}$) which are conserved tensors saturating the unitarity bound $\Delta =s$, as the result we have the simplification or holomorphic factorization.
This is also precisely what the Virasoro blocks reduce to at this order in $1/c$-expansion.

As explained in subsection \ref{sec:maldacena}, we need to extract contributions which can be obtained from lower-point functions. The first term  in $\eqref{G-+exp}$ corresponds to the contribution from the products of disconnected two-point functions,
\begin{align}
 \langle\, \mathcal{O}_-^\text{AdS} (\infty) \bar{\mathcal{O}}_-^\text{AdS} (0)\, \rangle  \langle\, \mathcal{O}_+^\text{AdS} (1) \bar{\mathcal{O}}_+^\text{AdS} (z)\, \rangle \, . 
\end{align}
More precisely at this order in $1/c$-expansion, the scaling dimension now consists of the tree level $\Delta_+^{(0)}$ and the extra one-loop correction $\Delta^{(1)}_+/c$. From the next line in \eqref{G-+exp}, we can extract the contribution of the form \eqref{A1d}. For this, we evaluate
\begin{align} \label{hatI}
\begin{aligned}
  &\frac{1}{|1 - z|^{2 (1+\lambda)} }   \hat{\mathcal{I}}_{s,0} (1-z,1-\bar z)  \\
  & \quad = \frac{k_{1-s,1} k_{s,0}}{\pi^2  B^{(s)}}\int d ^2 x\, d ^2 y\, 
\frac{     \langle\, \mathcal{O}_-^\text{AdS} (\infty)\, \bar{\mathcal{O}}_-^\text{AdS} (0)\, J_{(s)}^\text{AdS} (x)\, \rangle
   \langle\, J_{(s)}^\text{AdS} (y)\, \mathcal{O}_+^\text{AdS} (1)\, \bar{\mathcal{O}}_+^\text{AdS} (z) \, \rangle  }
         {{(x -y)^{2-2s} (\bar x - \bar y)^{2}}}  \, , 
\end{aligned}
\end{align}
where we have used the fact that the bi-tensorial projector simplifies in two dimensions.
As in \cite{Simmons-Duffin:2012juh}, it is convenient to define a shadow operator $\tilde \Phi_{1-h, 1 - \bar h}$ for a primary operator $\Phi_{h, \bar h}$ by
\begin{align}
    \tilde \Phi_{1-h,1- \bar h} (x) 
        =\frac{  k_{h,\bar h }}{\pi}
        \int d ^2 y\, \frac{1}{(x - y)^{2 - 2h} (\bar x - \bar y)^{2 - 2 \bar h}}\, \Phi_{h, \bar h} (y) \, . \label{defshadow}
\end{align}
The shadow operator $\tilde \Phi_{1-h, 1 - \bar h}$ satisfies the normalization condition:
\begin{align}
    \langle\, \tilde \Phi_{1 -h , 1 - \bar h} (x)\, \Phi_{h , \bar h} (y)\, \rangle 
        = \delta^{(2)}(x - y)
\end{align}
for $\langle \Phi_{h, \bar h} (z) \Phi_{h , \bar h} (0) \rangle = z^{-2h} \bar z^{-2 \bar h}$ and $k_{h , \bar h}$ defined in \eqref{K}.
In terms of the shadow operator, the integral \eqref{hatI} can be written as:
\begin{align}
     \begin{aligned}
      & 
    \frac{1}{|1 - z|^{2 (1+\lambda)} }\hat{\mathcal{I}}_{s,0} (1-z,1-\bar z)  \\
      & \quad 
      = \frac{\Gamma(2s)}{\pi B^{(s)}}\int d ^2 x\, \langle\, \mathcal{O}_-^\text{AdS} (\infty) \, \bar{\mathcal{O}}_-^\text{AdS} (0) \, J_{(s)}^\text{AdS}  (x)\, \rangle\,  \, \langle \, \tilde J_{(s)}^\text{AdS}  (x)\, \mathcal{O}_+^\text{AdS} (1)\, \bar{\mathcal{O}}_+^\text{AdS} (z) \, \rangle \, .
     \end{aligned}
 \end{align}
This is nothing but the definition of conformal partial wave up to an overall factor given as:
\begin{align}
 \hat{\mathcal{I}}_{s,0} (1-z, 1 - \bar z)  = (-1)^s\frac{C^{(s)}_- C^{(s)}_+}{B^{(s)}}\, \mathcal{I}_{s,0} (1-z , 1 - \bar z) \, .
\end{align}
The conformal partial wave $\mathcal{I}_{h,\bar h} (z , \bar z)$ is introduced in \eqref{cpw}, and in the current case it is written explicitly as
\begin{align}
\begin{aligned}
 \frac{\Gamma(s)^2}{\Gamma(2s) } \mathcal{I}_{s,0} (z , \bar z) 
& = \frac{\Gamma(s)^2}{\Gamma(2s) }z^s {}_2F_1 (s,s; 2s ; z) \\
& + \frac{\Gamma (2s -1)}{\Gamma(s)^2} z^{1-s} {}_2 F_1 (1-s,1-s;2-2s;z) \bar z {}_2F_1(1,1;2;\bar z) \, .
\end{aligned} \label{shadow}
\end{align}
Replacing the conformal blocks by the conformal partial waves, we have
\begin{align}
\label{G-+exps}
G_{- +} (z) & = \frac{1}{|1 - z|^{2 \Del_+^{(0)} +2 \Delta_{+}^{(1)}/c }}  \\
& +  \frac{1}{c}\frac{1}{|1 - z|^{ 2{ \Del_+^{(0)}}}}\left[ \sum_{s=2}^\infty (-1)^s\, \frac{C^{(s)}_- C^{(s)}_+}{B^{(s)} }\, \mathcal{I}_{s,0} (1 -z , 1 - \bar z) + ( \lambda^2 - 1 )\ln |z|^2   \right]+ \mathcal{O} (c^{-2}) \, , \nonumber
\end{align} 
see appendix \ref{sec:shadow} for its derivation.
The term proportional to $\ln |z|^2$ is the contribution which cannot be obtained from lower-point functions, i.e., four-point contact diagrams.

\subsubsection{Four-point functions \texorpdfstring{$G_{\pm\pm}(z)$}{Gpmpm}}

We next move to the other correlators listed in \eqref{4ptdS}.
These have not been dealt with in \cite{Chen:2022ozy} since expressions become more complicated and contain two possible exchange channels. However, we will see that similar analysis to the previous case can be applied to these correlators as well.
We use the four-point functions of scalar operators dual to scalar fields $\phi_\pm^\text{AdS}$ in Prokushkin-Vasiliev theory on AdS$_3$. The correlation functions are given by
\begin{align} \label{Gpmpm}
 G_{\pm\pm}(z) = \langle\, \mathcal{O}^\text{AdS}_\pm (\infty) \, \bar{\mathcal{O}}^\text{AdS}_\pm (1)\, \mathcal{O}^\text{AdS}_\pm (z)\, \bar{\mathcal{O}}^\text{AdS}_\pm (0 )\,\rangle \, .
\end{align}
They were also computed exactly in \cite{Papadodimas:2011pf} with finite $N,k$ as
\begin{align} \label{Gpp}
    \begin{aligned}
         G_{++}(z)&= \frac{1}{|z(1 -z)|^{2 \Delta_+}} \left | (1 -z)^{1 + \lambda}\, {}_2 F_1 \left( 1 + \frac{\lambda}{N} , - \frac{\lambda}{N} ; - \lambda  ; z \right) \right |^2  \\
        &  + \frac{\mathcal{N}_+ }{|z(1 -z)|^{ 2 \Delta_+}} \left | z^{1 + \lambda}\, {}_2 F_1 \left( 1 + \frac{\lambda}{N} , - \frac{\lambda}{N} ;  2 + \lambda  ; z \right ) \right |^2
    \end{aligned} 
\end{align}
with the parameter:
\begin{align} \label{Np}
\mathcal{N}_+ = - \frac{\Gamma \left(1 + \lambda - \frac{\lambda}{N}\right)\, \Gamma(-\lambda)^2\, \Gamma\left(2 + \lambda + \frac{\lambda}{N}\right)}{\Gamma\left(-1 - \lambda - \frac{\lambda}{N}\right) \, \Gamma\left(\frac{\lambda}{N} - \lambda\right) \Gamma(2 + \lambda)^2} \, ,
\end{align} 
and
\begin{align} \label{Gmm}
\begin{aligned}
G_{--}(z)&=\frac{1}{ |z(1 -z)|^{ 2 \Delta_-}} \left | (1 -z)^{1 - \lambda '}\, {}_2 F_1 \left( 1 - \frac{\lambda '}{N} ,  \frac{\lambda '}{N} ; \lambda '  ; z \right) \right |^2  \\
&  + \frac{\mathcal{N}_-}{ |z(1 -z)|^{ 2 \Delta_-}} \left | z^{1 -  \lambda '} \, {}_2 F_1 \left( 1 -  \frac{\lambda '}{N} ,  \frac{\lambda '}{N} ;  2 - \lambda '  ; z \right ) \right |^2 
\end{aligned}
 \end{align}
with the parameter:
\begin{align}\label{Nm}
\mathcal{N}_- = - \frac{\Gamma \left(1 - \lambda ' + \frac{\lambda '}{N}\right) \, \Gamma(\lambda ') ^2 \, \Gamma\left(2 - \lambda ' - \frac{\lambda '}{N}\right)}{\Gamma\left(-1 + \lambda ' + \frac{\lambda '}{N}\right)\, \Gamma\left( -\frac{\lambda '}{N} + \lambda '\right)\, \Gamma(2  - \lambda ')^2} \, .
\end{align}
Here we have set 
\begin{align*}
\lambda ' = \frac{N}{k + N +1} \, .
\end{align*}
Note that it becomes identical to $\lambda$ in \eqref{tHooft} at the leading 't Hooft limit. Even though it may not be obvious immediately, as shown in appendix \ref{app:s-t}, both $G_{\pm\pm}(z)$ are in fact invariant under the $s$-$t$ channel crossing transformation $(z, \bz) \to (1-z, 1-\bz)$. This also implies that both correlators contain $s$- and $t$-exchange channels to all orders of $k$ and $N$.

In order to apply the method introduced in subsection \ref{sec:maldacena}, we expand the four-point correlators in both of these channels.
As in the previous case, we first expand the four-point functions \eqref{Gpmpm} in terms of the global conformal blocks in both $s$- and $t$-channels and here we do this as (see \cite{Hikida:2017byl} and appendix \ref{sec:sym}) up to order $1/c$: 
\begin{align} \label{Gpmpmf}
\begin{aligned}
 G_{\pm \pm}(z) & = \frac{1}{|z|^{ 2\Delta_{\pm}^{(0)}+2 \Delta_{\pm}^{(1)}/c}} + \frac{1}{\left|1-z\right|^{ {2\Delta_{\pm}^{(0)}+2 \Delta_{\pm}^{(1)}/c }}}  \\
&+\frac{1}{c}
\Biggl[ \frac{1}{|z|^{2 \Delta_{\pm}^{(0)}}} \sum_{s=2}^\infty \frac{(C^{(s)}_\pm)^2}{ B^{(s)}}\, z^s\, {}_2 F_1 (s,s;2s;z) \\
& \quad \quad + \frac{1} {\left| 1-z\right|^{ 2 \Delta_{\pm}^{(0)}}}\sum_{s=2}^\infty  \frac{(C^{(s)}_\pm)^2}{B^{(s)}}\, (1-z)^s\, {}_2 F_1 (s,s;2s;1-z) + \text{c.c.}\Biggr]\\
&+ \frac{1}{c}\left[ \frac{1}{|z|^{ 2 \Delta_{\pm}^{(0)}} } g(\pm \lambda;\bar  z) + \text{c.c.} \right]
+ \mathcal{O} (c^{-2}) \, , 
\end{aligned}
\end{align}
with
\begin{align} \label{flz}
&g (\lambda ; \bar z) =   - \frac{\epsilon}{1 + \lambda}\left(\frac{\bar z}{1-\bar z}\right)^{ 1 + \lambda}\, , \quad \epsilon =(1 - \lambda^2)\Gamma(2 + \lambda)\, \Gamma(1-\lambda) \, .
\end{align}
On the right hand side, the first and second terms are respectively the products of disconnected two-point functions of scalar operators in $s$- and $t$-channels, while the third and fourth terms are respectively the exchange of spin-$s$ tensor field in $s$- and $t$-channels.  
The final line, when the both terms are taken together, becomes crossing invariant and can be attributed to the four point contact interactions. 
Notice that such a term does not appear for $G_{-+}(z)$ correlation function.
In order to precisely extract the parts which cannot be obtained from lower-point coefficient functions such as the contact interactions, we rewrite \eqref{Gpmpmf} in terms of conformal partial waves instead. 
Here we introduce the corresponding conformal partial wave:
\begin{align}\label{hatIpm} 
\begin{aligned}
& \frac{1}{|z|^{2 (1 \pm \lambda)}}\hat{\mathcal{I}}_{s,0}^\pm (z,\bar z) \\
&\quad = \frac{k_{1-s,1} k_{s,0}}{\pi^2 B^{(s)}}\int d ^2 x\, d ^2 y\, \frac{\langle\, \mathcal{O}_\pm^\text{AdS} (\infty)\, \bar{\mathcal{O}}_\pm^\text{AdS} (1)\, J_{(s)}^\text{AdS} (x)\, \rangle\, \langle\, J_{(s)}^\text{AdS} (y)\, \mathcal{O}_\pm^\text{AdS} (z)\, \bar{\mathcal{O}}_\pm^\text{AdS} (0) \, \rangle }{(x -y)^{2-2s} (\bar x - \bar y)^{2}}
        \\
        & \quad  = \frac{\Gamma(2s)}{\pi B^{(s)}}\int d ^2 x\, \langle\, \mathcal{O}_\pm^\text{AdS} (\infty) \, \bar{\mathcal{O}}_\pm^\text{AdS} (1) \, J_{(s)}^\text{AdS}  (x)\, \rangle\,  \langle \, \tilde J_{(s)}^\text{AdS}  (x)\, \mathcal{O}_\pm^\text{AdS} (z)\, \bar{\mathcal{O}}_\pm^\text{AdS} (0) \, \rangle \, ,
\end{aligned}
\end{align}
where we have substituted the definition of shadow operator $\tilde{J}_{(s)}^{\rm AdS}(x)$. The integral is proportional to the global conformal partial wave:
\begin{align}
 \hat{\mathcal{I}}_{s,0}^\pm (z,\bar z)  =  \frac{(C^{(s)}_\pm)^2}{B^{(s)}}\, \mathcal{I}_{s,0} (z , \bar z) \, .
\end{align}
Thus, we can rewrite \eqref{Gpmpmf} as
\begin{align} \label{Gpmpmp}
\begin{aligned}
 G_{\pm \pm}(z) &= \frac{1}{|z|^{2\Delta_{\pm}^{(0)}+2 \Delta_{\pm}^{(1)}/c} } + \frac{1}{\left|1-z\right|^{2\Delta_{\pm}^{(0)}+2 \Delta_{\pm}^{(1)}/c} } \\ 
&+ \frac{1}{c}  
\Biggl[ \frac{1}{|z|^{ 2\Delta_{\pm}^{(0)}}} \left( \sum_{s=2}^\infty \frac{(C^{(s)}_{\pm})^2}{B^{(s)}} \mathcal{I}^\pm_{s,0} (z, \bar z) +(\lambda^2 - 1 ) \ln (1 -  \bar z ) \right) 
\\ 
& \quad \quad + \frac{1}{| 1-z |^{ 2 \Delta_{\pm}^{(0)}}} \left(\sum_{s=2}^\infty  \frac{(C^{(s)}_{\pm})^2}{B^{(s)}}  \mathcal{I}^\pm_{s,0} (1-z,1 -\bar z) + (\lambda^2 - 1 ) \ln \bar z \right) + \text{c.c.}  \Biggr] \\
 &  + \frac{1}{c}\left[ \frac{1}{|z|^{ 2 \Delta_{\pm}^{(0)}} } g(\pm \lambda;\bar  z) + \text{c.c.} \right]+  \mathcal{O} (c^{-2}) \, . 
 \end{aligned}
\end{align}
The contact interaction terms which cannot be obtained from lower-point coefficient functions are thus summarized as
\begin{align}
\begin{aligned}
    \mathcal{A}^{\pm}_{4,c} (z , \bar z) &= 
\frac{1}{c}(\lambda^2 - 1 ) \left[  \frac{1}{|z|^{ 2\Delta_{\pm}^{(0)}}} \ln |1 - z |^2 +  \frac{1}{\left| 1-z\right|^{ 2\Delta_{\pm}^{(0)}}} \ln | z |^2 \right] \\
 &  - \frac{1}{c}\frac{\epsilon}{1 \pm \lambda} \left[\frac{1}{\left(z (1-\bar z )\right)^{ \Delta_{\pm}^{(0)} } }+\frac{1}{ \left(\bar z (1- z )\right)^{ \Delta_{\pm}^{(0)}}} \right] + \mathcal{O} (c^{-2})
\end{aligned} \label{A4c}
\end{align}
as extracted from boundary CFT.

Before ending the analysis in the case of AdS$_3$, we would like to make comments on the bulk interpretation of the four-point interaction terms, which should be of the form
\begin{align}
 \int_{\rm AdS_3} d X \sum_{s_1,s_2,s_3,s_4 =0}^\infty a_{s_1,s_2,s_3,s_4}\nabla^{s_1} \phi_\pm^\text{AdS} (X) \nabla^{s_2} \bar \phi_\pm^\text{AdS} (X) \nabla^{s_3} \phi_\pm^\text{AdS} (X) \nabla^{s_4} \bar \phi_\pm^\text{AdS} (X) 
\end{align}
where $s_{1,\dots 4}$ are non-negative integers and $a_{s_1,s_2,s_3,s_4}$ are the tensorial constant coefficients.
See appendix \ref{app:embedding} for notations. 
Note that, since a dimensionless combination is given by $\ell_\text{AdS} \nabla$, the higher derivative terms would contribute unless we consider highly curved regime with $\ell_\text{AdS} \sim 0$.
In principle, we can fix the coefficient constants $a_{s_1,s_2,s_3,s_4}$ such as to reproduce the CFT result. However, it is a cumbersome task due to the ambiguity associated with gauge fixing and equations of motion. As a future important problem, we would like to evaluate them. For a similar problem on AdS$_4$, see, e.g., \cite{Bekaert:2014cea,Bekaert:2015tva}.

\subsection{Analytic continuation from \texorpdfstring{AdS$_3$}{AdS3} to \texorpdfstring{dS$_3$}{dS3}}\label{Sec: 4pt exp}

In the previous subsection, we obtain the terms which cannot be obtained from lower-point coefficient functions, i.e., $n$-point AdS bulk contact diagrams. In order to obtain dS four-point coefficient functions from these terms, we have to perform an analytic continuation from AdS$_3$ to dS$_3$. 
We have already defined how individual operators are mapped from AdS$_3$ to dS$_3$ as in \eqref{omap} and \eqref{jmap} in general. We have also observed that we have extra $-i$ factor if correlation functions are proportional to $c$ due to the map which replaces $c$ by  $- i c^{(g)}$ with real $c^{(g)}$.
Up to now, we have considered two-point functions of scalar operators and conserved currents, and the three-point functions of scalar-scalar-higher spin currents. We have observed that all correlation functions are proportional to $c$, which induces $-i$ factor by mapping from AdS$_3$ to dS$_3$.

Let us consider connected $n$-point functions of certain operators in the 2d CFT dual to higher-spin AdS$_3$ gravity in general. Applying an operator product expansion (OPE), we can replace two neighboring local operators by another one. The coefficients of OPE are thus given by three-point functions divided by two-point function. Therefore, repeating OPEs, any connected $n$-point function can be written as a product of $(n-2)$ three-point functions divided by $(n-3)$ two-point functions, regardless of OPE channels. Thus, the connected $n$-point function is always proportional to $c$. After mapping from AdS$_3$ to dS$_3$, we have an extra phase factor $-i$ for all connected $n$-point function. This is consistent with the fact that the coupling constant in the bulk theory is organized by $k_\text{CS}$ or $- i \kappa$ after analytic continuation.

\subsubsection{Four-point correlator from \texorpdfstring{$G_{-+}(z)$}{G-+}}

We again begin with the first correlator in \eqref{4ptdS}.
According to the above arguments, the connected part of four-point correlation functions are related as
\begin{align}
\begin{aligned}
\langle\, \mathcal{O}_- (\infty) \, \mathcal{O}_+ (1)\, \tilde{\mathcal{O}}_+ (z)\, \tilde{\mathcal{O}}_- (0 )\, \rangle &= - i e^{ 2 \pi i(h_+ + h_-)}
\langle\, \mathcal{O}^\text{AdS}_- (\infty) \, \mathcal{O}^\text{AdS}_+ (1)\, \bar{\mathcal{O}}^\text{AdS}_+ (z)\, \bar{\mathcal{O}}^\text{AdS}_- (0 )\, \rangle \\
&= - i 
\langle\, \mathcal{O}^\text{AdS}_- (\infty) \, \mathcal{O}^\text{AdS}_+ (1)\, \bar{\mathcal{O}}^\text{AdS}_+ (z)\, \bar{\mathcal{O}}^\text{AdS}_- (0 ) \rangle \, .
\end{aligned}
\label{OOOO}
\end{align}
We have used \eqref{omap} and $h_+ + h_- = 1$ for $d=2$ to obtain the factor $e^{2 \pi i (h_+ + h_-)} =1$. Another factor $-i$ comes from the fact that the connected four-point correlators are of order $\mathcal{O}(c)$ as explained above.
The term proportional to $\ln |z|^2$ in \eqref{G-+exps} is a real function. Therefore, due to the $-i$ factor in \eqref{OOOO}, we conclude
\begin{align}
\langle \mathcal{O}^4 \rangle_c = 0
\end{align}
in \eqref{O4c}, which only contains real quantities by definition. 
Therefore, in this case, there are no terms corresponding to the bulk contact diagram as already mentioned in \cite{Chen:2022ozy}.
The {remaining} part $\langle \mathcal{O}^4 \rangle_d$ exclusively corresponds to the sum of bulk exchange diagrams. Using the expression of conformal partial waves given in \eqref{hatI}, we can obtain the resultant phase factor after the map from AdS$_3$ to dS$_3$. The result is
\begin{align}\label{Od4+-}
  \langle \mathcal{O}^4 \rangle_d =\sum_{s=2}^\infty  \frac{\lambda_{h_+,h_+,s}\lambda_{h_-,h_-,s}}{a_{h_+}^2 a_{h_-}^2 a_{(s)} c^{(g)}}  \frac{1}{|1 - z|^{2 (1 + \lambda)}} ( \hat{\mathcal{I}}_{s,0} (1-z,1-\bar z)  - \hat{\mathcal{I}}_{0,s} (1-z,1-\bar z) ) \, ,
\end{align}
where $a_{h_\pm}$, $a_{(s)}$  and  $\lambda_{h_\pm, h_\pm, s}$ are defined in \eqref{aDelta}, \eqref{as} and \eqref{phase3pt} respectively.
Applying \eqref{inverse1} as in \eqref{4pt}, we find
\begin{align}\label{phase4pt1}
\begin{aligned}
&\langle\, \phi_- (\infty)\,  \phi_+ (1)\, \bar \phi_+ (z)\, \bar \phi_- (0 )\,\rangle_c \\
& \quad =  \sum_{s=2}^\infty  \frac{\lambda_{h_+,h_+,s}\lambda_{h_-,h_-,s}}{ a_{(s)}}\frac{\mathcal{B}_{-+}(s) }{|1 - z|^{2 (1 - \lambda)}} ( \hat{\mathcal{I}}_{s,0} (1-z,1-\bar z)  - \hat{\mathcal{I}}_{0,s} (1-z,1-\bar z) )
\end{aligned}
\end{align}
with
\begin{align}
    \mathcal{B}_{-+}(s) = \frac{\Gamma(- \lambda)\Gamma(1 - \lambda) \Gamma(\lambda) \Gamma(1 + \lambda)}{\Gamma(1 - s + \lambda) \Gamma (s + \lambda)\Gamma(1 - s - \lambda) \Gamma (s - \lambda)}   \, .
\end{align}
The result is consistent with the generic expression, say, in (4.39) of \cite{Sleight:2021plv}.
{Here the overall factor $\mathcal{B}_{-+}(s)$ was inherited from the pre-factor in three-point inversion formula \eqref{3pt-inv}.}

\subsubsection{Four-point correlator from \texorpdfstring{$G_{\pm\pm}(z)$}{Gpmpm}}

We next consider the analytic continuation of the other more complicated correlation functions in \eqref{4ptdS}.
Applying the similar procedures as above, the relation among the connected part of four-point correlators is obtained as
\begin{align}
\langle\, \mathcal{O}_\pm (\infty)\, \tilde{\mathcal{O}}_\pm (1) \, \mathcal{O}_\pm (z)\, \tilde{\mathcal{O}}_\pm (0 )\, \rangle = - i  e^{4 \, i \, \pi h_\pm} \, 
\langle\, \mathcal{O}^\text{AdS}_\pm (\infty) \, \bar{\mathcal{O}}^\text{AdS}_\pm (1)\, \mathcal{O}^\text{AdS}_\pm (z)\, \bar{\mathcal{O}}^\text{AdS}_\pm (0 )\, \rangle \, , \label{OOOO2}
\end{align}
which leads to the contributions of the four point contact diagram being:
\begin{align}
    &\langle \mathcal{O}^4 \rangle_c = 
    - 2 \sin (4 \pi h_\pm)  \mathcal{A}^\pm_{4,c} (z , \bar z) \, ,
\end{align}
where $\mathcal{A}^\pm_{4,c}(z, \bz)$ is given in \eqref{A4c}.
The remaining exchange part \eqref{O4d} can be obtained from the expression of conformal partial waves in \eqref{hatIpm} as
\begin{align}
\begin{aligned}
  \langle \mathcal{O}^4 \rangle_d&  = \Biggl[  \frac{1}{|z|^{2 (1 \pm \lambda)} }\sum_{s=2}^\infty
  \frac{(\lambda_{h_\pm,h_\pm,s})^2}{a_{h_\pm}^4 a_{(s)}c^{(g)}}
  ( \hat{\mathcal{I}}_{s,0}^\pm (z,\bar z)  - \hat{\mathcal{I}}_{0,s}^\pm (z,\bar z) ) \\
  & +\frac{1}{|1-z|^{2 (1 \pm \lambda)}} \sum_{s=2}^\infty 
  \frac{(\lambda_{h_\pm,h_\pm,s})^2}{a_{h_\pm}^4 a_{(s)}c^{(g)}}
  (\hat{\mathcal{I}}_{s,0}^\pm (1-z,1-\bar z)  - \hat{\mathcal{I}}_{0,s}^\pm (1-z,1-\bar z) ) \Biggr]   \, .
  \end{aligned}
\end{align}
The bulk four-point correlators in late time dS$_3$ can thus be divided into two parts as
\begin{align}
\begin{aligned}
    &\langle\, \phi_\pm (\infty)\,  \bar \phi_\pm (1)\, \phi_\pm (z)\, \bar \phi_\pm (0 )\,\rangle   = \langle \phi^4 \rangle_c + \langle \phi^4 \rangle_d \, .
    \end{aligned}
\end{align}
The first part can be rewritten in terms of the following inversion integral:
\begin{align} \label{4ptcontact}
   \langle \phi^4 \rangle_c  =  - 2 \sin (4 \pi h_\pm)  a_{h_\pm}^4 \int \left[ \prod_{i=1}^4 d^2z_i ' K(z_i ,z_i ') \right]\frac{|z'|^{2 (1 \pm \lambda)} }{|z_{1'2'}|^{2 (1 \pm \lambda)} |z_{3' 4'}|^{2 (1 \pm \lambda)}}  \mathcal{A}^\pm_{4,c}(z')
\end{align}
with cross ratio $z' = (z_{1 '2 '} z_{3 '4 '} )/(z_{1 '3 '} z_{2 '4 '})$ 
and
$(z_1,z_2,z_2,z_4) = (\infty,1,z,0)$. It is possible to perform the integration over $z_i'$, but we do not do so since it is not so illuminating.

For the second part, the integration over $z_i'$ in \eqref{4pt}
can be easily performed as 
\begin{align} \label{phase4pt2}
  \langle \phi^4 \rangle_d
 & =  \Biggl[  \frac{1}{| z|^{2 (1 \mp \lambda)}} \sum_{s=2}^\infty
 \frac{(\lambda_{h_\pm,h_\pm,s})^2}{ a_{(s)} c^{(g)}} \mathcal{B}_\pm(s)
 ( \hat{\mathcal{I}}_{s,0}^\pm (z,\bar z)  - \hat{\mathcal{I}}_{0,s}^\pm (z,\bar z) ) \nonumber \\
  &  +\frac{1}{|1-z|^{2 (1 \mp \lambda)}} \sum_{s=2}^\infty 
  \frac{(\lambda_{h_\pm,h_\pm,s})^2}{ a_{(s)} c^{(g)}} \mathcal{B}_\pm(s)
  (\hat{\mathcal{I}}_{s,0}^\pm (1-z,1-\bar z)  - \hat{\mathcal{I}}_{0,s}^\pm (1-z,1-\bar z) ) \Biggr]   
\end{align}
with
\begin{align}
    \mathcal{B}_{\pm}(s) = \left[ \frac{\Gamma(\mp \lambda)\Gamma(1 \mp \lambda) }{\Gamma(1 - s \pm \lambda) \Gamma (s \pm \lambda)}  \right]^2 \, .
\end{align}
This is again consistent with the result of \cite{Sleight:2021plv}. Note that we have now the explicit expression of the contribution from bulk four-point contact diagram \eqref{4ptcontact}, which is not available from generic arguments previously.

Let us end this section about the four-point correlation functions up to the first order in ${1}/{c}$- or ${1}/{c^{(g)}}$-expansions by outlining the explicit construction of their holographic dual configurations. In particular, we focus on the terms, which are further expanded in terms of the conformal partial waves for the exchange of holomorphic conserved tensor operators. These terms correspond to the tree level bulk exchange diagrams.
In \cite{Chen:2017yia}, inspired by the split representation of the AdS harmonic functions, so-called geodesic three-point Witten diagram was introduced, whose interaction vertex moves along the geodesic connecting two of the three boundary points. Such an object is purely kinematical and differs from the usual three-point Witten diagram by an overall factor which can be attributed to the strong coupling OPE coefficient. We can thus use the result in appendix \ref{sec:hgauge}, which applies the bulk to boundary propagator in the holographic gauge to reproduce the three-point correlation function of scalar-scalar-higher spin currents including the desired OPE coefficient, see \eqref{AdS-3pt}. In appendix \ref{Sec:hblock}, the holomorphically factorizable two-dimensional global conformal block was further expressed in terms of an integral over two copies of three-point geodesic Witten diagrams, utilizing the holomorphic polarization vector, see \eqref{3pt-sGWD}. 
Combining these results, along with the analytic continuation of the bulk to boundary propagators in going from AdS to dS (c.f. \cite{Sleight:2021plv}), we can directly reconstruct the holographic configurations in both AdS and dS spacetimes for these correlation functions. 

\section{Example of other late-time correlator}
\label{sec:kto-2}

Up to now, we have only examined late-time correlation functions on the pure dS$_3$ spacetime at the leading order in the Newton constant $G_N$ or the inverse of central charge $1/c$ by taking the 't Hooft limit. Our explicit formulation however is not restricted to these cases, and it can be applied to more generic situations. In order to illustrate this, we examine two interesting example, though full analysis is left as a future problem.
In this section, we analyze the two-point correlator of scalar field on dS$_3$ conical geometry. 
In general, the scalar two-point correlator on a geometry can be used to extract properties of the geometry, and hence the analysis should be useful for the purpose.
In the next section, we consider another example that is a type of one-loop correction to two-point correlator on pure dS$_3$ spacetime. We would like to work on the other types of one-loop correction in near future. In any case, it is quite important problem to deal with $1/c$-corrections systematically and a preliminary work is done in subsection \ref{Sec:4pt exp2}.

In order to compute such bulk correlators, it is convenient to utilize the limit used in \cite{Hikida:2021ese,Hikida:2022ltr}. 
So far, we have examined the `t Hooft limit where $N,k \gg 1$ but keep $\lambda = N/(N+k)$ of order $\mathcal{O}(N^0)$.
Instead of this, we consider dS$_3$ holography at the regime where $k \sim -1 - N$ in this section.%
\footnote{The limit taken in \cite{Hikida:2021ese,Hikida:2022ltr} is $k \to - N$, which is slightly different from \eqref{limitdS}. If we write $k$ in terms of $c$, then there are two branches and the difference appears due to the choice of branch, see, e.g., \cite{Perlmutter:2012ds}.} 
In AdS$_3$ holography, a similar limit was considered in \cite{Castro:2011iw,Gaberdiel:2012ku,Perlmutter:2012ds}. In the case, the higher-spin gravity is described by  $\text{SL}(N) \times \text{SL}(N)$ Chern-Simons gauge theory. Due to the truncation of spin, the gravity theory is more tractable than that with infinite dimensional gauge algebra $\mathfrak{hs}[\lambda]$ appeared in the case with 't Hooft limit \eqref{tHooft}. Moreover, we can argue that some states of the coset CFT \eqref{cosetCFT} in the limit $k \to - 1 - N$ have conformal dimensions of order $\mathcal{O}(c)$, which have been identified with conical defect geometry in \cite{Castro:2011iw,Perlmutter:2012ds} for AdS$_3$ and in \cite{Hikida:2021ese,Hikida:2022ltr} for dS$_3$. 
In the next subsection, we review the dS$_3$/CFT$_2$ correspondence with the critical limit of the coset \eqref{cosetCFT} introduced in \cite{Hikida:2021ese,Hikida:2022ltr}.
In subsection \ref{sec:conical} we compute bulk two-point correlators on dS$_3$ conical defect geometry at late time.

\subsection{Higher-spin \texorpdfstring{dS$_3$/CFT$_2$}{dS3/CFT2} with truncated spin}

We begin by reviewing the higher-spin AdS$_3$ holography with the limit of \cite{Castro:2011iw,Gaberdiel:2012ku,Perlmutter:2012ds} and its analytic continuation to the higher-spin dS$_3$ holography  as in \cite{Hikida:2021ese,Hikida:2022ltr}.
We consider the coset model \eqref{cosetCFT} with finite $N$, which possesses $W_N$-algebra symmetry with spin-$s$ currents $(s=2,3,\ldots ,N)$.
The central charge is bounded as $c < N-1$ for integer $k$, which can be seen from \eqref{central}. In order to be dual to classical gravity, the central charge should be large. In \cite{Castro:2011iw,Gaberdiel:2012ku,Perlmutter:2012ds}, the large central charge is realized by performing an analytic continuation of $k$ as
\begin{align}
k = - 1 -N + \frac{N(N^2 -1)}{c} + \mathcal{O} (c^{-2}) \, . \label{limit}
\end{align}
Because of this analytic continuation, the unitarity of the theory is lost.

The conformal dimensions of operators $\mathcal{O}_\pm$ are given in \eqref{Deltapm}.
In the limit of \eqref{limit}, they become
\begin{align}
\Delta_+ = 1 -N - \frac{(-1+N^2)^2}{c} + \mathcal{O} (c^{-2}) \, , \quad 
\Delta_- = - \frac{c}{N^2} + \mathcal{O}(c^0) \, . \label{Deltalimit}
\end{align}
Note that the scaling dimensions are negative, which is a consequence of non-unitarity.
The operator $\mathcal{O}_+$ is dual to a perturbative matter, while the other operator $\mathcal{O}_-$ is dual to a non-perturbative gravity solution.
More generically, the operators of the coset are labeled by two Young diagrams $(\Lambda_+  ; \Lambda_-)$ as explained in subsection \ref{sec:GG}.
At the limit \eqref{limit}, the conformal dimension of operator labeled by $(0;\Lambda_-)$ become
\begin{align}
\Delta_{(0;\Lambda_-)} = 2 h_{(0;\Lambda_-)} = - \frac{2 c \, C_2 (\Lambda_-)}{N (N^2 -1)} + \mathcal{O}(c^0) \, ,
\end{align}
where \eqref{conformalweight} has been used and $C_2(\Lambda_-)$ defines the eigenvalue of the quadratic Casimir operator.

For real $c$, the dual gravity is supposed to be $\text{SL}(N) \times \text{SL}(N)$ Chern-Simons gauge theory described by the action \eqref{CSaction} with real $k_\text{CS}$. A complex matter with mass
\begin{align}
\ell^2_\text{AdS} m^2 = N^2 -1
\end{align}
is coupled with the gauge fields. The coset operator labeled by $(0;\Lambda_-)$ is dual to conical defect geometry with the same label $\Lambda_-$ \cite{Castro:2011iw,Perlmutter:2012ds}. Generic coset operator labeled by $(\Lambda_+;\Lambda_-)$ is then dual to a bound state of the conical geometry labeled by $\Lambda_-$ and composites of matter fields corresponding to $\Lambda_+$.

As in \cite{Hikida:2021ese,Hikida:2022ltr}, we perform further analytic continuation in order to map from AdS$_3$/CFT$_2$ to dS$_3$/CFT$_2$. In the CFT side, we have proposed to take the limit \eqref{limit} but with $c = - i c^{(g)}$ $(c^{(g)} \in \mathbb{R})$. This is motivated by the fact that higher-spin dS$_3$ gravity can be obtained by taking $k_\text{CS} = - i \kappa$ with $\kappa \in \mathbb{R}$ \cite{Witten:1988hc} and the asymptotic symmetry near the boundary is given by $W_N$-algebra with central charge $c= 6 k_\text{CS} = - i 6 \kappa$. In this way, we could propose a dS$_3$/CFT$_2$ correspondence. The gravity side is 
the $\text{SL}(N) \times \text{SL}(N)$ Chern-Simons gauge theory with level $- i \kappa$ $(\kappa \in \mathbb{R})$. dS$_3$ is realized as a solution to the equations of motion, and the level is written as $\kappa = \ell /4 G_N$. The CFT side is given by
the coset model \eqref{cosetCFT} with the limit
\begin{align}
k = - 1 -N + i  \frac{N(N^2 -1)}{ c^{(g)}} + \mathcal{O} (c^{(g)-2}) \, . \label{limitdS}
\end{align}
As evidence, gravity partition functions were computed from the both sides and the two computations were found to agree with each other \cite{Hikida:2021ese,Hikida:2022ltr}. In these papers, other supporting arguments were also provided.

\subsection{Two-point correlator on \texorpdfstring{dS$_3$}{dS3} conical defect}
\label{sec:conical}

In this subsection, we compute bulk two-point function of scalar field
\begin{align}
\langle \phi_+ (z_1) \bar \phi_+ (z_2) \rangle_{\Lambda_-}
\label{2ptconical}
\end{align} 
at late time but on dS$_3$ conical defect geometry instead from dual CFT \eqref{cosetCFT}. The conical defects are obtained by the solutions of $\text{SL}(N) \times \text{SL}(N)$ Chern-Simons gravity. The conical defects on Euclidean AdS$_3$ have been classified in \cite{Castro:2011iw}, and those on Lorentzian dS$_3$ could be obtained by an analytic continuation.

We first consider dS$_3$ solution corresponding to the principal embedding of $\mathfrak{sl}(2)$ into $\mathfrak{sl}(N)$.
Following the notation in \cite{Hikida:2021ese,Hikida:2022ltr}, the generators of $\mathfrak{su}(2)$ are expressed by
\begin{align}
L_3 = \sum_{i=1}^N \rho_i e_{i,i} \, , \quad \text{tr}\,  L_3 L_3 = \frac{N(N^2-1)}{12}  \, (\equiv \epsilon_N )
\end{align}
and $L_{1},L_2$ satisfying $[L_a,L_b] = \frac{i}{2} \epsilon_{abc} L_{c} $. Here we use $N \times N$ matrix $(e_{i,j})_k^{~l} = \delta_{i,k} \delta_{j}^{~l}$. In this notation, the level of Chern-Simons theory is related to Newton constant as $\kappa = \ell/(8 G_N \epsilon_N)$. The metric can be read off from gauge fields as
\begin{align}
g_{\mu \nu} = - \frac{\ell^2}{4 \epsilon_N} \text{tr} \, [ (A_\mu - \bar A_\mu) (A_\nu - \bar A_\nu)] \, .
\end{align}
Fixing a gauge, we can put the gauge fields into the form 
\begin{align}
A = e^{ - i  \theta L_0}  a (z) e^{i \theta L_0} + i L_0 d \theta  \, , \quad \bar A = e^{i  \theta L_0} \bar a (\bar z)e^{-  i  \theta L_0} - i L_0 d \theta  
\end{align}
with $a = a_+ (\d \phi + i  \d \tau) $ and $\bar a =  \bar a_- ( \d \phi - i  \d \tau)$. With the gauge configuration
\begin{align}
a_+ =  i  L_{1}  \, , \quad \bar a_- = i  L_{1} \, , 
\end{align}
we find
\begin{align}
d s^2 = \ell^2 (d \theta^2  - \cos^2 \theta d \tau^2 + \sin ^2 \theta d \phi^2  )\, .
\end{align}
This is a metric for a static patch of dS$_3$. If we perform a Wick rotation as $\tau \to - i t_\text{E}$, then the metric is for three-sphere as utilized in \cite{Hikida:2021ese,Hikida:2022ltr}.

We then consider more generic gauge fields but corresponding to geometry without any rotation.
Following \cite{Hikida:2021ese,Hikida:2022ltr} (motivated by the AdS case in \cite{Castro:2011iw}), we assign the condition of trivial holonomy around $\phi$-cycle around $\theta =0$. We define the holonomy matrix by
\begin{align}
\text{Hol}_\phi (a_+) =  \mathcal{P} \exp \left( \int_{\theta=0} a_+ \d \phi \right) \, , 
\end{align}
where $\mathcal{P}$ represents the path ordering.
We would like to require the trivial holonomy condition. The gauge group is now SL$(N,\mathbb{C})$, so the center of the group is $Z_N$. Therefore, we require that
\begin{align} \label{trivial}
\text{Hol}_\phi (a_+) \propto e^{- \frac{2 \pi i m}{N}} \mathbbm{1} \, , \quad  m \in \mathbb{Z}_N \, .
\end{align}
The situation is the same as in the case of Euclidean AdS$_3$ analyzed in \cite{Castro:2011iw}, since the Lorentzian dS$_3$ is related to the Euclidean AdS$_3$ by a simple analytic continuation $\ell^\text{AdS} = - i \ell$ \eqref{map} or $k_\text{CS} = - i \kappa$ \eqref{cg}. Note that the conical defects analyzed in \cite{Hikida:2021ese,Hikida:2022ltr} are solutions in $\text{SU}(2) \times \text{SU}(2)$ Chern-Simons theory, thus the trivial holonomy condition is different from the current case. For the difference between Lorentzian and Euclidean AdS$_3$ cases, see \cite{Castro:2011iw}.

Denoting the eigenvalues of $a_+$ as $i n_j$, the triviality condition becomes
\begin{align}
 n_j = m_j - \frac{m}{N}\, , \quad m_j \in \mathbb{Z} \text{ for all }j \, .
\end{align}
Due to the diagonalizability, all $n_j$ should be distinct and we can set $n_1 > n_2 > \cdots > n_N$. These numbers represent  a Young diagram with $r_j$ box in $j$-th row, and relation of parameters is
\begin{align}
n_j = r_j + \frac{N+1}{2} - \frac{B}{N} - j \, , \quad B = \sum_j r_j \, .
\end{align}
In this way, we have constructed conical defect geometry satisfying the trivial holonomy condition as in the case of AdS analyzed in \cite{Castro:2011iw}, and the geometry is labeled by a Young diagram.

We would like to next evaluate bulk two-point correlator \eqref{2ptconical} on dS$_3$ conical geometry at late time. For this, we need the AdS counterpart of the correlator. This is given by four-point function with insertions of $\mathcal{O}_+^\text{AdS}$ and $\mathcal{O}_{(0;\Lambda_-)}^\text{AdS}$ and their conjugates, i.e.,
\begin{align} \label{4ptconical}
G(z, \bar z) = \langle \mathcal{O}^\text{AdS}_{(0;\Lambda_-)} (\infty) \bar{\mathcal{O}}^\text{AdS}_+ (1)  
\mathcal{O}^\text{AdS}_+ (z) \bar{\mathcal{O}}^\text{AdS}_{(0;\Lambda_-)} (0) \rangle \, .
\end{align}
In the limit \eqref{limit}, it becomes a heavy-heavy-light-light correlator. Therefore, the correlator can be computed by the bulk scalar propagator on the conical defect labeled by $\Lambda_-$ as \cite{Hijano:2013fja}
\begin{align}
G(z , \bar z) =\left| (N-1) ! z^{\frac{N-1}{2}} \sum_{j=1}^N \frac{z^{n_j}}{\prod_{l \neq j} (n_l - n_j) }\right| ^2 \, .
\end{align}
We can obtain the above expression from dual CFT as well.
Restoring the coordinate dependence, we have:
\begin{align}
\begin{aligned}
G(z_i, \bar z_i) &= \langle \mathcal{O}^\text{AdS}_{(0;\Lambda_-)} (z_1) \bar{\mathcal{O}}^\text{AdS}_+ (z_2)  
\mathcal{O}^\text{AdS}_+ (z_3) \bar{\mathcal{O}}^\text{AdS}_{(0;\Lambda_-)} (z_4) \rangle \\
 &= \left| \frac{z}{z_{12} z_{34}} \right|^{\Delta_- + (1 - N)} \left|\frac{z_{24} z_{13}}{z_{14}^2} \right|^{\Delta_- - (1 -N)}  G(z , \bar z)
\end{aligned}
\end{align}
with 
\begin{align}
z = \frac{z_{12} z_{34}}{z_{13} z_{24}} \, , \quad \bar z =  \frac{\bar z_{12} \bar z_{34}}{\bar z_{13} \bar z_{24}} \, .
\end{align}

Now we would like to move to the case with dS$_3$ conical defect. For this, we take the limit \eqref{limitdS}. The gravity dual of $\mathcal{O}_{(0;\Lambda)}$ is given by the dS$_3$ conical defect. On the other hand, the gravity dual of $\mathcal{O}_{+}$ is a bulk perturbative scalar $\phi_+$. Therefore, from the four-point function \eqref{4ptconical}, we can read off the information about the two-point coefficient function on dS$_3$ conical defect. Applying the inverse functions \eqref{K}, the correlator \eqref{2ptconical} can be expressed as
\begin{align}
\langle \phi_+ (z_2 ') \phi_+ (z_3 ') \rangle_{\Lambda_-} 
= \frac{a_{h_+}^2 k_{h_+,h_+}^2}{\pi^2}  \int d^2 z_2 d^2 z_3 \frac{1}{|z_{2 ' 2}|^{2(1 + N)} 
 |z_{ 3 ' 3 } |^{2 (1 + N)}} G(z_i, \bar z_i)  \, .
\end{align}
We put the conical defect from $z_4 \to 0$ to $z_1 \to \infty$ and the scalar propagator from $z_2 ' \to z_1$ to $z_3 ' \to z_2$. 
We may observe divergence from $a_{h_+}^2$ at $2h_+ = 1 - N$. However, the conformal dimension $\Delta_+$ receives quantum corrections as mentioned above, so we may slightly shift $N$ as $N + \epsilon$ with $\epsilon \to 0$ and absorb $1/\epsilon$ by the normalization of scalar field. 
We are now examining its bulk interpretation and would like to report on the result in near future.

\section{Comments on the loops in dS spacetime} \label{Sec:dS loop}

In this section, we present some partial new results on the loop computations in the dS spacetime and higher order $1/c$-expansions of the dual CFT correlation functions. We are planing to continue systematic investigations on higher order corrections in $1/c$ and the results we obtain in this section hopefully will become relevant in the subsequent studies.

\subsection{One-loop corrections to two-point correlator}
\label{sec:oneloop0}

In this subsection, we consider the two-point correlator of scalar field $\phi_+$ on pure dS$_3$ at late time,
\begin{align}
\begin{aligned}
\langle \phi_+ (\vec x) \bar{\phi}_+ (0) \rangle &= \frac{\mathcal{C}_0 + \delta \mathcal{C}}{|x^2|^{2 - \Delta_0 - \delta \Delta}} \\
&= \frac{\mathcal{C}_0}{|x^2|^{2-\Delta_0}} + \delta \Delta \frac{\mathcal{C}_0}{|x^2|^{2- \Delta_0}} \ln |x^2| + \frac{\delta \mathcal{C}}{|x^2|^{2-\Delta_0}} + \mathcal{O} (c^{-2}) \, , 
\end{aligned}
\end{align}
where $\Delta_0 =2 h_+ = 1 -N$ and $\delta \Delta, \delta \mathcal{C}$ is assumed to be of order $\mathcal{O}(c^{-1})$. In the following, we only keep the term proportional to $\delta \Delta$ and neglect the term proportional to $\delta \mathcal{C}$.
Typically, a divergence arises due to the infinite sum over spin $s=2,3,\ldots,\infty$ for higher-spin gauge fields running over the loop. 
In order to avoid the difficulty, we first work on the holography with the limit \eqref{limit}, where there is an upper bound of spin as $s \leq N$ for tensor fields. We then discuss the case with the limit \eqref{tHooft}.

For the current analysis, it is convenient to move to the momentum basis. We introduce Fourier transform formula:
\begin{align}
     \int \frac{d^2 x}{|x^2|^\Delta} e^{i x \cdot p} = \frac{\Gamma(1 - \Delta)}{\Gamma(\Delta)} \left |\frac{4}{p^2}\right |^{1 -\Delta} 
    \, .
\end{align}
In the momentum basis, the two-point function is thus written as
\begin{align}
\begin{aligned}
    \langle \mathcal{O}^\text{AdS}_+ (\vec p) \bar{\mathcal{O}}^\text{AdS}_+ (- \vec p) \rangle '
    &= \frac{\Gamma(1-\Delta)}{\Gamma(\Delta)} \left |\frac{4}{p^2}\right |^{1 - \Delta} \\
    &= \frac{\Gamma(1-\Delta_0)}{\Gamma(\Delta_0)} \left |\frac{4}{p^2}\right |^{1 - \Delta_0} +  \delta \Delta  \frac{\Gamma(1-\Delta_0)}{\Gamma(\Delta_0)} \left |\frac{4}{p^2}\right |^{1 - \Delta_0} \ln |p^2|  + \cdots \, . 
\end{aligned} \label{AdS2ptfull}
\end{align}
Here $\langle * \rangle '$ indicates that the divergent factor $(2 \pi)^2 \delta^{(2)} (0)$ is removed. 
If we put the leading order value as $\Delta_0 = 2h_+ =  1 -N$, then there appears divergence from $\Gamma(1-N)$ in the denominator for positive integer $N$. The divergence is harmless since it could be removed by changing overall normalization of $\mathcal{O}_+$ (or $\phi_+$) as discussed above.
The two-point correlator of scalar field on dS$_3$ at late time is then obtained via the inversion formula as:%
\footnote{At the 't Hooft limit \eqref{tHooft} with $h_\pm = (1 \pm \lambda)/2$, we obtain $  \text{Re} \langle \mathcal{O}_+ (\vec p) \tilde{\mathcal{O}}_+ (- \vec p) \rangle ' =  - \sin \pi (1 \pm  \lambda) \frac{\Gamma (\mp \lambda)}{\Gamma(1 \pm \lambda)} |\frac{4}{p^2}|^{\mp \lambda} =  - \frac{ \pi }{\Gamma(1 \pm \lambda)^2} |\frac{4}{p^2}|^{\mp \lambda} < 0$. This ensures the positivity of bulk two-point function and the convergence of the Gaussian path-integral for \eqref{ms2pt}, see footnote \ref{ft:sign} as well.}
\begin{align}
\begin{aligned}
    \langle \phi_+ (\vec p) \bar{\phi}_+ (- \vec p) \rangle '
   & = - \frac{1}{2 \text{Re}     \langle \mathcal{O}_+ (\vec p) \tilde{\mathcal{O}}_+ (- \vec p) \rangle '} \\
   & = a_{h_+}   \frac{\Gamma(1 -N)}{\Gamma(N)} \left |\frac{p^2}{4}\right |^{N} + \delta \Delta\,   a_{h_+}   \frac{\Gamma(1 -N)}{\Gamma(N)} \left |\frac{p^2}{4}\right |^{N} \ln |p^2| + \cdots \, .
   \end{aligned}
\label{tree2pt}
\end{align}
We stress that the aim of this exploratory subsection is merely to show that our method can be applied to one-loop corrections to two-point function and not to provide a complete analysis of quantum corrections.

Here we only consider a single type of contribution as in fig.\,\ref{fig:2pt},
where a spin-$s$ field runs over the upper half of the loop and a scalar field runs over the lower half of the loop. This thus implies that we only need to consider scalar-scalar-spin vertices.
More possible exchange loop diagrams need to be included, and complete analysis is left as a future work.
Applying the split representation of bulk two-point functions as in \eqref{bulkpropagator} with \eqref{AdSharmonic} together with the inversion formula for the two-point functions, a $1/c$-contribution in \eqref{AdS2ptfull} can be compute from
\begin{align}
    \begin{aligned}
        &\frac{k_{1 -h_+ , 1 - h_+} k_{1-s,1}}{4 \pi^2}\int d ^2 z_2\, d ^2 z_2 '\, d^2 z_3 \, d^2 z_3 ' \, \langle\, \mathcal{O}_+^\text{AdS} (z_1)\, \bar{\mathcal{O}}_+^\text{AdS} (z_2)\, J_{(s)}^\text{AdS} (z_3)\, \rangle\, \\
       & \quad \times K_{h_+,h_+} (z_2 , z_2 ') \frac{K_{s,0}(z_3,z_3 ')}{c B^{(s)}}\, \langle\, J_{(s)}^\text{AdS} (z_3 ')\, \mathcal{O}_+^\text{AdS} (z_2 ')\, \bar{\mathcal{O}}_+^\text{AdS} (z_4) \, \rangle \\
        & \quad =\frac{k_{1 -h_+ , 1 - h_+} }{4 \pi} \int d^2 z_2 \, d^2 z_2 ' \, \frac{1}{|z_{1 2} |^{4 h_+} |z_{2 ' 4} |^{4 h_+}} \frac{(C_+^{(s)})^2}{c B^{(s)}} \mathcal{I}_{s,0} (z)  K_{h_+,h_+} (z_2 , z_2 ') \, , 
    \end{aligned}
\end{align}
where the cross ratio is given by
\begin{align}
z = \frac{z_{12} z_{2 ' 4}}{z_{1 2'} z_{2 4}} \, , \quad \bar z =  \frac{\bar z_{12} \bar z_{2 ' 4}}{\bar z_{1 2'} \bar z_{2 4}}\, . 
\end{align}
We may express the result above through the following integral:
\begin{align}
\begin{aligned}
D_{n,0} &\equiv \int d^2 z_2 \, d^2 z_2 ' \, \frac{1}{|z_{1 2}|^{4 h_+} |z_{2 ' 4} |^{4 h_+}} z^n K_{h_+,h_+} (z_2 , z_2 ')  \\
& = \frac{k_{h_+,h_+}}{\pi}\int d^2 z_2 \, d^2 z_2 ' \, \frac{1}{ (z_{4 2 '})^{2 h_+ - n} (z_{1 2 '})^n  (z_{2 2 '})^{2 - 2h_+} } \frac{1}{   (\bar z_{2 2 '})^{2 - 2h_+}(\bar z_{4 2 '})^{2 h_+ }}\\
 & \quad \times \frac{1}{(z_{1 2})^{2 h_+ - n} (z_{4 2})^n (\bar z_{1 2})^{2 h_+ }} \, .
\end{aligned}
\end{align}
The integration over $z_2'$ can be performed by applying the conformal integral formula
\cite{Dolan:2011dv}
\begin{align}
    \frac{1}{\pi} \int d^2 z \prod_{i=1}^3\frac{1}{(z - z_i)^{h_i}(\bar z - \bar z_i)^{\bar h_i}} 
    = K_{123} z_{12}^{h_3 -1} z_{23}^{h_1 -1} z_{31}^{h_2 -1} \bar z_{12}^{\bar h_3 -1} \bar z_{23}^{\bar h_1 -1} \bar z_{31}^{\bar h_2 -1} \, , \label{conformalintengral}
\end{align}
where 
\begin{align}
    K_{123} = \prod_{i=1}^3 \frac{\Gamma(1 - h_i)}{\Gamma(\bar h_i)}
    = \prod_{i=1}^3 \frac{\Gamma(1 - \bar h_i)}{\Gamma(h_i)} \, .
\end{align}
Here we assumed 
\begin{align}
    \sum_{i=1}^3 h_i = \sum_{i=1}^3 \bar h_i = 2 \, ,
\end{align}
for the conformality condition to be satisfied.
We find that:
\begin{align}
D_{n,0} &= \frac{\Gamma (1 - 2 h_+  + n) }{\Gamma (2 h_+ ) \Gamma ( n)} \frac{1}{|z_{14}|^{4 h_+ -2}} \int d^2 z_2 \, \frac{1}{|z_1 - z_2|^2 |z_4 - z_2|^2} \, .
\end{align}
Applying a formula analogous to (1.5) of \cite{Giombi:2017hpr} and neglecting the terms not including $\log z_{14}$, 
the integration over $ z_2$ leads to
\begin{align}
D_{n,0} = 4 \pi \frac{\Gamma (1 - 2 h_+  + n) }{\Gamma (2 h_+ ) \Gamma ( n)} \frac{1}{|z_{14}|^{4 h_+ }} \log z_{14} + \cdots  \, .
\end{align}
We thus have
\begin{align}
&\frac{k_{1 - h_+ , 1- h_+}}{4 \pi} \frac{(C_+^{(s)})^2}{c B^{(s)}} \frac{\Gamma (2s)}{\Gamma (s)^2} \sum_{n=0}^\infty \frac{\Gamma (s + n)^2}{\Gamma (2 s + n) n!}D_{n+s,0} \nonumber \\
&\quad = \frac{ N  (1 - N^2) (2 s -1)\Gamma (s - N) }{c \Gamma(1 - N) \Gamma (s + N)}
\frac{1}{|z_{14}|^{4 h_+ }} \log z_{14}\sum_{n=0}^\infty \frac{\Gamma (s + n)^2}{\Gamma (2 s + n) n!} \frac{\Gamma ( N  + n + s) }{ \Gamma ( n + s)}  + \cdots \nonumber  \\
&\quad = \frac{(2 s -1)(N^2 -1) }{c  }
\frac{1}{|z_{14}|^{4 h_+ }} \log z_{14} + \cdots  \, . \label{2pt1loop}
\end{align}
Here we have used the identity:
\begin{align}
\sum_{n=0}^\infty \frac{\Gamma (s + n)^2}{\Gamma (2 s + n) n!} \frac{\Gamma ( N  + n + s) }{ \Gamma ( n + s)} = \frac{\Gamma ( - N ) \Gamma (s + N )}{\Gamma (s - N )} \, .
\end{align}
From the expression, we can read off the $1/c$-contribution from the loop diagram in fig.\,\ref{fig:2pt} as
\begin{align} \label{AdSano}
    \delta \Delta_s^\text{AdS} = - \frac{(2 s-1) (N^2 -1)}{c} + \mathcal{O} (c^{-2}) \, . 
\end{align}
In the current case, there are conserved currents with spin $s=2,3,\ldots,N$.
This means that we need to take a sum over the spin as
\begin{align}
\delta \Delta^\text{AdS} =  \sum_{s=2}^N \delta \Delta_s^\text{AdS} = -  \frac{(N^2 -1)^2}{c} + \mathcal{O}(c^{-2})\, . \label{extra}
\end{align}
As one can read off from \eqref{Deltalimit}, this actually reproduces the anomalous dimension at the next leading order in $1/c$.
The result is consistent with the analysis on loop-corrections from Wilson line networks as in discussion section of \cite{Hikida:2017ehf}.
However, as seen in fig.~12 of \cite{Giombi:2017hpr}, there are many other possible one-loop Witten diagrams  contributing to the two-point function.
Currently, we do not have any good explanation why only the one-loop Witten diagram of the form in fig.~\ref{fig:2pt} can explain the anomalous dimension of scalar operator at the order.

We may map the result from AdS$_3$ to dS$_3$.
A type of one-loop contribution to two-point coefficient function is given by
\begin{align}
  \frac{\lambda_{h_+ , h_+ , s}^2}{a_{h_+} a_{(s)}} \frac{(2 s-1) (N^2 -1)}{2 c^{(g)}} \frac{1}{|z^2|^{1-N}} \log z + \cdots \, . 
\end{align}
Mapping to the momentum space, this type of one-loop correction to the bulk two-point function becomes
\begin{align}
 \frac{(N^2 -1) a_{h_+}^3}{ i c^{(g)}} \frac{\Gamma(1- N)}{\Gamma(N)} (-1)^s  (2 s - 1) \sin^2 [(1 - N + s/2)\pi]\left | \frac{p^2}{4} \right|^{N}\log p + \cdots  \, .
\end{align}
Here we do not sum over $s$ since there seems other types of one-loop contributions as mentioned above in this case.
Again the investigation of its bulk interpretation in the dS$_3$ spacetime is postponed as a future work.

We would like to also comment on the case of $1/c$-expansion with the 't Hooft parameter $\lambda$ fixed as in \eqref{tHooft}. In this case, the anomalous dimension is given by
\begin{align}
   \delta \Delta_ + = \frac{\lambda^2 -1}{c} + \mathcal{O} (c^{-2}) \label{Deltap}
\end{align}
as in \eqref{Deltaexp}. In the 't Hooft limit \eqref{tHooft}, the expression \eqref{AdSano} becomes
\begin{align}
\delta \Delta_s^\text{AdS} =  \frac{(2 s -1)(1 - \lambda^2) }{c} + \mathcal{O} (c^{-2}) \, .
\end{align}
In \eqref{extra}, the sum over spin is finite as $s=2,3,\ldots,N$, while in the current case the infinite sum over $s=2,3,\ldots$ looks divergent. Applying the zeta functional regularization, we have
\begin{align}
\delta \Delta^\text{AdS} =  \sum_{s=2}^N \delta \Delta_s^\text{AdS} =  \frac{2( \lambda^2 - 1)}{3 c} + \mathcal{O} (c^{-2}) \, .
\end{align}
The anomalous dimension is very close to \eqref{Deltap} but there is an extra factor $3/2$.
Instead of this, we may first sum over $s=2,3,\ldots$. Then the computation reduces to
\begin{align}
\begin{aligned}
&  \frac{(1 - \lambda^2)\Gamma(1+\lambda)}{2 c \pi \Gamma(-\lambda)}  \sum_{n=2}^\infty \left[ - \frac{1}{n} + \frac{\Gamma(1 - \lambda) \Gamma(n)}{\Gamma(n - \lambda)} \right ] D_{n,0}\\
& \quad =  \frac{(1 - \lambda^2)\Gamma(1+\lambda)}{ c \Gamma(-\lambda)}  \frac{1}{|z_{14}|^{4 h_+ }} \log z_{14} \sum_{n=2}^\infty \left[ - \frac{1}{n} + \frac{\Gamma(1 - \lambda) \Gamma(n)}{\Gamma(n - \lambda)} \right]  \frac{\Gamma ( - \lambda  + n) }{\Gamma (1 + \lambda) \Gamma ( n)} + \cdots  \, .
\end{aligned}
\end{align}
After the zeta functional regularization, we find
\begin{align}
\frac{(1 - \lambda^2)(2 + \lambda )}{2 c} \frac{1}{|z_{14}|^{4 h_+ }} \log z_{14}  + \cdots \, .
\end{align}
It again does not reproduce \eqref{Deltap} but with an extra factor $-(2 + \lambda)/2$. Moreover, the sum over $s$ and the zeta functional regularization do not commute with each other. 

We may obtain the anomalous dimension for the 't Hooft expansion in the AdS$_3$ case by applying the triality relation \cite{Gaberdiel:2012ku}
\begin{align} \label{triality}
 W_\infty [N] \simeq W_\infty \left[\frac{N}{N+k}\right] \simeq W_\infty \left[ - \frac{N}{N+k+1} \right] \, .
\end{align}
As in \eqref{extra}, we have reproduced the conformal dimension with finite $N$ from the gravity loop calculation,
\begin{align} \label{anoN}
    \Delta^\text{AdS} = 1 - N + \delta \Delta^\text{AdS} = 1 - N - \frac{(N^2 -1)^2}{c} + \mathcal{O} (c^{-2}) \, .
\end{align}
If we replace $N$ by $N/(N+k)$, then \eqref{anoN} becomes
\begin{align} 
    \Delta^\text{AdS} = 1 - \lambda  - \frac{(\lambda^2 -1)^2}{c} + \mathcal{O} (c^{-2}) 
\end{align}
in $1/c$-expansion with fixed $\lambda = N/(N+k)$.
This reproduces $\Delta_-$ in \eqref{Deltaexp2}. On the other hand, if we replace $N$ by $- N/(N+k+1)$, then \eqref{anoN} becomes
\begin{align} 
    \Delta^\text{AdS} = 1 + \lambda + \frac{\lambda^2 -1}{c} + \mathcal{O} (c^{-2}) 
\end{align}
in $1/c$-expansion with fixed $\lambda = N/(N+k)$.
This reproduces $\Delta_+$ in \eqref{Deltaexp}. 
In this way, we can reproduce the anomalous dimension even for the 't Hooft expansion. However, if we apply the triality relation \eqref{triality}, then it becomes unclear how to map the AdS result to the dS one.

\subsection{The expansion of four-point correlation functions at \texorpdfstring{$\mathcal{O}(c^{-2})$}{O(c-2)}}\label{Sec:4pt exp2}
In this subsection, we extend our analysis in section \ref{Sec:4pt-AdS3} to consider the expansion of four-point CFT correlation functions at the order $\mathcal{O}(c^{-2})$ in 't Hooft limit. The purpose of including some partial results here is to facilitate the comparisons with the various bulk loop corrected four-point function in AdS$_3$ hence dS$_3$ in the subsequent work.  
Our strategy here will be similar as previous sections; start with the $1/N$-expansion and use the definition \eqref{central}
to obtain the inverted relation:
\begin{equation}\label{Nc-conversion}
\frac{1}{N}=\frac{1}{c}(1-\lambda^2)-\frac{1}{c^2}(1-\lambda^2)(1-\lambda^3)+O(c^{-3})
\end{equation}
and make the appropriate substitutions.

Starting with the relatively simple exact functional forms of $G_{-+}(z)$ given in \eqref{G-+}, the direct expansion in $1/N$ yields:
\begin{equation}
\begin{aligned}
G_{-+}(z)
&=\frac{1}{|1-z|^{2\Delta_+}}\Big\{1
+\frac{1}{N}\left(\frac{1-z}{z}+\log z+{\rm c. c.}\right) \\
&+\frac{1}{N^2}\left[\frac{1}{2} \left(\frac{1-z}{z}+\log z+{\rm c. c.}\right)^2-\frac{1}{2}\left(\left(\frac{1-z}{z}\right)^2+{\rm c.c.}\right)\right]+O(N^{-3})\Big\} \, .
\end{aligned}
\end{equation}
We next apply the relation \eqref{Nc-conversion} to substitute $1/c$ for $1/N$, and expand the correlation function into series of $1/c$ by the form:
\begin{equation}
\begin{aligned}\label{G-+exp2form}
G_{-+}(z)&=\frac{1}{|1-z|^{2(\Delta_{+}^{(0)}+\Delta_{+}^{(1)}/c+\Delta_{+}^{(2)}/c^2)}}Q_{-+}^{(0)}(1-z)
+\frac{1}{c}\frac{1}{|1-z|^{2(\Delta_{+}^{(0)}+\Delta_{+}^{(1)}/c)}}Q_{-+}^{(1)}(1-z)\\
&+\frac{1}{c^2}\frac{1}{|1-z|^{2\Delta_{+}^{(0)}}}Q_{-+}^{(2)}(1-z)+O(c^{-3}) \, .
\end{aligned}
\end{equation}
As in \eqref{G-+exp}, we single out the expansion of conformal dimension $\Delta_{+}$. It corresponds to the loop corrections on the external legs of the diagrams, and their dependence on the 't Hooft coupling $\lambda$ at each loop order are given by:
\begin{equation}\label{Delta+exp2}
\Delta_+^{(0)}=1+\lambda \, , \quad
\Delta_+^{(1)}=\lambda^2-1 \, , \quad 
\Delta_+^{(2)}=(1-\lambda^2)(1-\lambda)\, ,
\end{equation}
and for completeness we include the similar expansion for $\Delta_{-}$:
\begin{equation}\label{Delta-exp2}
\Delta_{-}^{(0)}=1-\lambda \, ,\quad
\Delta_{-}^{(1)}=-(1-\lambda^2)^2 \, ,\quad
\Delta_{-}^{(1)}=(1-\lambda^2)^2(1+\lambda-2\lambda^3) \, .
\end{equation}
Finally, in the expansion of $G_{-+}(z)$ \eqref{G-+exp2form} can be expressed by power series around $z=1$ as
\begin{align}
Q_{-+}^{(0)}(1-z)&=1 \, , \nonumber \\
Q_{-+}^{(1)}(1-z)&=(1-\lambda)^2\sum_{r=1}^{\infty}\frac{1}{r}(r-1)(1-z)^{r}+{\rm c.c.} \, , \label{G+-exp2}\\
Q_{-+}^{(2)}(1-z)
&=\frac{1}{2}Q_{-+}^{(1)}(1-z)^2-(1-\lambda^3)Q_{-+}^{(1)}(1-z)-\frac{(1-\lambda^2)^2}{2}\left(\sum_{r=2}^{\infty}(r-1)(1-z)^r+\rm{c. c.}\right) \, . \nonumber
\end{align}
We can now comment on the identifications of some of the terms listed in \eqref{G-+exp2form} with the possible Witten diagrams in AdS$_3$ bulk, facilitating the explicit quantitative comparisons in the subsequent work. 
The first term in \eqref{G-+exp2form} corresponds to the contribution of a pair of boundary to boundary propagators
corrected up to two-loop order. 
As shown in section \ref{Sec:4pt-AdS3}, $Q_{-+}^{(1)}(1-z)$ can be further expanded into a summation of  global conformal blocks for the conserved tensor operators, and we can respectively associate them with the four-point tree level exchange Witten diagrams in AdS$_3$.
It is thus natural to expect that the second term in \eqref{G-+exp2form} comes from replacing one of the external bulk to boundary propagators with the loop corrected one in the corresponding tree level Witten diagrams respectively. Finally the last term would be attributed to certain combination of the exchange diagrams now with one-loop corrected three-point vertex, or with one-loop corrected bulk to bulk exchange propagator. The appearance of $Q_{-+}^{(1)}(1-z)^2$ can perhaps be explained by the fact that all of these diagrams can be constructed by fusing a pair of tree level four-point exchange diagrams using the split representation of the AdS Harmonic function. However to make the identification precise at this level, it would be necessary to combine the result from the higher order $1/c$ expansion of Virasoro conformal block, see e.g. \cite{Bombini:2018jrg}, as well as the actual AdS loop computations.

We then examine more complicated examples of $G_{\pm\pm}(z)$ in \eqref{Gpp} with \eqref{Np} and \eqref{Gmm} with \eqref{Nm}.
The main difference for expanding $G_{\pm\pm}(z)$ to order $1/N^2$ is that they contain both $s$- and $t$-channels, i.e. the correlation function is invariant under the exchange $z\leftrightarrow 1-z$ as shown in appendix \ref{app:s-t}.
This implies that we need to expand around $z=0$ and $z=1$ separately. We expand $G_{++}(z)$ as a demonstration of some key features. First before the expansion, we use the identities of hypergeometric functions as in appendix \ref{sec:sym} to transform the second term in \eqref{Gpp} into: 
\begin{equation}
\begin{aligned}
G_{++}(z)
=&\frac{1}{|z(1-z)|^{2\Delta_+}}\big|(1-z)^{1+\lambda}{}_2F_1(1+\frac{\lambda}{N},-\frac {\lambda}{N};-\lambda;z)\big|^2\\
&+\CN_{+}A^2\frac{1}{|z(1-z)|^{2\Delta_+}}\big|z^{1+\lambda}{}_2F_1(1+\frac{\lambda}{N};-\frac{\lambda}{N};-\lambda;1-z)\big|^2\\
&+\CN_{+}B^2\frac{1}{|z(1-z)|^{2\Delta_+}}\big|(1-z)^{1+\lambda}{}_2F_1(1+\frac{\lambda}{N},-\frac{\lambda}{N};2+\lambda;1-z)\big|^2\\
&+\CN_{+}AB\frac{1}{|z(1-z)|^{2\Delta_+}}\big[z^{1+\lambda}(1-\bar{z})^{1+\lambda}{}_2F_1(1+\frac{\lambda}{N};-\frac{\lambda}{N};-\lambda;1-z)
\\&{}_2F_1(1+\frac{\lambda}{N},-\frac{\lambda}{N};2+\lambda;1-\bar{z})+\rm{c.c.}\big]
\end{aligned}
\end{equation}
with
\begin{equation}
\begin{aligned}
 A&=\frac{\Gamma(2+\lambda)\Gamma(1+\lambda)}{\Gamma(1+\lambda-\frac{\lambda}{N})\Gamma(2+\lambda+\frac{\lambda}{N})} \, , \\
B&=\frac{\Gamma(2+\lambda)\Gamma(-1-\lambda)}{\Gamma(1+\frac{\lambda}{N})\Gamma(-\frac{\lambda}{N})}=\frac{\sin(-\frac{\pi\lambda}{N})}{\sin\pi(-1-\lambda)}=\frac{-\sin(\frac{\pi\lambda}{N})}{\sin(\pi\lambda)}=\frac{-\pi\lambda}{\sin(\pi\lambda)}\frac{1}{N}+O(N^{-3}) \, .
\end{aligned}
\end{equation}
Hence we can ``symmetrize'' the correlation function into manifestly $s-t$ crossing symmetric form suitable for the $1/c$-expansion in each channel:
\begin{align}
G_{++}(z)=&\frac{1}{2}[G_{++}(z)+G_{++}(1-z)] \nonumber \\
=&\frac{1}{|z(1-z)|^{2\Delta_+}}\Big\{(1-\frac{1}{2}B^2)\big|(1-z)^{1+\lambda}{}_2F_1(1+\frac{\lambda}{N},-\frac {\lambda}{N};-\lambda;z)\big|^2 \nonumber \\
&+\CN_{+}\frac{B^2}{2}\big|(1-z)^{1+\lambda}{}_2F_1(1+\frac{\lambda}{N},-\frac{\lambda}{N};2+\lambda;1-z)\big|^2 \label{G++sym}\\
&+\frac{1}{2}\CN_{+}AB\big[z^{1+\lambda}(1-\bar{z})^{1+\lambda}{}_2F_1(1+\frac{\lambda}{N};-\frac{\lambda}{N};-\lambda;z){}_2F_1(1+\frac{\lambda}{N},-\frac{\lambda}{N};2+\lambda;\bar{z})+{\rm c.c.}\big] \nonumber \\
&+(z\leftrightarrow 1-z)\Big\} \nonumber \, .
\end{align}
We can then expanding the above equation up to order $1/c^2$ and arrange it into the following form:
\begin{align}
G_{++}(z)&=\left[\frac{1}{|z|^{2\Delta_{+}^{(0)}+2\Delta_{+}^{(1)}/c+2\Delta_{+}^{(2)}/c^{2}}}Q_{++}^{(0)}(z)
+\frac{1}{c}\frac{1}{|z|^{2\Delta_{+}^{(0)}+2\Delta_{+}^{(1)}/c}}Q_{++}^{(1)}(z)
+\frac{1}{c^2}\frac{1}{|z|^{2\Delta_{+}^{(0)}}}Q_{++}^{(2)}(1-z)\right ] \nonumber \\
&\quad +\Big[
\frac{1}{(\bar{z}(1-z))^{\Delta_{+}^{(0)}+\Delta_{+}^{(1)}/c+\Delta_{+}^{(2)}/c^{2}}}h^{(0)}_{++}(z)
+\frac{1}{c}\frac{1}{(\bar{z}(1-z))^{\Delta_{+}^{(0)}+\Delta_{+}^{(1)}/c}}h^{(1)}_{++}(z)\label{G++exp2form}
\\& \quad +\frac{1}{c^2}\frac{1}{(\bar{z}(1-z))^{\Delta_{+}^{(0)}}}h^{(2)}_{++}(z)+{\rm c.c.}
\Big]+(z\leftrightarrow 1-z)+O(c^{-3}) \, . \nonumber
\end{align}
Again as in \eqref{G-+exp2form}, we single out the loop corrections of conformal dimension, such that the first term in the first line of \eqref{G++exp2form} corresponds to the a pair of disconnected  loop corrected boundary to boundary propagators.
Furthermore, in this case, we separate terms which have distinct channel structure $Q_{++}(z)$ with those related to the contact diagram $h_{++}(z)$. And the final result can be found by
\begin{align}
&Q^{(0)}_{++}(z)=1 \, ,\quad
Q^{(1)}_{++}(z)=(1-\lambda^2)\left[\sum_{n=1}^{\infty}(\frac{\Gamma(1-\lambda)\Gamma(n)}{\Gamma(n-\lambda)}-\frac{1}{n})z^{n}+{\rm c.c}\right] \, , \nonumber \\
&Q^{(2)}_{++}(z)=\frac{Q^{(1)}_{++}(z)^2}{2}-(1-\lambda^{3})Q^{(1)}_{++}(z)+\frac{\pi^2\lambda^2(1-\lambda^2)^2}{2\sin^2(\pi\lambda)} \nonumber \\&\quad +\left\{-\frac{(1-\lambda^2)^2}{2}\left[\sum_{n=1}^{\infty}\frac{\Gamma(1-\lambda)\Gamma(n)}{\Gamma(n-\lambda)}z^{n}\right]^2+\lambda(1-\lambda^2)^2\sum_{n=1}^{\infty}\frac{1}{n}\left[\frac{\Gamma(1-\lambda)\Gamma(n)}{\Gamma(n-\lambda)}-1\right]z^{n}
+{\rm c.c}\right\} \, , \nonumber \\
&h^{(0)}_{++}(z)=0 \, ,\quad
h^{(1)}_{++}(z)=-\frac{\pi\lambda(1-\lambda^2)}{2\sin(\pi\lambda)} \, , \label{G++exp2}\\
&h^{(2)}_{++}(z)=h^{(1)}_{++}(z)\big[(\lambda-1)(\lambda^2+1)+Q^{(1)}_{++}(z)+(1-\lambda^2)\sum_{n=1}^{\infty}\frac{1}{n}[z^{n}-(1-z)^{n}] \nonumber \\
&\quad -(1-\lambda^2)\sum_{n=1}^{\infty}[\frac{\lambda\Gamma(2+\lambda)\Gamma(n)}{\Gamma(2+\lambda+n)}+\frac{1}{n}]\bar{z}^n\big] \nonumber \, .
\end{align}
As noted in section \ref{Sec:4pt-AdS3}, in contrast with the expansion for $G_{-+}(z)$, the main additional feature here is the additional terms which can be attributed to the manifestly crossing symmetric bulk four-point contact diagrams as denoted by function $g(z)$ \eqref{flz} earlier. We can thus readily identify $h_{++}^{(1)}(z)$
as the four-point contact diagram with an external bulk to boundary propagator replaced by loop corrected one, while $h_{++}^{(2)}(z)$ can be attributed to possible one-loop correction to the four-point vertex or so-called box diagram. 
It would be interesting and non-trivial to verify these predictions from 2d CFT with explicit bulk loop computations in AdS$_3$ and hence dS$_3$ spacetimes.

\subsection*{Acknowledgements}

We are grateful to Thomas Creutzig, Toshifumi Noumi, Kenta Suzuki, Tadashi Takayanagi, Yusuke Taki, Seiji Terashima, and Takahiro Uetoko for useful comments.
We would like to particularly thank Tatsuma Nishioka for the collaboration at the early stage. 
The work of H.\,Y.\,C. and S.\,C. are supported in part by Ministry of Science
and Technology (MOST) through the grant 110-2112-M-002-006-.
This work of Y.\,H. was supported by JSPS Grant-in-Aid for Scientific Research (B) No.19H01896, Grant-in-Aid for Scientific Research (A) No.\,21H04469, Grant-in-Aid for Transformative Research Areas (A) ``Extreme Universe'' No.\,21H05187.

\appendix

 \section{Partial wave (conformal block) expansions}
\label{app:cb}

In this appendix, we prove the crossing symmetric property of the $G_{\pm\pm}(z)$ correlation functions and collect some detailed computations associated with the expansions by conformal blocks or conformal partial waves.

\subsection{Crossing symmetry under \texorpdfstring{$s$-$t$}{s-t} channel exchange}\label{app:s-t}

Starting with $G_{++}(z, \bz)$ given in \eqref{Gpp}, here we parameterize it in terms of the following combinations for the convenience:
\begin{equation}
G_{++}(z, \bz) = \frac{\left(|Q_+(z)|^2+\CN_{+} |P_+(z)|^2\right)    }{|z(1-z)|^{2\Del_+}}
\end{equation}
where we have defined the functions:
\begin{equation}
Q_+(z) =   (1 -z)^{1 + \lambda}\, {}_2 F_1 \left( 1 + \frac{\lambda}{N} , - \frac{\lambda}{N} ; - \lambda  ; z \right), ~~
P_+(z) = z^{1 + \lambda}\, {}_2 F_1 \left( 1 + \frac{\lambda}{N} , - \frac{\lambda}{N} ;  2 + \lambda  ; z \right ),
\end{equation}
and the constant $\mathcal{N}_+$ is given in \eqref{Np}.
Now under the \texorpdfstring{$s$-$t$}{s-t} channel crossing transformation $(z, \bz) \to (1-z, 1-\bz)$, using the hypergeometric function identities, we have:
\begin{align}
Q_+(1-z) &=+\frac{\sin \pi \frac{\lambda}{N}}{\sin \pi \lambda} Q(z)+ \CN_+ \frac{\Gamma(-\lambda)\Gamma(-(1+\lambda))}{\Gamma(-1-\lambda-\frac{\lambda}{N})\Gamma(-\lambda+\frac{\lambda}{N})}P(z)\, , \\
P_+(1-z) &=-\frac{\sin \pi \frac{\lambda}{N}}{\sin \pi \lambda} P(z)+  \frac{\Gamma(1+\lambda)\Gamma(2+\lambda)}{\Gamma(1+\lambda-\frac{\lambda}{N})\Gamma(2+\lambda+\frac{\lambda}{N})}Q(z) \, .
\end{align}
Now if we evaluate:
\begin{equation}
G_{++}(1-z, 1-\bz) =\frac{\left(|Q_+(1-z)|^2+\CN_{+} |P_+(1-z)|^2\right) }{|z(1-z)|^{2\Del_+}}
\end{equation}
and notice that the coefficients for $|Q_+(z)|^2$ and $|P_+(z)|^2$, after repeatedly using the $\Gamma$-function identity $\Gamma(x)\Gamma(1-x)=\frac{\pi}{\sin\pi x}$, are given by:
\begin{eqnarray}
|Q_+(z)|^2 &:& \frac{\sin^2 \pi \frac{\lambda}{N}}{\sin^2 \pi \lambda} + \frac{\sin \pi (\lambda+\frac{\lambda}{N})\sin \pi (\lambda-\frac{\lambda}{N})}{\sin^2 \pi \lambda} = 1 \, , \\
|P_+(z)|^2 &:& \CN_+\frac{\sin^2 \pi \frac{\lambda}{N}}{\sin^2 \pi \lambda} + \CN_+\frac{\sin \pi (\lambda+\frac{\lambda}{N})\sin \pi (\lambda-\frac{\lambda}{N})}{\sin^2 \pi \lambda} =\CN_+ \, ,
\end{eqnarray}
while the coefficients for the crossed terms $Q(z) P(\bz)$ and $Q(\bz)P(z)$ vanish identically.
We thus established that:
\begin{equation}
G_{++}(1-z, 1-\bz) = G_{++}(z, \bz)   
\end{equation}
to the all order of ${1}/{N}$ or $1/c$ expansions. We can similarly prove the crossing symmetric property for $G_{--}(z, \bz)$.

\subsection{Expansions of \texorpdfstring{$G_{\pm\pm}(z)$}{Gpmpm}}
\label{sec:sym}

In this subsection, we derive \eqref{Gpmpmf} but focusing on the $G_{++}(z)$ case \eqref{Gpp}.
The first term has been analyzed in \cite{Hikida:2017byl}, but let us repeat the analysis for completeness. We consider $1/N$-expansion instead of $1/c$-expansion as in \cite{Hikida:2017byl} since the expressions become simpler.
However, we can easily relate them by consistently make the replacement $1/N \to (1 - \lambda^2)/c$ at the order we are interested in as before.

Since the conformal dimension $\Delta_+$ is given by
\begin{align}
    \Delta_+ = \frac{(N-1) (2 N +1 + k)}{N (N+k)} = 
    1 + \lambda - \frac{1}{N} + \mathcal{O} (N^{-2}) \, ,
\end{align}
we have
\begin{align}
 (1-z)^{- \Delta_+ + 1 + \lambda} 
 = 1 - \frac{1}{N} \sum_{l=1}^\infty\, \frac{1}{l}\, z^l + \mathcal{O} (N^{-2}) \, .
\end{align}
Moreover, we can expand
\begin{align}
    \begin{aligned}
    {}_2 F_1 \left(1 + \frac{\lambda}{N} , - \frac{\lambda}{N} ; - \lambda ; z \right)
        &= 
            \frac{\Gamma ( -\lambda)}{ \Gamma\left(1 + \frac{\lambda}{N}\right)\, \Gamma \left( - \frac{\lambda}{N}\right)}
            \sum_{n=0}^\infty\, \frac{\Gamma \left(1 + \frac{\lambda}{N} + n\right)\, \Gamma \left( - \frac{\lambda}{N} + n\right)}{\Gamma (- \lambda + n)}\, \frac{z^n}{n!} \\
        &= 
            1 + \frac{\Gamma (1 - \lambda)}{N} \sum_{n=1}^\infty\, \frac{\Gamma (n)}{\Gamma (n - \lambda)}\, z^n + \mathcal{O} (N^{-2}) \, .
    \end{aligned}
\end{align}
Thus the first line in the right hand side of \eqref{Gpp} becomes
\begin{align}
   |z|^{2 \Delta_+} G_{++(1)}(z) = 1 + \left[  \frac{1}{N} \sum_{n=1}^\infty 
    \left( - \frac{1}{n} + \frac{\Gamma (1 - \lambda)\, \Gamma (n)}{ \Gamma (n - \lambda)}\right) z^n + \text{c.c}\right] + \mathcal{O} (N^{-2}) \, .
\end{align}
The vacuum conformal block is
\begin{align}
    \mathcal{V}_0 = 1 + \frac{1}{2} \left(\frac{1 + \lambda}{1-\lambda} \right) z^2\, {}_2 F_1 (2,2;4;z) + \mathcal{O} (N^{-2}) \, ,
\end{align}
while the general block with an intermediate state with conformal weight $h_p$ is
\begin{align}
    \mathcal{V}_p = z^{h_p}\, {}_2 F_1 (h_p , h_p ; 2 h_p ; z) + \mathcal{O} (N^{-1}) \, .
\end{align}
Let us write down
\begin{align}
   |z|^{2 \Delta_+} G_{++(1)} (z) = 1 + \left[ \frac{1}{N} \sum_{s=2}^\infty \left(D^{(s)}_+\right)^2\, z^s\, {}_2 F_1 (s,s;2s;z) + \text{c.c} \right] + \mathcal{O}(N^{-2}) \, . \label{Gppexp}
\end{align}
Comparing the $z$-expansions in the both sides, we have the equations
\begin{align} \label{cond1}
 - \frac{1}{n} + \frac{\Gamma (1 - \lambda)\, \Gamma (n)}{\Gamma (n - \lambda)} = \Gamma (n)^2\, \sum_{s=2}^n \frac{\Gamma(2s)\, \left(D^{(s)}_+\right)^2}{\Gamma (s)^2 \,\Gamma (s + n)\, (n -s)!} \, .
\end{align}
Here we have used
\begin{align}
    z^s\, {}_2 F_1 (s,s;2s;z)
    = \frac{\Gamma(2s)}{\Gamma(s)^2}\, \sum_{n=0}^\infty \frac{\Gamma (s +n)^2}{\Gamma (2 s + n)}\, \frac{z^{n+s}}{n!} \, .
\end{align}
The non-trivial solutions to \eqref{cond1} were found to be \cite{Hikida:2017byl}
\begin{align}
 D^{(s)} = \eta^s_+\, \sqrt{\frac{\Gamma(s)^2\, \Gamma (1 - \lambda)\, \Gamma (s + \lambda)}{\Gamma (2 s-1)\, \Gamma (1 + \lambda)\, \Gamma(s - \lambda)}}
 = {\sqrt{\frac{N}{c B^{(s)}}}}\, C_+^{(s)}\, , 
\end{align}
where $\eta^s_+ = \pm 1$.

Next, we consider the second term in \eqref{Gpp}.
The coefficient $\mathcal{N}_+$ can be expanded as
\begin{align}
    \mathcal{N}_+ = 1+ \frac{1}{N}\frac{2 \lambda}{1+\lambda}+\mathcal{O} (N^{-2}) \, . \label{Npexp}
\end{align}
Applying a well-known formula for hypergeometric function:
\begin{align}
    \begin{aligned}
    &{}_2 F_1 (\alpha , \beta ; \gamma ; z)
        =
            \frac{\Gamma (\gamma)\, \Gamma (\gamma - \alpha - \beta)}{\Gamma (\gamma - \alpha) \, \Gamma (\gamma - \beta)}\, {}_2 F_1 (\alpha , \beta ; \alpha + \beta - \gamma + 1 ; 1 -z)\\
        & \quad
            + \frac{\Gamma(\gamma)\,\Gamma (\alpha + \beta - \gamma)}{\Gamma (\alpha)\, \Gamma (\beta)}\, (1 - z)^{\gamma - \alpha - \beta}\, {}_2 F_1 (\gamma - \alpha , \gamma - \beta ; \gamma - \alpha - \beta + 1 ; 1 - z) \, , 
    \end{aligned}
\end{align}
we can rewrite 
\begin{align} 
&z^{- \Delta_+ + 1 + \lambda}\, {}_2 F_1 \left( 1 + \frac{\lambda}{N} , - \frac{\lambda}{N} ;  2 + \lambda  ; z \right ) \nonumber \\
        & \quad =
            z^{- \Delta_+ + 1 + \lambda}\,\frac{\Gamma (2 + \lambda)\,\Gamma (1 + \lambda)}{ \Gamma \left(1 + \lambda - \frac{\lambda}{N}\right)\, \Gamma \left(2 + \lambda + \frac{\lambda}{N}\right)}\, {}_2 F_1\left( 1 + \frac{\lambda}{N} , - \frac{\lambda}{N} ;  - \lambda  ; 1-z \right ) \label{1-z}\\
        & \qquad
            + z^{- \Delta_+ + 1 + \lambda}\,(1 -z)^{1 + \lambda}\, \frac{\Gamma (2 + \lambda)\, \Gamma (-1-\lambda)}{\Gamma \left(1 + \frac{\lambda}{N}\right)\, \Gamma\left( - \frac{\lambda}{N}\right)}\,{}_2 F_1\left( 1 + \lambda - \frac{\lambda}{N} , 2 + \lambda + \frac{\lambda}{N} ;  2 + \lambda  ; 1-z \right ) \, . \nonumber 
\end{align}
Since we can expand
\begin{align}
 \frac{\Gamma (2 + \lambda)\,\Gamma (1 + \lambda)}{ \Gamma \left(1 + \lambda - \frac{\lambda}{N}\right)\, \Gamma \left(2 + \lambda + \frac{\lambda}{N}\right)}  = 1- \frac{1}{N}\frac{\lambda}{1 +\lambda} +\mathcal{O} (N^{-2})\, ,
\end{align}
the first term in the right hand side of \eqref{1-z} can be expressed as
\begin{align} 
 1 + \sum_{s=2}^\infty \frac{(C^{s}_\pm)^2}{B^s}\, (1-z)^s\, {}_2 F_1 (s,s;2s;1-z)  - \frac{1}{N}  \frac{\lambda}{\lambda +1} + \mathcal{O} (c^{-2}) \, , \nonumber
\end{align}
where we have used \eqref{Gppexp} but replacing $z$ by $1-z$. Further applying
\begin{align}
    {}_2F_1 (\alpha , \beta ; \gamma ; z) = (1 - z)^{\gamma - \alpha - \beta}\, {}_2 F_1 (\gamma - \alpha ,\gamma- \beta ; \gamma ; z ) \, , 
\end{align}
the second term can be written as
\begin{align}
\begin{aligned}
   & (1 -z)^{1 + \lambda}\, \frac{\Gamma (2 + \lambda)\, \Gamma (-1-\lambda)}{\Gamma \left(1 + \frac{\lambda}{N}\right) \, \Gamma\left( - \frac{\lambda}{N}\right)}\,{}_2 F_1\left( 1 + \lambda - \frac{\lambda}{N} , 2 + \lambda + \frac{\lambda}{N} ;  2 + \lambda  ; z \right )\\
    & \quad = z^{-1 - \lambda}\, (1 -z)^{1 + \lambda}\, \frac{\Gamma (2 + \lambda) \, \Gamma (-1-\lambda)}{\Gamma \left(1 + \frac{\lambda}{N}\right) \, \Gamma\left( - \frac{\lambda}{N}\right)}\, {}_2 F_1\left( 1 + \frac{\lambda}{N} , - \frac{\lambda}{N} ;  2 + \lambda  ; z \right ) \\
    & \quad = - \frac{\lambda \pi \csc (\lambda \pi)}{N}\, z^{-1 - \lambda}\,(1 - z)^{1 + \lambda} + \mathcal{O} (N^{-2}) \, .
    \end{aligned}
\end{align}

\subsection{Shadow blocks for conserved currents}
\label{sec:shadow}

In this appendix, we consider the conformal wave expansions of 
$\langle \mathcal{O}_{h_1} (\infty)\mathcal{O}_{h_2} (1) \mathcal{O}_{h_3} (z)\mathcal{O}_{h _4} (0) \rangle$ {in terms of the} exchange operator with conformal weights $(h, \bar h)$.
For the application to our examples, we set $h_1 = h_2$ and $h_3 = h_4$.
We use the following formula given in \cite{Dolan:2011dv,Osborn:2012vt} as
\begin{align}
 & \frac{\Gamma (\bar h ) \Gamma (1 - \bar h )}{\Gamma(1 - h ) \Gamma (h )} \frac{1}{\pi} \int d ^2 z \mathcal{F}_{12}^{h, \bar h} (z , \bar z) \mathcal{F}_{34}^{1-h ,1-\bar h} (z , \bar z) = \frac{1}{z_{12}^{2 h_1} z_{34}^{2 h_3}} 
  \frac{1}{\bar z_{12}^{2 \bar h_1 } \bar z_{34}^{2 \bar h_3 }} \mathcal{I} (z , \bar z) \label{intcpw}
\end{align}
with cross ratios
\begin{align}
    z = \frac{z_{12} z_{34}}{z_{13} z_{24}} \, , \quad 
  \bar    z = \frac{\bar z_{12} \bar z_{34}}{\bar z_{13} \bar z_{24}} \, .
\end{align}
The conformal partial wave is
\begin{align} \label{cpw}
  \frac{(-1)^{2 (h - \bar h)}}{k_{1 -h , 1- \bar h}}  \mathcal{I}_{h , \bar h} (z , \bar z) = \frac{(-1)^{2 (h - \bar h)}}{k_{1 -h , 1- \bar h}} \frac{\Gamma (\bar h )^2 }{\Gamma (1 - h )^2 } F_{h , \bar h} (z , \bar z)
    + \frac{1}{k_{h, \bar h}} \frac{\Gamma(1 - \bar h  )^2 }{\Gamma (h)^2 }F_{1-h , 1-\bar h} (z , \bar z) \, , 
\end{align}
where we have defined
\begin{align}
F_{h, \bar h}(z, \bar z) = z^h {}_2F_1(h  , h  ; 2h ; z) \bar z^{\bar h} {}_2F_1(\bar h , \bar h  ; 2 \bar h ; \bar z) \, . 
\end{align}

We then examine a four-point conformal partial wave of exchange of conserved spin-$s$ current for $\langle \mathcal{O}_{h_1} (\infty)\mathcal{O}_{h_1} (1) \mathcal{O}_{h_3} (z)\mathcal{O}_{h_3} (0) \rangle$. 
It is given by a sum of conformal block of conserved spin-$s$ current and its shadow block as in \eqref{shadow}, see \cite{Haehl:2019eae} for some properties. In general, a conformal partial wave is a sum of conformal block and its shadow block such that there are no monodromy around, say, $z=0$ and $z=1$. In the current case, it is obvious that there are no monodromy around $z=0$, thus we focus on the behavior near $z=1$.
As in (4.6) of \cite{Heemskerk:2009pn}, the first term of \eqref{shadow} (after exchanging $1-z$ by $z$) can be written as
\begin{align}
z^s {}_2 F_1 (s,s;2s;z) = -  \log (1 -z) z^s \frac{\Gamma (2 s)}{\Gamma (s)^2} {}_2 F_1 (s,s;1;1-z)  + \text{holomorphic at $z =1$} \, .
\end{align}
On the other hand, because of
\begin{align}
\bar z {}_2 F_1 (1,1;2 ; \bar z) = - \log (1 - \bar z) \, ,
\end{align}
the second term is equal to
\begin{align}
-  \log (1 -\bar z) \frac{\Gamma (2 s)}{\Gamma (s)^2} {}_2 z^s F_1 (s,s;1;1-z) \, .
\end{align}
Therefore, the sum of two terms does not have any monodromy at $z =1$.
This is actually an important property of conformal partial wave, which might be seen from the integral representation \eqref{intcpw}.
It might be also useful to rewrite the shadow block as
\begin{align}
\begin{aligned}
\frac{\Gamma (2s) \Gamma (2s -1)}{\Gamma (s)^4} z^{1-s} {}_2 F_1 (1-s ,1-s ; 2-2s ;z)   = \frac{\Gamma (2s)}{\Gamma (s)^2} \sum_{n=0}^{s-1} (-1)^n \frac{\Gamma (2s - n -1)}{\Gamma (s-n)^2 } \frac{z^{n + 1-s}}{n!} \, .
\end{aligned}
\end{align}

We begin by studying $G_{-+}(z)$ in \eqref{G-+}. The coefficients in front of conformal blocks of spin-$s$ current exchange are obtained by solving the constraint equations
\begin{align}
 \frac{n-1}{n} = \frac{\Gamma (s)^2}{\Gamma (2s)}\sum_{s=2}^n \frac{(D^{(s)}_{-+})^2 \Gamma (n)^2}{\Gamma (s + n) (n-s)!} 
\end{align}
for $n=2,3,\ldots$, see (4.14) of \cite{Hikida:2017byl}.
Here we have
\begin{align}
    (D^{(s)}_{-+})^2 = \frac{\Gamma (s)^2 }{\Gamma(2s -1)} = \frac{(-1)^s C_+^{(s)} C_-^{(s)} N}{c B^{(s)}} 
\end{align}
for $s = 2,3,\ldots$.
From the sum of shadow blocks over $s=2,3,\ldots$, we find
\begin{align}
 \left[ \sum_{n=0}^\infty \sum_{s=n+1}^\infty (-1)^{s - n - 1} \frac{(2 s-1) \Gamma (s +n)}{\Gamma (n+1)^2} \frac{(1-z)^{-n}}{(s - n - 1)!} \right] - 1 = -1 \, .
\end{align}
Multiplying $- \log \bar z$, we obtain the corresponding term in \eqref{G-+exps}.
The result may be the expected one. The sum of conformal blocks over $s=2,3,\ldots$ has the term $\frac1N \log z$ for the holomorphic part as in \eqref{G-+}. Thus the sum of shadow blocks over $s=2,3,\ldots$ should be $\tfrac1N \log \bar z$, which makes the sum of the two parts to have no monodromy at $z=1$.

We then move to $G_{++}(z)$ in \eqref{Gpp}. 
Since the computation of \eqref{O4d} leads to a sum of conformal partial waves, we would like to compute
\begin{align}
   \sum_{s=2}^\infty \frac{(C_+^{(s)})^2}{c B^{(s)}} \mathcal{I}_{s,0} \, .
\end{align}
As explained in appendix \ref{sec:sym}, the sum over conformal blocks of spin-$s$ exchange is given by
\begin{align}
(1 - z)^{\frac{1}{N}} {}_2 F_1 (1 + \tfrac{\lambda}{N} , - \tfrac{\lambda}{N} ; - \lambda ; z)  - 1 + \mathcal{O}(N^{-2}) \, .
\end{align}
From the arguments of monodromy around $z=1$, we can deduce the sum of shadow blocks as in the previous example. Namely, the sum of shadow blocks should be
\begin{align}
     \frac{1}{N}\log (1 - \bar z) \, . \label{shadowform}
\end{align}

\section{Bulk interpretations}
\label{sec:bulk}

In the main context, we mainly work on the dual CFT and did not deal with the bulk theory in explicit details. However, things sometimes become intuitively clearer from the bulk viewpoints. In this appendix, for completeness, we will summarize the essential details about embedding formalism for Euclidean AdS$_{d+1}$ following \cite{Costa:2014kfa}.
We have also utilized specific features in 2d CFT, such as, holomorphic factorization. We show that the property can be seen from the bulk viewpoints as well.

\subsection{Embedding formalism for Euclidean \texorpdfstring{AdS$_{d+1}$}{AdSd+1}}
\label{app:embedding}
In this work, we compute bulk correlators in terms of dual CFT.
Even so, it is illustrative to see which bulk computations correspond to boundary ones.  In order to explain the bulk computations, it is useful to introduce embedding formulation, where Euclidean AdS$_{d+1}$ space is expressed by a hypersurface,
\begin{align}
X^2 = -1 \, , \quad X^0 > 0 \, , 
\end{align}
in $(d+2)$ dimensional Minkowski space $\mathbb{M}^{d+2}$. 
Here we set the AdS radius to be one as $\ell_\text{AdS} = 1$.
We may use the light-cone coordinates $X^A = (X^+ , X^- , \vec X)$ with the metric
\begin{align}
X^2 = \eta_{AB} X^A X^B = - X^+ X^- + \vec X \cdot \vec X \, .
\end{align}
The Poincar\'e coordinates \eqref{AdSmetric} are given by $X = \frac{1}{y} (1 , y^2 + \vec x^2 , \vec x)$. 
The AdS boundary is located at $y=0$, which may be expressed by light rays
\begin{align}
P^2 = 0 , \quad P = \lambda P
\end{align}
with $\lambda \in \mathbb{R}$.

We consider a symmetric traceless tensor field $\sigma_{i_1  \cdots i_s}$ on AdS$_{d+1}$, which is related to a symmetric traceless tensor field $\Sigma_{A_1 \cdots A_s}$ on $\mathbb{M}^{d+2}$ as
\begin{align}
\sigma_{i_1 \cdots i_s} = \frac{\partial X^{A_1}}{\partial x^{i_s}} \cdots  \frac{\partial X^{A_s}}{\partial x^{i_s}} \Sigma_{A_1 \cdots A_s} (X) \, . 
\end{align}
We define $\Sigma_{A_1 \cdots A_s}$ on the surface $X^2 =-1$ and assign the transverse condition
\begin{align}
 X^{A_1} \Sigma_{A_1 \cdots A_s} (X) = 0 \, .
\end{align}
Introducing bulk polarization vectors $W^A$, we define
\begin{align}
\Sigma (X,W) = W^{A_1} \cdots W^{A_s} \Sigma_{A_1 \cdots A_s} (X) \,  , 
\end{align}
where we assign $W^2 =0$, $X \cdot W = 0$ corresponding to the traceless and transverse conditions, respectively.
On the AdS boundary described by $P^2 = 0$, we also define a symmetric traceless operator $\Xi_{A_1  \cdots A_s} (P)$.
We require $\Xi(\lambda P) = \lambda^{- \Delta} \Xi(P)$ with conformal dimension $\Delta$.  Introducing polarization vectors $Z^A$, we define
\begin{align}
\Xi(P,Z) = Z^{A_1} \cdots Z^{A_J} \Xi_{A_1 \cdots A_J} (P) 
\end{align}
with $Z^2 =0$, $P \cdot Z = 0$. 
We also assign $P^{A_1} \Xi_{A_1 \cdots A_J} = 0$, which can be encoded by requiring $\Xi(P,Z+ \alpha Z) = \Xi(P,Z)$.

We then consider a bulk to bulk propagator of spin-$s$ symmetric traceless field with dual scaling dimension $\Delta$.
The propagator between two AdS bulk points $X_1,X_2$ and polarization vectors $W_1,W_2$ is denoted by $G_{\Delta, s} (X_1 ,X_2 ; W_1 ,W_2)$. Similarly, a bulk-to-boundary propagator is denoted by $K_{\Delta,s} (X,P;W,Z)$. The AdS harmonic function is defined by
\begin{align}
\begin{aligned}
&\Omega_{\nu , s} (X_1 , X_2 ; W_1 , W_2) \\
& \quad    = \frac{\nu^2}{\pi s! (\frac{d}{2} -1)_s} \int d P K_{\frac{d}{2} + i \nu , s} (X_1 , P ; W_1 , D_Z) 
K_{\frac{d}{2} - i \nu , s} (X_2 , P ; W_2 ,Z) \, . \label{AdSharmonic}
\end{aligned}
\end{align}
Here $(a)_n = a (a+1) \cdots (a + n)$ and
\begin{align}
D_Z = \left( \frac{d}{2} - 1 + Z \cdot \frac{\partial}{\partial Z} \right) \frac{\partial}{\partial Z_A} - \frac12 Z^A \frac{\partial^2}{\partial Z \cdot \partial Z} \, .
\end{align}
The bulk to bulk propagator is now expressed in terms of the AdS harmonic function as
\begin{align}
    G_{\Delta , s} (X_1 , X_2 ; W_1 , W_2)
    = \sum_{l=0}^s \int d \nu  a_l (\nu) ((W_1 \cdot \nabla_1) (W_2 \cdot \nabla_2))^{s-l} \Omega_{\nu , l} (X_1 , X_2 ; W_1 , W_2) \, ,  \label{bulkpropagator}
\end{align}
where
\begin{align}
    a_s (\nu) = \frac{1}{\nu^2 + (\Delta - \frac{d}{2})^2}
\end{align}
for the highest spin contribution.
See \cite{Costa:2014kfa} for more detailed analysis.

\begin{figure}
  \centering
  \includegraphics[width=3.7cm]{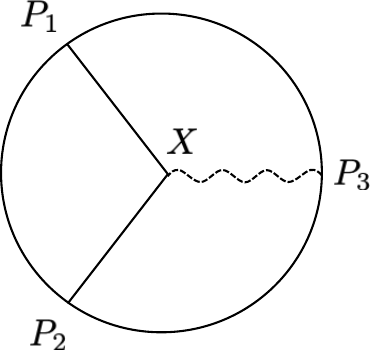}
 \caption{Witten diagram dual to three-point function of scalar-scalar-higher spin current.}
  \label{fig:bulk3pt}
\end{figure}

We will consider three-point function of scalar-scalar-higher spin current in a CFT. Let us denote dual fields by $\phi^\text{AdS}_1$, $\phi^\text{AdS}_2$, $\sigma^\text{AdS}_{i_1 \cdots i_s}$ and introduce a type of interaction
\begin{align}
    g \int_\text{AdS} dx \sqrt{g} ( \phi^\text{AdS}_1 \nabla_{i_1} \cdots \nabla_{i_s} \phi^\text{AdS}_2) \sigma_\text{AdS}^{i_1 \cdots i_s} \, .
\end{align}
Then the three-point correlation function in the dual CFT can be computed via Witten diagram (see fig.~\ref{fig:bulk3pt}) as \cite{Costa:2014kfa} 
\begin{align} \label{bulk3pt}
\begin{aligned}
  &  \left \langle \mathcal{O}_1^\text{AdS} (P_1) \mathcal{O}_2^\text{AdS} (P_2) J^\text{AdS} (P_3,Z) \right \rangle  \\
 &   = \frac{g}{\sqrt{\mathcal{C}_{\Delta_1,0} \mathcal{C}_{\Delta_2,0} \mathcal{C}_{\Delta,s} }}
    \int d X K_{\Delta_2,0} (X , P_2) 
    \frac{K_{\Delta,s} (X, P_3 ; K ,Z) (W \cdot \nabla)^s K_{\Delta_1 ,0} (X , P_1)}{s! (\frac{d-1}{2})_s} \\
 &   = \frac{g}{\sqrt{\mathcal{C}_{\Delta_1,0} \mathcal{C}_{\Delta_2,0} \mathcal{C}_{\Delta,s}}}
 b(\Delta_1 , \Delta_2 , \Delta ,s) \frac{((Z \cdot P_1) P_{23} - (Z \cdot P_2) P_{13})^s}{P_{12}^{\frac{\Delta_1 + \Delta_2 - \Delta +s}{2}} P_{13}^{\frac{\Delta_1 + \Delta - \Delta_2 + s}{2}} P_{23}^{\frac{\Delta + \Delta_2 - \Delta_1 + s}{2}}} \, . \end{aligned}
\end{align}
Here $\mathcal{C}_{\Delta,s}$ is the coefficient of two-point function and $P_{ij} = - 2 P_i \cdot P_j$. Moreover, $K$ is a projection operator defined by
\begin{align}
\begin{aligned}
    K &= \frac{d-1}{2} \left[ \frac{\partial}{\partial W^A} + X_A \left(X \cdot \frac{\partial}{W} \right) \right] + \left( W \cdot \frac{\partial}{\partial W} \right) \frac{\partial}{\partial W^A} \\
    & \quad + X_A \left(W \cdot \frac{\partial}{\partial W}  \right)\left(X \cdot \frac{\partial}{\partial W}  \right) - \frac{1}{2}W_A \left[ \frac{\partial^2}{\partial W \cdot \partial W} + \left( X \cdot \frac{\partial}{\partial W} \right)^2\right] \, .
    \end{aligned}
\end{align}
The coefficient function was computed in \cite{Costa:2014kfa} as
\begin{align}
\begin{aligned}
   & b(\Delta_1 , \Delta_2 , \Delta , s) \\
   & \quad = \mathcal{C}_{\Delta_1,0} \mathcal{C}_{\Delta_2 ,0} \mathcal{C}_{\Delta,s}
    \frac{\pi^{\frac{d}{2}} \Gamma(\frac{\Delta_1 + \Delta_2 + \Delta - d + s}{2})\Gamma(\frac{\Delta_1 + \Delta_2 - \Delta + s}{2})\Gamma(\frac{\Delta + \Delta_1 - \Delta_2 + s}{2})\Gamma(\frac{\Delta + \Delta_2 - \Delta_1 + s}{2})}{2^{1-s} \Gamma (\Delta_1) \Gamma (\Delta_2) \Gamma(\Delta + s)} \, .
    \end{aligned}
\end{align}

Finally, we consider four-point correlation function of scalar operators
\begin{align}
   \left \langle \mathcal{O}_1 (P_1) \mathcal{O}_2 (P_2) \mathcal{O}_3 (P_3) \mathcal{O}_4 (P_4) \right \rangle \, .
\end{align}
We focus on the contribution coming from the Witten diagram with the exchange of a spin-$s$ field. Then the four-point correlation functions can be computed from the bulk by
\begin{figure}
  \centering
  \includegraphics[width=10cm]{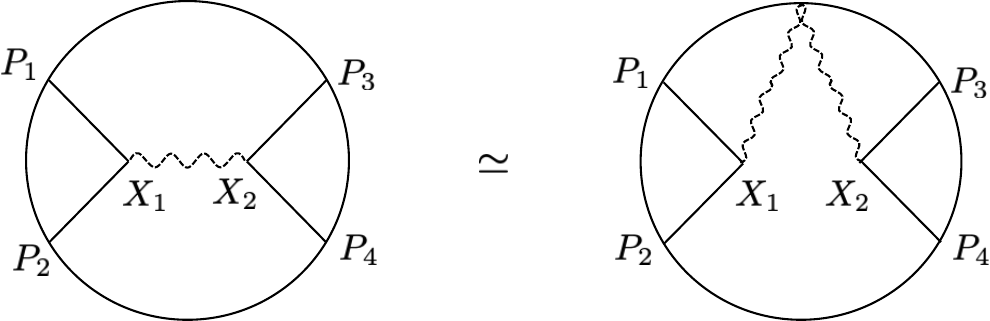}
 \caption{Witten diagram dual to a contribution to four-point function of scalar operators. The diagram can be written as a product of two three-point functions by applying the split representation of bulk to bulk propagator.}
  \label{fig:Witten4pt}
\end{figure}
\begin{align}
\begin{aligned}
    &\frac{g^2}{(s! (\frac{d-1}{2})_s)^2} \int d X_1 dX_2 K_{\Delta_1 ,0} (P_1 , X_1) \left[ (K_1 \cdot \nabla_1)^s K_{\Delta_2 ,0} (P_2 , X_1)\right]\\
    &\quad \times K_{\Delta_3 ,0} (P_3 , X_2) \left[ (K_2 \cdot \nabla_2)^s K_{\Delta_4 ,0} (P_4 , X_2)\right] \sum_{l=0}^s \int d \nu a_l (\nu) \Omega_{\nu ,s} (X_1, X_2 , W_1 ,W_2) \, .
    \end{aligned}
\end{align}
For the bulk to bulk propagator, we have used \eqref{bulkpropagator}. If we apply the split representation of AdS harmonics \eqref{AdSharmonic}, then the above expression is reduced to
\begin{align} \label{cpweg}
  \left \langle \mathcal{O}_1 (P_1) \mathcal{O}_2 (P_2) \mathcal{O}_3 (P_3) \mathcal{O}_4 (P_4) \right \rangle = \sum_{l=0}^\infty \int d \nu c_{l} (\nu) F_{\nu ,l}(P_i)
\end{align}
with
\begin{align} \label{cpwg}
  F_{\nu ,l}(P_i) =  \int d P_5
    \left \langle \mathcal{O}_1^\text{AdS} (P_1) \mathcal{O}_2^\text{AdS} (P_2) J^\text{AdS}_{\frac{d}{2} + i\nu ,l} (P_5,D_Z) \right \rangle
     \left \langle J^\text{AdS}_{\frac{d}{2} - i\nu ,l} (P_5,Z) \mathcal{O}_3^\text{AdS} (P_3) \mathcal{O}_4^\text{AdS} (P_4)  \right \rangle \, . 
\end{align}
Here we have used the bulk expression of three-point function in \eqref{bulk3pt}. The expressions  \eqref{cpwg} and \eqref{cpweg} are definitions of conformal partial waves and their expansions of four-point function, respectively, see fig.~\ref{fig:Witten4pt}.

\subsection{Three-point function in the holographic gauge}
\label{sec:hgauge}

In this appendix, we obtain the coordinate dependence of three-point function \eqref{3pt} from the bulk theory.
We will see that the holographic gauge for higher-spin gauge fields
is suitable to make holomorphic factorization transparency. 
For a scalar field, we use the bulk to boundary propagator in the global coordinates:
\begin{align}
K_\Delta (y , z ; z') = C_\Delta^\text{AdS} \left( \frac{y}{y^2 + |z - z'|^2 }\right)^\Delta \, .
\end{align}
For a spin-$s$ field $\varphi_{\mu_1 \cdots \mu_s}$, we use a holographic gauge with $\varphi_{\mu_1 \cdots \mu_s} = 0$ if at least one of $\mu_l = y$. For two dimensions, only physical fields are $\varphi_{z \cdots z}$ and $\varphi_{\bar z \cdots  \bar z}$. 
Then the bulk to boundary propagator may be obtained as \cite{Das:2018ajg}
\begin{align}
K_{l,l} (y , z ; z') = C_{l}^\text{AdS} y^{-l}K_{l} (y , z ; z')  \, .
\end{align}

Let us assume the bulk coupling $g \bar \phi \nabla^{\mu_1} \cdots \nabla^{\mu_s}  \phi \varphi_{\mu_1 \cdots \mu_s}$.
In the holographic gauge, only the non-trivial contribution is $g ( \bar \phi (\nabla^{z} )^s \phi \varphi_{z \cdots z} + \phi (\nabla^{\bar z} )^s \phi \varphi_{\bar z \cdots \bar z})$. 
 Let us focus only on one term, which leads to
\begin{align}
A (z_1,z_2,z_3) = - g 
\int \frac{d w d \bar w d y}{y^{3}} y^{-l}K_{l,l}(y,w;z_1) y^{2l} \partial_{\bar w}^l K_{\Delta}(y,w;z_2) K_{\Delta}(y,w;z_3) \, .
\end{align}
As in \cite{Freedman:1998tz}, we utilize the conformal symmetry of CFT under $z \to z '  = -1/z$ and $\bar z \to \bar z ' = -1/\bar z$.
The symmetry corresponds to the AdS isometry associated with $y \to y ' =  y |z|^{-2}$.
The computation becomes simplified as
\begin{align}
\begin{aligned}
A (z_1,z_2,0) &= - g C_{\Delta}^\text{AdS}  \frac{1}{{\bar z}_1^{2l} |z_2|^{2\Delta}}  \int 
\frac{ d w ' d \bar w ' d y '}{(y ' )^{3}} K_{l,l}(y ',w ';z_1 ')  \partial_{\bar w '}^l K_{\Delta}(y ',w ';z_2 ' )  (y')^{\Delta + l}\, .
\end{aligned}
\end{align}
Here notice that
\begin{align}
\varphi_{z \cdots z} =\left(\frac{\partial z '}{ \partial z} \right)^l \varphi_{z ' \cdots z '} = \frac{1}{z^{2l}}  \varphi_{z ' \cdots z '}  \, .
\end{align}
According to (22) and (23) of  \cite{Freedman:1998tz}, we may conclude that the integral is proportional to $(\bar z_1' - \bar z_2 ')^{-l}$. Thus, we have
\begin{align}
A (z_1,z_2,0) \propto \frac{1 }{{\bar z_1}^{4l} |z_2|^{2\Delta}}  \frac{1}{(\bar z_1 ' - \bar z_2 ')^l} = \frac{\bar z_2^l}{\bar z_1 ^l(\bar z_2 - \bar z_1)^l }  \frac{1}{|z_2|^{2\Delta}} \, .
\end{align}
In order to compute the coefficient, we need to evaluate the integral. For this, we rewrite it as
\begin{align}
\begin{aligned}
  & \lim_{\epsilon \to 0} (-1)^l \partial_{\bar z_2 '}^l \int 
\frac{ d w ' d \bar w  ' d y ' }{(y ' )^{3}} K_{l,l}(y ' ,w ' ;z_1 ' )   K_{\Delta}(y ' ,w ' ;z_2 '  )  ( y ' )^{\Delta + l + 2 \epsilon} \\
& \quad =   \lim_{\epsilon \to 0} (-1)^l \partial_{\bar z_2 ' }^l C_{l}^\text{AdS} C_{\Delta}^\text{AdS}I[2\Delta +2 l + 2 \epsilon -3,l,\Delta,2] |\bar z_1 ' - \bar z_2 '|^{2 \epsilon} \, .
\end{aligned}
\end{align}
Here $I(a,b,c,d)$ is given by (23) of \cite{Freedman:1998tz} as
\begin{align}
\begin{aligned}
I[2 \Delta + 2 l +2 \epsilon -3,l,\Delta,2] 
&= \frac{\pi}{2} \frac{\Gamma (\Delta + l + \epsilon -1 ) \Gamma (- \epsilon)}{\Gamma (l)\Gamma(\Delta)} \frac{\Gamma (\Delta + \epsilon) \Gamma ( l + \epsilon )}{\Gamma(\Delta + l + 2 \epsilon)} \\
&= - \frac{1}{\epsilon}\frac{\pi}{2 (\Delta + l -1)} + \mathcal{O} (\epsilon^0) \, . 
\end{aligned}
\end{align}
In total, we find
\begin{align}
\begin{aligned}\label{AdS-3pt}
A (z_1,z_2,0) &= g (C_{\Delta}^\text{AdS})^2C_{l}^\text{AdS}  \frac{(-1)^{l+1} \Gamma(l)\pi}{2 (\Delta + l -1)}\frac{\bar z_2^l}{\bar z_1 ^l(\bar z_2 - \bar z_1)^l } \frac{1}{|z_2|^{2\Delta}} \, .
\end{aligned}
\end{align}

\subsection{Geodesic Witten diagrams and two-dimensional global conformal blocks}\label{Sec:hblock}
\paragraph{}
In \cite{Osborn:2012vt}, an expression for two-dimensional global conformal partial wave with arbitrary external twists was derived, which can be written as:
\begin{equation}\label{2dCPW}
W^{(s)}_{h, \bh}(z_i) = \frac{1}{\pi}\int d z_0 d \bz_0 \CF_{12}^{h\bh}(z_0, \bz_0) \CF_{34}^{1-h 1-\bh}(z_0, \bz_0)=
K_{h_i}^{(s)}(z_i) K_{\bh_i}^{(s)}(\bz_i) \hat{\CI}_{h, \bh}(\eta, \bar{\eta}) \, , 
\end{equation}
with
\begin{equation}\label{2d-3pt}
   \CF_{12}^{h\bh}(z_0, \bz_0) = \frac{1}{(z_{12}^2)^{\frac{h_1+h_2-h_0}{2}}(z_{10}^2)^{\frac{h_0+h_1-h_2}{2}}(z_{20}^2)^{\frac{h_0+h_2-h_1}{2}}} \frac{1}{(\bz_{12}^2)^{\frac{\bh_1+\bh_2-\bh_0}{2}}(\bz_{10}^2)^{\frac{\bh_0+\bh_1-\bh_2}{2}}(\bz_{20}^2)^{\frac{\bh_0+\bh_2-\bh_1}{2}}} 
\end{equation}
where the holomorphic and anti-holomorphic twists are related to the scaling dimensions $\{\Delta_i\}$ and spins $\{s_i\}$ via $h_i+\bh_i = \Del_i, h_i-\bh_i=s_i$ and $s_0+s_1+s_2 \in \mathbb{Z}$, while the overall kinematic factor $K_{h_i}^{(s)}(z)$ is given by:
\begin{equation}
K_{h_i}^{(s)}(z_i) = \frac{1}{(z_{12}^2)^{\frac{h_{12}^+}{2}}(z_{34}^2)^{\frac{h_{34}^+}{2}}} \left(\frac{z_{24}^2}{z_{14}^2}\right)^{h_{12}^-}  \left(\frac{z_{14}^2}{z_{13}^2}\right)^{h_{34}^-}, \quad h_{ij}^{\pm} =h_i\pm h_j \, ,
\end{equation}
similarly for $K_{\bh_i}(\bz_i)$.
The kinematic integral over product of three-point functions can be performed and yields:
\begin{align}
    \begin{aligned}\label{I-CPW}
 \hat{\CI}^{(s)}_{h, \bh}(\eta, \bar{\eta})
 &=\frac{\Gamma(1-2h)}{\Gamma(2\bh)}\frac{\Gamma(h-h_{12}^-)\Gamma(\bh+\bh_{12}^-)}{\Gamma(1-h-h_{12}^-)\Gamma(1-\bh+\bh_{12}^-)}   F_{h, \bh}^{(s)}(\eta, \bar{\eta}) \\
 & \quad + \frac{\Gamma(2h-1)}{\Gamma(2-2\bh)}\frac{\Gamma(1-h-h_{34}^-)\Gamma(1-\bh+\bh_{34}^-)}{\Gamma(h-h_{34}^-)\Gamma(\bh+\bh_{34}^-)}   F_{1-h, 1-\bh}^{(s)}(\eta, \bar{\eta})
 \end{aligned}
\end{align}
and 
\begin{equation}
F_{h, \bh}^{(s)}(\eta, \bar{\eta}) =  
\eta^{h}\, {}_2 F_1 \left( h-h_{12}^- , h+h_{34}^+;  2h  ; \eta \right ) 
\bar{\eta}^{\bh}\, {}_2 F_1 \left( \bh-\bh_{12}^- , \bh+\bh_{34}^+;  2\bh  ; \bar{\eta} \right ) \, . 
\end{equation}
We can then expand the four-point correlation function of primary operators $\varphi_{h_i,\bh_i }(z_i, \bz_i)$ in terms of resultant global conformal block:
\begin{equation}
\left \langle \prod_{i=1}^4 \varphi_{h_i,\bh_i }(z_i, \bz_i) \right \rangle =     K_{h_i}^{(s)}(z_i) K_{\bh_i}^{(s)}(\bz_i)\sum_{h, \bh} a_{h, \bh} F_{h, \bh}^{(s)}(z, \bz)  \, .
\end{equation}

If we now focus on the factorizable three-point function \eqref{2d-3pt} of scalar-scalar-higher spin current in two-dimensional CFTs such that $2h_{1,2} = 2\bar{h}_{1,2} =\Delta_{1,2}$, it can also be expressed in terms of the embedding formalism by choosing the following embedding coordinates in $\mathbb{M}^{1+3}$:
\begin{equation}
P_{i}^{A} = (P_i^{+}, P_i^{-}, P_i, \bar{P}_i) = (1, z_i \bz_i, z_i, \bz_i),\quad i =0, 1, 2    
\end{equation}
where we have chosen complex coordinates $(z_i, \bz_i)$ in the last two spatial embedding coordinates. 
For such coordinate choice, the polarization vector $Z_i$ satisfying transverse condition $P_i\cdot Z_i  =0$ in \eqref{bulk3pt} can be parameterized as:
\begin{equation}
Z^A_i = (0, w_i \bz_i+\bar{w}_i z_i, w_i, \bar{w}_i ) \, ,   
\end{equation}
where $(w_i, \bar{w}_i)$ parameterize the otherwise arbitrary two dimensional polarization vector.
Since it is only used to perform tensor index contraction, we can treat $w_i$ and $\bar{w}_i$ as independent complex variables, and for our case if we can make the following choice:
\begin{equation}\label{hol-gauge}
  Z^A_0 = \mathcal{Z}^A_0 = \left(1, -2\frac{\hat{z}_{12}z_0}{\hat{z}_{10}\hat{z}_{20}} , -2\frac{\hat{z}_{12}}{\hat{z}_{10}\hat{z}_{20}}, 0 \right) \, , \quad \hat{z}_{ij} = \frac{z_{ij}}{|z_{ij}|} \, , \quad i, j=0,1,2 \, . 
\end{equation}
The direct substitutions and simple calculations then shows that \eqref{2d-3pt} is indeed proportional to \eqref{bulk3pt} for $d=2$.

This choice of polarization vector also leads us to directly rewrite \eqref{2d-3pt} in terms of a single integral giving rise to geodesic Witten diagram. Starting with the scalar-scalar-scalar case, we have the following identity:
\begin{align}
\begin{aligned}
&\frac{1}{|z_{12}|^{{\Delta_1+\Delta_2-\Delta_0}} |z_{10}|^{{\Delta_0+\Delta_1-\Delta_2}} |z_{20}|^{{\Delta_0+\Delta_2-\Delta_1}} } \\
& \quad = \frac{2}{{\rm B}(\frac{\Delta_0+\Delta_1-\Delta_2}{2}, \frac{\Delta_0-\Delta_1+\Delta_2}{2})}\int^{\infty}_{-\infty} d\lambda \prod_{i=0}^2 \frac{1}{(-2P_i \cdot X(\lambda))^{\Delta_i}}
\end{aligned}
\end{align}
where $B(x, y)$ is the usual beta function and
\begin{equation}
    X^A(\lambda) = \frac{P_1^A e^{\lambda}+P_2^A e^{\lambda}}{(-2P_1\cdot P_2)^{\frac{1}{2}}} \, , \quad -\infty < \lambda < +\infty
\end{equation}
can be regarded geometrically as a AdS$_{3}$ bulk coordinate restricted to move along the geodesic connecting the boundary points $P_1$ and $P_2$.
We can next apply the differential identity given in equation (130) of \cite{Costa:2014kfa}:
\begin{equation}\label{3pt-sGWD}
\frac{1}{\left(\frac{\Delta_0+\Delta_1'-\Delta_2}{2}\right)_{s_0}} \frac{2}{{\rm B}(\frac{\Delta_0+\Delta_1'-\Delta_2}{2}, \frac{\Delta_0-\Delta_1'+\Delta_2}{2})}\int^{\infty}_{-\infty} d\lambda
(D_{01}\vert_{Z_0 =\mathcal{Z}_0})^{s_0}
\prod_{i=0}^2 \frac{1}{(-2P_i \cdot X(\lambda))^{\Delta_i}}\vert_{\Delta_1 \to \Delta_1'}
\end{equation}
where the differential operator is given by:
\begin{equation}
D_{01} = (P_1\cdot Z_0)\left(Z_0 \frac{\partial }{\partial Z_0}-P_0\frac{\partial}{\partial P_0}\right)+(P_1\cdot P_0)\left(Z_0 \frac{\partial}{\partial P_0}\right)    
\end{equation}
and we have made the replacements of the scaling dimension $\Delta_1 \to \Delta_1' = \Delta_1-s_0$ and the polarization vector $Z_0 \to \mathcal{Z}_0$ in $D_{01}$ after the differentiation. We can then identify the resultant geodesic integral with the three point scalar-scalar-spin interaction along the geodesic. 
We can next apply similar construction in \cite{Chen:2017yia} which utilizing the split representation of the AdS harmonic function reviewed earlier to construct the holographic dual configuration to global conformal partial wave \eqref{2dCPW}.


\providecommand{\href}[2]{#2}\begingroup\raggedright\endgroup

\end{document}